\documentclass[11pt]{article}
\usepackage{epsfig}
\usepackage{amsfonts,amsmath,amsthm,latexsym,amssymb}
\usepackage{graphicx}
\usepackage{subfigure}
\usepackage{hyperref,cite}
\usepackage[left=2.5cm,right=2.5cm,top=2cm,bottom=3cm,a4paper]{geometry}

\def\be{\begin{eqnarray}}\def\ba{\begin{eqnarray}}
\def\ee{\end{eqnarray}}\def\ea{\end{eqnarray}}
\def\ben{\begin{enumerate}}\def\bitem{\begin{itemize}}
\def\een{\end{enumerate}}\def\eitem{\end{itemize}}
\def\no{\nonumber\\}

\def\Tr{{\mbox{Tr}}}
\def\del{\partial}

\def\roughly#1{\mathrel{\raise.3ex\hbox{$#1$\kern-.75em%
\lower1ex\hbox{$\sim$}}}}

\def\vx{{\vec x}}

\def\A0{A_0}
\def\bq{\begin{equation}}
\def\eq{\end{equation}}
\def\la{\langle}\def\ra{\rangle}
\def\K0{K^0}

\def\half{\frac{1}{2}}

\def\lb{\label}
\def\eq#1{Eq. (\ref{#1})}

\def\ph{\phi}

\def\k{\kappa}

\def\pa{\partial}

\def\la{\langle}\def\ra{\rangle}
\def\lb{\left[}

\def\rb{\right]}

\def\Tr{{\rm Tr}\,}
\def\det{{\rm det}}

\def\no{\nonumber\\}
\def\CL{{\cal L}}
\def\CN{{\cal N}}

\def\CR{{\cal R}}
\def\CY{{\cal Y}}

\def\bfe{{\bf e}}
\newcommand{\tr}{\mbox{tr}}

\def\Tr{{\mbox{Tr}}}
\def\vx{{\vec x}}

\def\la{\langle}\def\ra{\rangle}

\makeatletter \@addtoreset{equation}{section}

\begin{document}
\begin{titlepage}
\begin{center}

\vskip 1.5cm

{\Large \bf Holography at work for nuclear and hadron physics }
\vskip 1. cm
  {\large Youngman Kim\footnote{e-mail: ykim@apctp.org}  and Deokhyun Yi\footnote{e-mail: dada@postech.ac.kr} }

\vskip 0.5cm
{\it Asia Pacific Center
for Theoretical Physics and Department of Physics, Pohang University
of Science and Technology, Pohang, Gyeongbuk 790-784, Korea}

\end{center}

\vskip 1cm
\vspace{1.0cm plus 0.5cm minus 0.5cm}

\begin{abstract}
The purpose of this review is to provide  basic ingredients of holographic QCD to non-experts in string theory
and to summarize its interesting achievements in nuclear and hadron physics.
We focus on results from a less stringy bottom-up approach and review a stringy top-down  model with some calculational details.
\end{abstract}

\end{titlepage}

\tableofcontents

\section{Introduction}

The approaches based on the Anti de Sitter/conformal field theory (AdS/CFT) correspondence
\cite{Maldacena:1997re,Gubser:1998bc, Witten:1998qj} find many interesting
possibilities to explore strongly interacting systems.
The discovery of D-branes in string theory \cite{Polchinski:1995mt} was a crucial  ingredient to put the correspondence on a firm footing.
Typical examples of the strongly interacting systems
are dense baryonic matter, stable/unstable nuclei, strongly interacting quark gluon plasma, and condensed matter systems.
The morale is to introduce an additional space, which roughly corresponds to the
energy scale of  4D boundary field theory, and try to construct a 5D
holographic dual model that captures certain non-perturbative aspects
of strongly coupled field theory, which are highly non-trivial to analyze in conventional quantum field theory based on perturbative techniques.
There are in general two different
routes to modeling holographic dual of quantum chromodynamics (QCD). One
way is a top-down approach based on stringy D-brane configurations. The other way is so-called a bottom-up approach to a holographic, in which a 5D
holographic dual is constructed from QCD. Despite the fact that this bottom-up
approach is somewhat  ad hoc, it reflects some important features of the
gauge/gravity duality and is rather successful in describing properties of hadrons.
However, we should keep in mind that a usual simple,
 tree-level analysis in the holographic dual model, both top-down and bottom-up,
is capturing the leading $N_c$ contributions, and we are bound to suffer from sub-leading
corrections.

The goal of this review is twofold.
First, we will assemble results mostly  from simple bottom-up models in nuclear and hadron physics.
Surely we cannot have them all here. We will devote to selected physical quantities discussed in the bottom-up model.
The selection of the topics is based on authors' personal bias.
Second, we present some basic materials that might be useful to understand some aspects of AdS/CFT and D-brane models.
We will focus on the role of the AdS/CFT in low energy QCD.
Although the correspondence between QCD and gravity theory is not known,
we can obtain much insights on QCD  by the gauge/gravity duality.

We organize this review as follows.
Section \ref{Sec:AdS/CFT}  reviews the gauge/gravity.
Section \ref{Sec:hQCD} briefly discuss developments of holographic QCD and demonstrates how to build up a bottom-up model using the AdS/CFT dictionary.
After discussing the gauge/gravity duality and modeling in the bottom-up approach,
we proceed with selected physical quantities. In each section, we show results mostly from the bottom-up approach and list some from the top-down model.
Section \ref{vacuumS} deals with vacuum condensates of QCD in holographic QCD. We will mainly discuss the gluon condensate and the quark-gluon mixed condensate.
Section \ref{Sec:SpecForm} collects some results on hadron spectroscopy and form factors from the bottom-up model. Contents are glueballs, light mesons, heavy quarkonium, and hadron form-factors.
Section \ref{Sec:QCDphase} is about QCD at finite temperature and density. We consider QCD phase transition and dense matter.
Section \ref{Sec:remarks} is devoted to some general remarks on holographic QCD and to list
 a few topics that are not discussed properly in this article.
Due to our limited knowledge, we are not able to cover all interesting works done in holographic QCD.
To compensate this defect partially, we will list some recent review articles on holographic QCD.

In Appendix, we look back on some basic materials that might be useful for non-experts in string theory to work in holographic QCD.
\ref{Sec:MvsO}: we review  the relation between the bulk mass and boundary operator dimension.
\ref{D3/D7}: we present a  D3/D7 model and axial U(1) symmetry in the model.
\ref{SSmodel}: we discuss non-Abelian chiral symmetry based on D4/D8/$\overline{\rm D8}$ model.
\ref{BH_how_to}\&\ref{BH}: we describe how to calculate the Hawking temperature of an AdS black hole.
\ref{HP_app}:  we encapsulate the Hawking-Page transition and sketch how to calculate
Polyakov loop expectation value in thermal AdS and AdS black hole.

We close this section with a cautionary remark.
Though it is tempting to argue that holographic QCD is dual to real QCD,
what we mean by QCD here might be mostly QCD-like or a cousin of QCD.

\section{Introduction to the AdS/CFT correspondence \label{Sec:AdS/CFT}}

The AdS/CFT correspondence, first suggested by Maldacena~\cite{Maldacena:1997re},
is a duality between gravity theory in  anti de Sitter space (AdS) background and
conformal field theory (CFT).
The original conjecture states that there is a correspondence between
a weakly coupled gravity theory (type IIB string theory) on $AdS_5\times S^5$
and the strongly coupled ${\cal N} = 4$ supersymmetric Yang-Mills theory
on the four-dimensional boundary of $AdS_5$.
The strings reside in a higher-dimensional curved spacetime and
there exists some well-defined  mapping between the objects
in the gravity side and and the dual objects in the four-dimensional gauge theory.
Thus,  the conjecture  allows the use of non-perturbative methods
for  strongly coupled theory through its gravity dual.

\subsection{D$p$ brane dynamics}
\begin{figure}[!htb]
\centering
\mbox{%
\subfigure[D3 branes sweep $(3\!+\!1)$ dimensions in $(9\!+\!1)$ space time.]{%
\includegraphics[width=5.5cm]{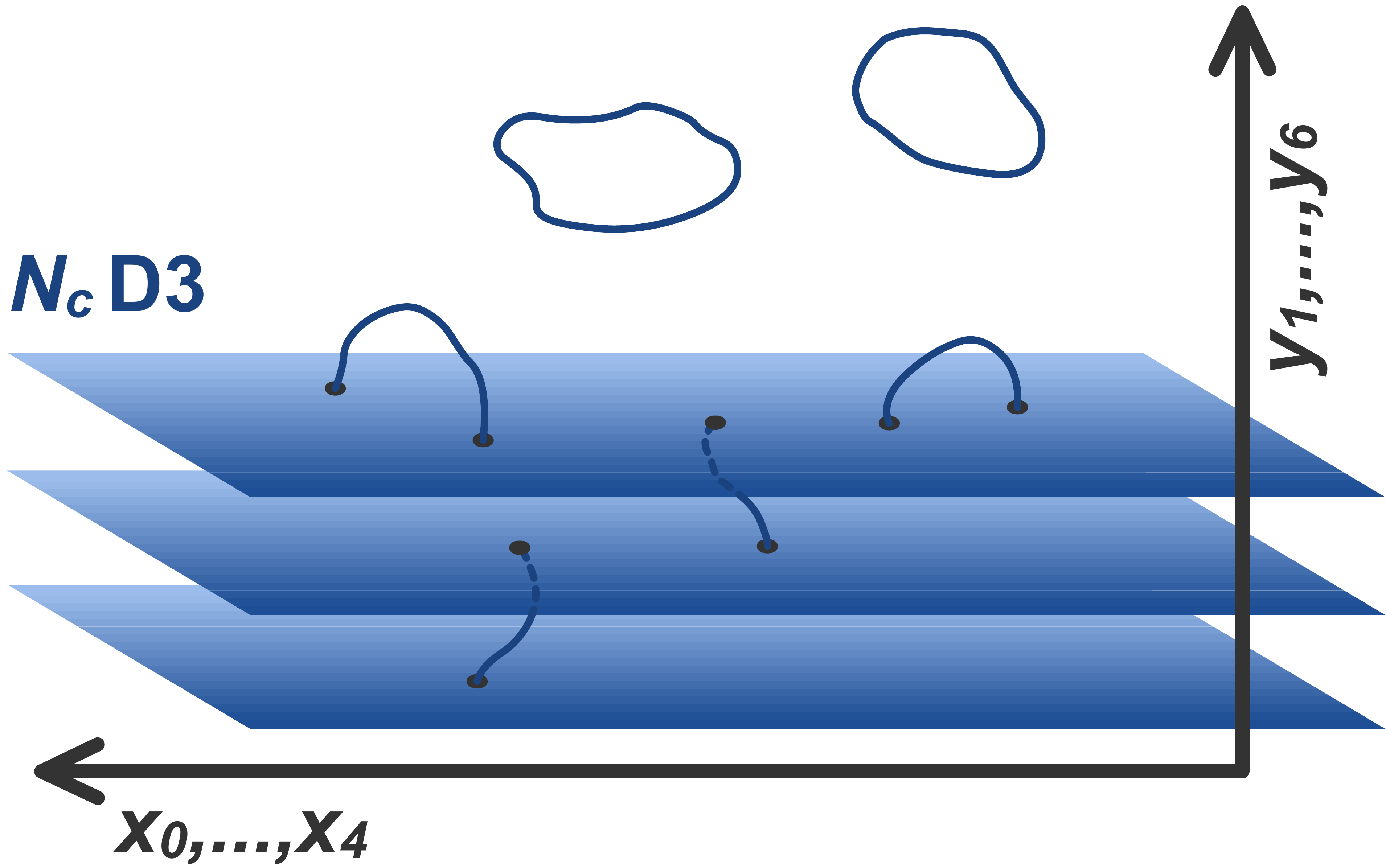}
\label{fig:NcD3}
}\hspace{1.2cm}
\subfigure[$N_c\!=\!3$ stack of D3 branes and all the possible classes of open strings.]{%
\includegraphics[width=8cm]{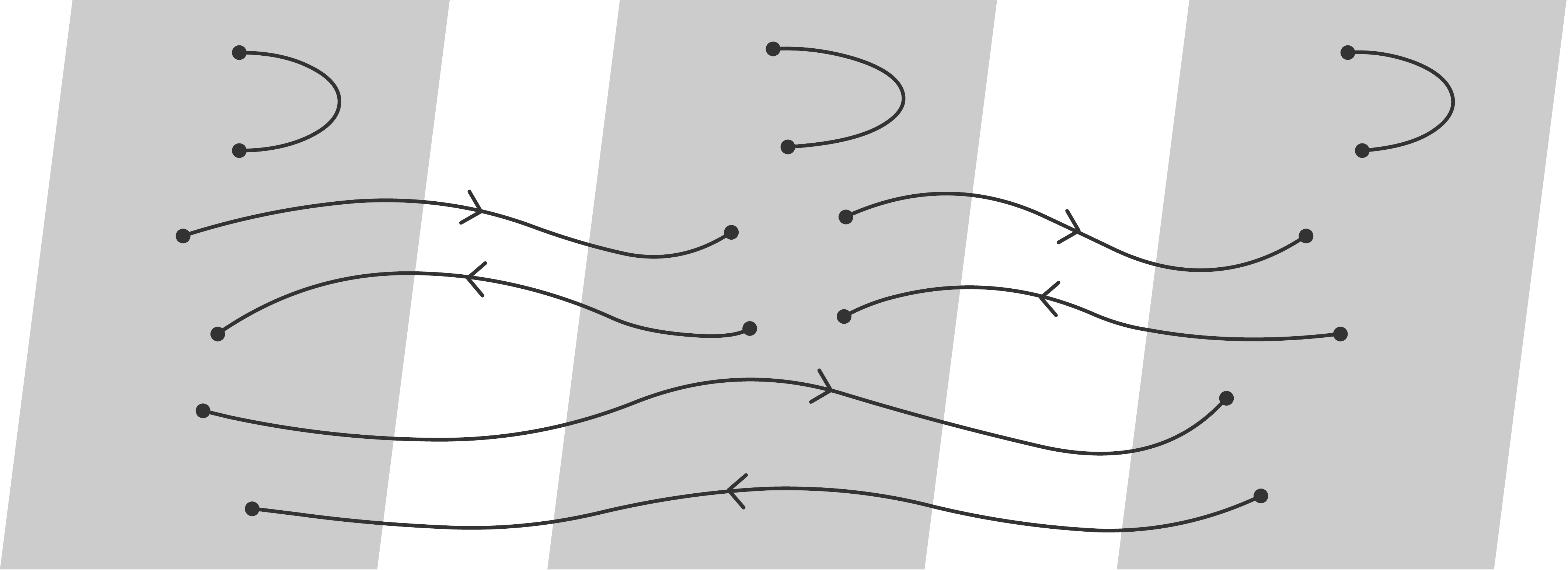}
\label{fig:U(3)}
}}
\caption{The configurations of $N_c$ stack of D3 branes in $10d$ spacetime.}
\end{figure}

The duality emerges from a careful consideration of the D-brane dynamics.
A D$p$ brane sweeps out $(p+1)$ world-volume in spacetime. Introducing D branes gives
open string modes whose endpoints lie on the D branes and the open string spectrum
consists of a finite number of massless modes and also an infinite tower of massive modes.
The open string end points can move only in the parallel $(p+1)$ directions of the brane,
see Figure~\ref{fig:NcD3}, and a D$p$ brane can be seen as a point along its transverse directions.
The dynamics of the D$p$ brane is described by the Dirac-Born-Infeld (DBI) action \cite{Leigh:1989jq} and Chern-Simons term,
\begin{equation}\label{eq:DpDBI}
S_{D_p} = -T_p\int d^{p+1}x\,e^{-\phi}
\sqrt{-\det\left(P[g]_{ab}+2\pi\alpha' F_{ab}\right)}
+  S_{CS}
\end{equation}
with a dilaton $e^{-\phi}$. Here $g_{ab}$ is the induced metric on $D_p$.
$P$ denotes the pullback and $F_{ab} $ is the world-volume field strength.
$T_{p}$ is the tension of the brane which has the form
\begin{equation}\label{eq:Dptension}
T_{p}=\frac{1}{(2\pi)^pg_sl_s^{p+1}}=\frac{1}{(2\pi)^pg_s \alpha'^{(p+1)/2}}
\end{equation}
and it is the mass per unit spatial volume.
Here $g_s$ is the string coupling and $l_s$ is the string length.
$\alpha'$ is the Regge slope parameter and related to the string length scale as $l_s=\sqrt{\alpha'}$.
In general,  states in the closed string spectrum contain a finite number of massless modes and an
infinite tower of massive modes with masses of order $m_s=l_s^{-1}=\alpha'^{-1/2}$.
Thus, at low energies $E\ll m_s$, the higher order corrections come in powers of $\alpha'E^2$
from integrating out the massive string modes.
If there are a $N_c$ stack of multiple D branes, the open strings between different branes
give a non-Abelian $U(N_c)$ gauge group, see Figure~\ref{fig:U(3)}.
In the low energy limit, we can integrate out the massive modes to obtain non-Abelian gauge theory
of the massless fields.

Now, we take $p=3$ and  consider $N_c$ D3 brane stacks in type IIB theory.
The low energy effective action of this configuration gives a non-Abelian gauge theory
with $U(N_c)$ gauge group.
In addition, this gauge group can be factorized into $U(N_c)=SU(N_c)\times U(1)$
and the $U(1)$ part, which describes the center of mass motion
of the D3 branes, can be decoupled by the global translational invariance.
The remaining subgroup $SU(N_c)$ describes the dynamics of branes from each other.
Therefore we see that in the low energy limit, the massless open string modes on $N_c$ stacks of D3-branes
constitute ${\cal N}=4$ $SU(N_c)$ Yang-Mills theory \cite{Witten:1995im} with 16 supercharges in $(3+1)$ spacetime.
From (\ref{eq:DpDBI}) for $p=3$,  we obtain the effective Lagrangian
at low energies  up to the two derivative order
\begin{equation}\label{eq:SYM}
{\cal L}=-\frac{1}{4\pi g_s}\Tr\left(
\frac{1}{4}F_{\mu\nu}F^{\mu\nu}+\frac{1}{2}D_\mu\phi^iD^\mu\phi^i-\frac{1}{4}\left[\phi^i,\phi^j\right]^2
+\frac{i}{2}\bar{\Psi}^I \Gamma^\mu D_\mu \Psi_I-\frac{i}{2}\bar{\Psi}^I\Gamma^i\left[\phi_i,\Psi_I\right]
\right)
\end{equation}
with a gauge field $A_\mu$, six scalar fields $\phi^i$ and four Weyl fermions $\Psi^I$.

In fact, the original system also contains closed string states.
The higher order derivative corrections for the Lagrangian (\ref{eq:SYM}) come both in powers of $\alpha'E^2$
from the massive modes and powers of the string coupling $g_sN_c$ for loop corrections.
It is known that the string coupling constant $g_s$ is related by the 10-dimensional gravity constant as
$G^{(10)}\sim g_s^2l_s^8$ and thus the dimensionless string coupling is of order $G^{(10)}E^8$, which is negligible
in the low energy limit. Therefore, at low energies closed strings are decoupled from
open strings and the physics on the $N_c$ D3 branes is described by the massless
${\cal N}=4$ super Yang-Mills theory with gauge group $SU(N_c)$.

\subsection{$AdS_5\times S^5$ geometry}

Now we view the same system from a different angle.
Since D branes are massive and carry energy and Ramond-Ramond (RR) charge,
 $N_c$ D3 branes deforms the spacetime around them to make a curved geometry.
Note that the total mass of a D3 brane is infinite because it occupies the infinite
world-volume of its transverse directions, but the tension, or the mass per unit three-volume of the D3 brane
\begin{equation}\label{eq:D3tension}
T_{3}=1/(2\pi)^3g_sl_s^4
\end{equation}
 is finite.

\begin{figure}[!htb]
\centering
\mbox{%
\subfigure[Flat spacetime for $r\gg R$.]{%
\includegraphics[width=7.5cm]{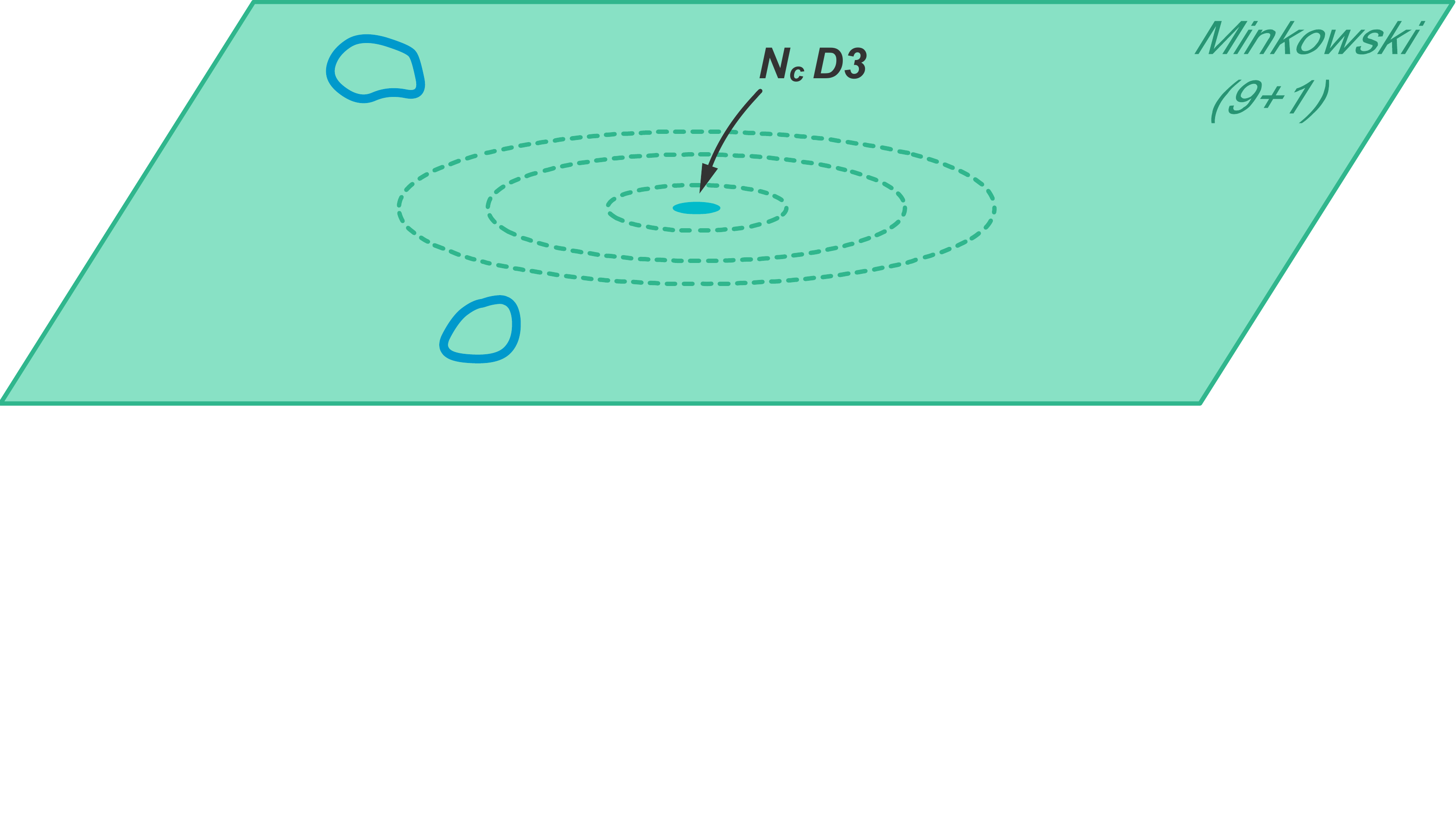}
\label{fig:D3AdSf}
}\hspace{.5cm}
\subfigure[$N_c$ D3 branes deform the spacetime.]{%
\includegraphics[width=7cm]{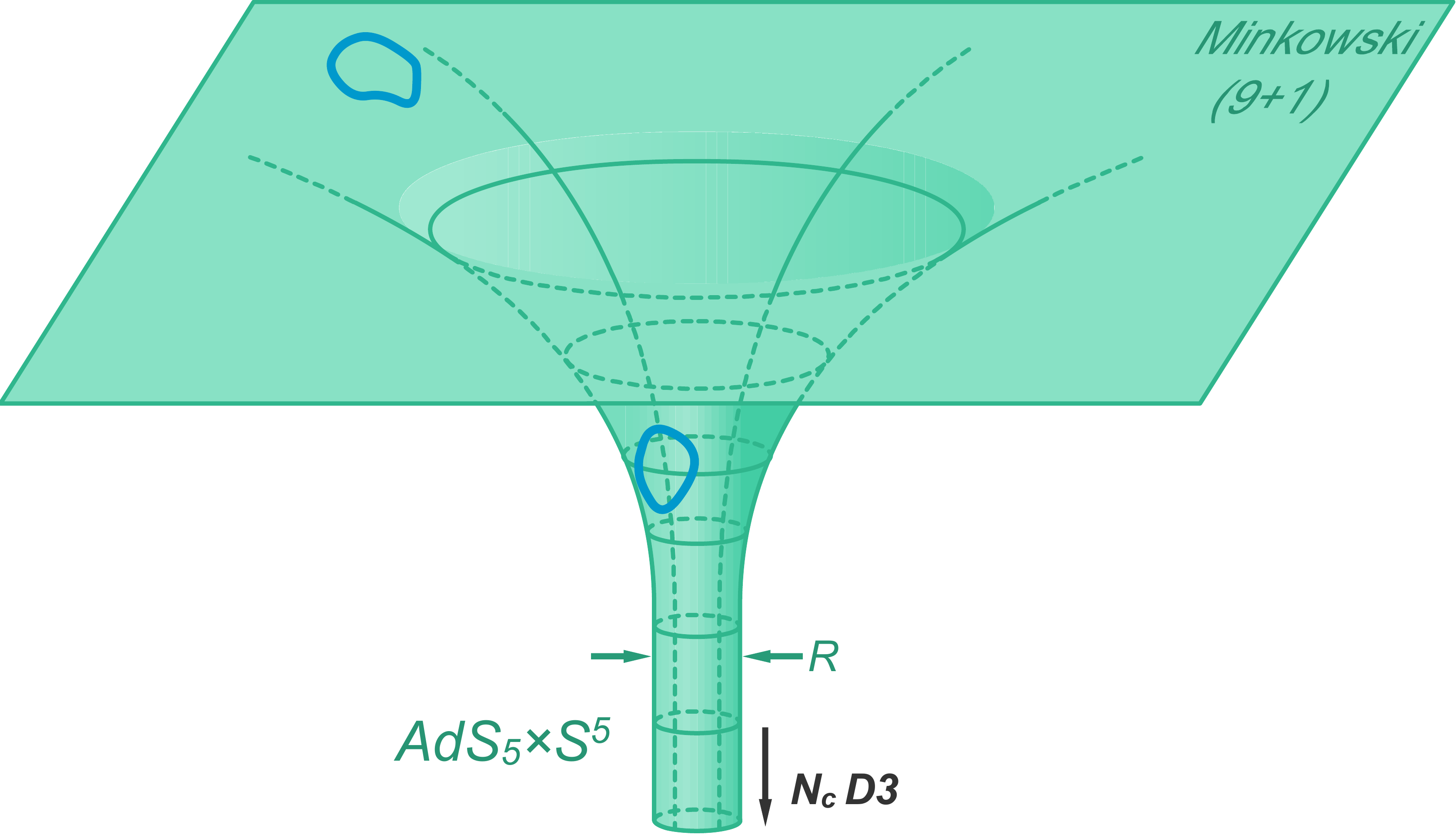}
\label{fig:figD3AdS}
}}
\caption{The two descriptions of the $N_c$ D3 configuration.}
\end{figure}

In the flat spacetime, the circumference of the circle surrounding an origin
at a distance $r$ is $2\pi r$, and it simply shrinks to zero if one approaches the origin.
But if there is a stack of D3 branes, it deforms the spacetime and makes throat geometry along its transverse directions.
Thus, near the D3 branes, the radius of a circle around the stack approaches a constant $R$,
 an asymptotical infinite cylinder structure, or $AdS_5\times S^5$, see Figure~\ref{fig:figD3AdS}.
 The $N_c$ D3 brane stack is located at the infinite end of the throat
and this infinite end is called the ``horizon''.
In the near horizon geometry,
a D3 brane is surrounded by a five-dimensional sphere $S^5$.

To be more specific, let us start with type IIB string theory for $p=3$.
We find a black hole type solution which is carrying charges with respect to the RR four-form potential.
The theory has magnetically charged D3 branes, which are electrically charged under the potential $dA_4$
and  it is self-dual $*F_5=F_5$.
The low energy effective action is
\begin{equation}
S=\frac{1}{(2\pi)^7l_s^8}\int d^{10}x \sqrt{-g}\left(e^{-2\phi}(R+4(\nabla\varphi)^2)-
\frac{2}{5!}F^2_5 \right)\, .
\end{equation}
We assume that the metric is spherically symmetric in seven-dimensions with the RR source at the origin,
then the $N_c$ parameter appears in terms of the five-form field RR-field strength on the five-sphere as
\begin{equation}
\int_{S^5}\, *F_5 = N_c
\end{equation}
where $S^5$ is the five-sphere surrounding the source for a four-form field $C_4$.
Now by using the Euclidean symmetry we get the curved metric solution
\cite{Gibbons:1987ps,Garfinkle:1990qj,Horowitz:1991cd} for the D3 brane
\begin{equation}
ds^2 = f(r)^{-1/2}\eta^{\mu\nu}dx_\mu dx_\nu
+ f(r)^{1/2} \left( dr^2 + r^2\, d\Omega_5^2\right)\, ,
\end{equation}
where
\begin{equation}
f(r)=1 + \frac{R^4}{r^4}
\end{equation}
with the radius of the horizon $R$
\begin{equation}\label{eq:R}
R^2=\sqrt{4\pi g_sN_c}\alpha'=\sqrt{4\pi g_sN_c}l_s^2\, .
\end{equation}
$d\Omega_5$ is the five-sphere metric.
For $r\gg R$ we have $f(r)\simeq 1$ and the spacetime becomes flat
with a small correction $R^4/r^4 = 4\pi g_s N_cl_s^4/r^4$. This factor can be interpreted as
a gravitational potential since $G^{(10)}\sim g_s^2l_s^8$ and $M_{D3}\sim N_cT_{3}\sim N_c/g_sl_s^4$
thus $R^4/r^4 \sim G M_{D3}/r^4$. In the near horizon limit, this gravitational effect become strong
and the metric changes into
\begin{equation}\label{eq:AdS5S5}
ds^2 = \frac{r^2}{R^2}\eta^{\mu\nu}dx_\mu dx_\nu
+ \frac{R^2}{r^2}\left(dr^2 +r^2d\Omega_5^2\right).
\end{equation}
This is $AdS_5\times S^5$.

The geometry by the D3 branes is  sketched in Figure~\ref{fig:figD3AdS}.
Far away from the D3 brane stacks, the spacetime is flat $(9+1)$ dimensional Minkowski spacetime
and the only modes which survive in the low energy limit are the  massless closed string (graviton) multiplets, and
they decouple from each other due to  weak interactions. On the other hand, close to the D3 branes,
the geometry takes the form $AdS_5 \times S^5$ and the whole tower of massive modes exists there.
This is because the excitations seen from an observer at infinity are close to the horizon and
a closed string mode in a throat should go over a gravitational potential to meet the asymptotic flat region.
Therefore, as we focus the lower energy limit, the excitation modes should be originated deeper in the throat,
and then they decouple from the ones in the flat region. Thus in the low energy limit the interacting sector
lives in $AdS_5\times S^5$ geometry.

\subsection{The gauge/gravity duality}
So far,
we have considered two seemingly different descriptions of the $N_c$ D3 brane configuration. As we mentioned,
each of the D3 branes carries the gravitational degrees of freedom in terms of its tension,
or the string coupling $g_s$ as in (\ref{eq:D3tension}). So the  strength of
the gravity effect due to $N_c$ stacks of D3 branes depends on the parameter $g_sN_c$.

If $g_s N_c \ll 1$,  from (\ref{eq:R}) we see that $R\ll l_s$ and therefore the throat geometry
effect is less than string length scale. Thus the spacetime is nearly flat and
the fluctuations of the D3 branes are described by open string states.
In this regime the string coupling $g_s$ is small and the closed strings are decoupled from the open strings.
Here the closed string description is inapplicable since one needs to know
about the geometry below the string length scale.
If we take the low energy limit, the effective  theory,
which describes the open string modes, is  ${\cal N}=4$ super Yang-Mills theory
with $SU(N_c)$ gauge group.

On the other hand, if $g_s N_c \gg 1$, then the back-reaction of the branes on the background
becomes important and spacetime will be curved.
In this limit the closed string description reduces to classical gravity
which is  supergravity theory in the near horizon geometry.
Here the open string description is not feasible because $g_sN_c$ is related with
the loop corrections and one has to deal with the strongly coupled open strings.
Again, if we take the low energy limit, the interaction is described by
the type IIB string theory in the near-horizon geometry, $AdS_5\times S^5$.

\begin{figure}[!ht]
\begin{center}
  \includegraphics[width=11cm]{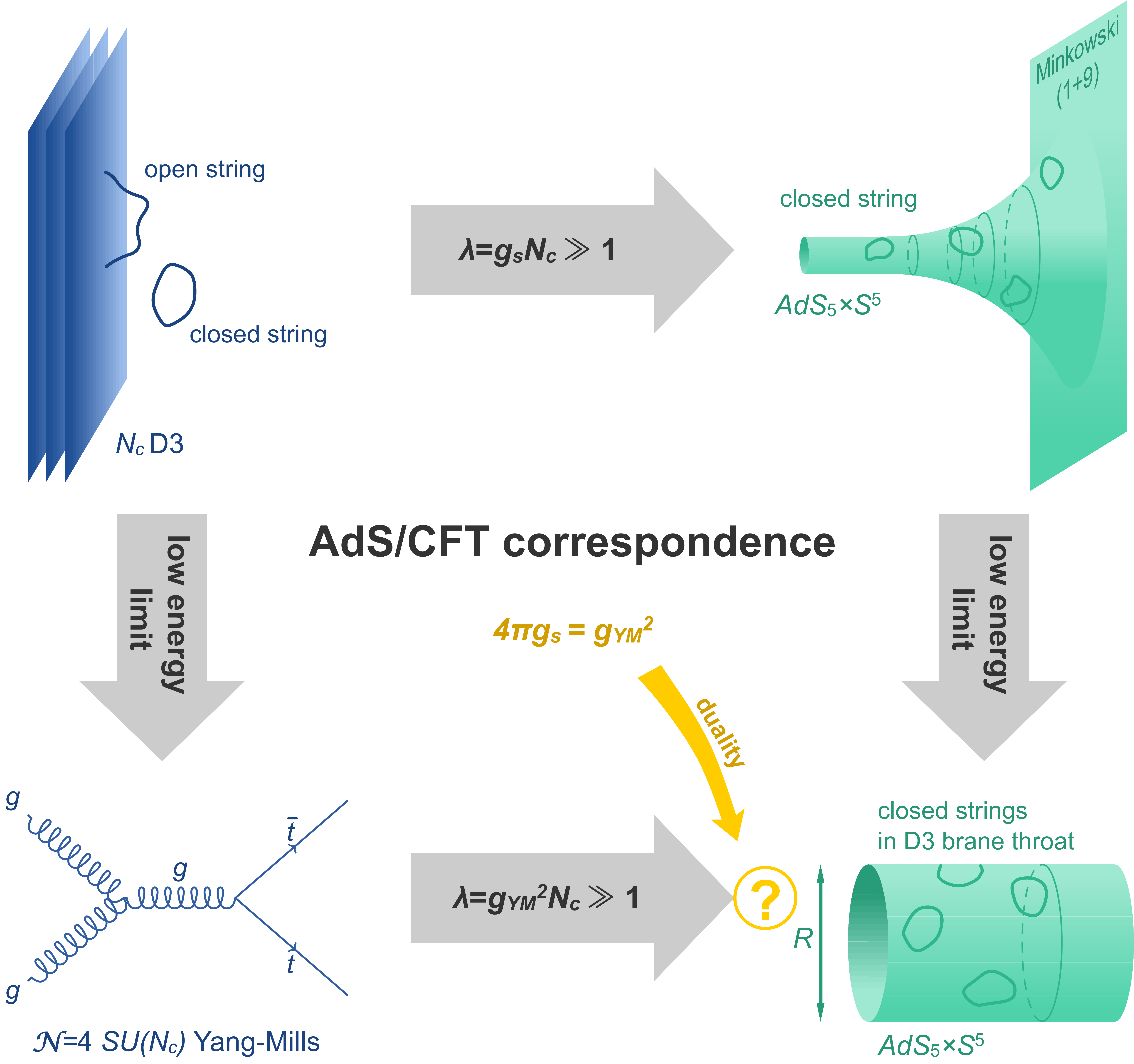}
  \caption{The sketch of the AdS/CFT correspondence.}\label{fig:AdSCFT}
\end{center}
\end{figure}

The gauge/gravity correspondence is nothing but the conjecture connecting these
two descriptions of $N_c$ D3 branes in the low energy limit.
It is a duality between the ${\cal N}=4$ super Yang-Mills theory with gauge group $SU(N_c)$ and
the type IIB closed string theory in $AdS_5\times S^5$, see Figure~\ref{fig:AdSCFT}.
The relation between the Yang-Mills coupling $g_{YM}^{}$
and the string coupling strength $g_s$ is given by
\begin{equation}
g_{YM}^2=4\pi g_s,\qquad (R/l_s)^4=4\pi g_sN_c
\end{equation}
Then, the 't~Hooft coupling $\lambda=g_{YM}^2N_c$ can be expressed in terms of
the string length scale
\begin{equation}
\lambda = R^4/l_s^4.
\end{equation}
Therefore, the dependence of $g_sN_c$ becomes the question of whether
the 't~Hooft coupling is large or small, or the gauge theory is
strongly or weakly coupled.

The two  descriptions can be viewed as two extremes of $r$.
For the sake of convenience, we use the coordinate $z=R^2/r$. Then, the $AdS_5\times S^5$ metric (\ref{eq:AdS5S5}) becomes
\begin{equation}\label{eq:AdS5S5 with z}
ds^2 = \frac{R^2}{z^2}\left(\eta^{\mu\nu}dx_\mu dx_\nu +dz^2\right)+R^2d\Omega_5^2,
\end{equation}
which shows the conformal equivalence between $AdS_5$ and flat spacetime more clearly.
In (\ref{eq:AdS5S5 with z}), each $z$-slice of $AdS_5$ is isometric to four-dimensional Minkowski spacetime.
In this coordinate, $z= 0$ is the boundary of $AdS_5$, where Yang-Mills theory lives, with identifying $x^\mu$ as the coordinates of the gauge theory.
If $z\rightarrow \infty$,  the determinant of the metric goes to zero and it is the Poincar{\' e} horizon.
Here the factor $R^2/z^2$ also has some relation with the energy scales.
If the gauge theory side has a certain energy scale $E$,  the corresponding energy in the gravity side
is $(z/R)E$.
In order words, a gauge theory object with an energy scale $E$ is involved with a bulk side one
localized in the z-direction at $z\sim 1/E$ \cite{Maldacena:1997re,Susskind:1998dq,Peet:1998wn}.
Therefore, the UV or high energy limit corresponds to $z\rightarrow 0$ (or $r\rightarrow \infty$)
and the IR or low energy limit corresponds to $z\rightarrow \infty$ (or $r\rightarrow 0$).

The operator-field correspondence between operators in the four-dimensional gauge theory
and corresponding dual fields in the gravity side was given in \cite{Gubser:1998bc,Witten:1998qj}.
Then the AdS/CFT correspondence can be stated as follows,
\begin{equation}\label{eq:GKP-W}
\left\langle T e^{\int \! d^4 x \, \phi_0(x) {\cal O}(x)} \right\rangle_{{\rm CFT}}
= Z_{{\rm sugra}}
\end{equation}
where  $\phi_0(x)=\phi(x,u\rightarrow\infty)$
and the string theory partition function $Z_{{\rm sugra}}$ at the boundary specified by $\phi_0$ has the form
\begin{equation}
Z_{{\rm sugra}}=e^{ -S_{{\rm sugra}}\left(\phi(x,u)\right)}\Big|_{u\rightarrow\infty}\, .
\end{equation}
The relation (\ref{eq:GKP-W}) implies that the generating functional of
gauge-invariant operators in CFT can be matched with the generating functional
for tree diagrams in supergravity.

\section{Holographic QCD\label{Sec:hQCD}}
Ever since the advent of the AdS/CFT correspondence, there have been many efforts,
based on the correspondence, to study non-perturbative physics
of strongly coupled gauge theories in general and QCD in particular.

Witten proposed \cite{Witten:1998zw} that we can
extend the correspondence to non-supersymmetric theories by considering the AdS black hole and showed that
this supergravity treatment qualitatively well describes strong coupled QCD (or QCD-like) at finite temperature:
for instance,  the area law behavior of Wilson
loops, confinement/deconfinement transition of  pure gauge theory through the Hawking-Page transition, and the mass gap for glueball states.
In \cite{Klebanov:1999tb} symmetry breaking by expectation values of scalar fields were analyzed in the context of the AdS/CFT correspondence,
which is essential to encode the spontaneous breaking of chiral symmetry in a holographic QCD model.
Regular supergravity backgrounds with less supersymmetries corresponding to
dual confining ${\cal N} = 1$  super Yang-Mills theories were proposed in \cite{Klebanov:2000hb, Maldacena:2000yy}.
It has been shown by Polchinski and Strassler~\cite{Polchinski:2001tt}
 that the scaling of high energy
QCD scattering amplitudes can be obtained from a gravity dual description in a sliced AdS geometry whose
IR cutoff is determined by  the mass of the lightest glueball.
Important progress towards flavor physics of QCD has been made by adding flavor degrees of freedom in the fundamental
representation of a gauge group to the gravity dual description \cite{Karch:2002sh}.
Chiral symmetry breaking and meson spectra were studied in a nonsupersymmetric gravity model dual to
large $N_c$ nonsupersymmetric gauge theories
\cite{Babington:2003vm}, where flavor quarks are introduced by a D7-brane probe on deformed AdS backgrounds.
Using a D4/D6 brane configuration,
the authors of \cite{Kruczenski:2003uq} explored the meson phenomenology of large $N_c$ QCD
 together with U(1)$_{\rm A}$ chiral symmetry breaking. They showed that the chiral condensate scales as $1/m_q$ for large $m_q$.
A remarkable observation made in \cite{Kruczenski:2003uq} is that in addition to the confinement/deconfinement phase transition
the model exhibits a possibility that another transition set by $T_{fund}$ could happen in deconfined phase, $T>T_{deconf}$,
where $T_{deconf}=M_{KK}/(2\pi)$. Since the value of $M_{KK}$ is around $1$ GeV, we can estimate $T_{deconf}\sim 160$ MeV.
In this case for $T_{deconf}<T<T_{fund}$ there exist free unbound quarks and meson bound sates of heavy quarks and above $T_{fund}$
the meson states dissociate into free quarks, which  in some sense mimics the dissociation of heavy quarkonium in quark-gluon plasma (QGP).
However, we should note that meson bound
states in Dp/Dq systems are deeply bound, while the heavy quarkonia in QCD are shallow bound states.
In this sense the bound state that disappears above $T_{fund}$ could be that of strange quarks rather than charmonium or bottomonium  \cite{Mateos:2007vn}.

To attain a realistic gravity dual description of (large $N_c$) QCD, non-Abelian chiral symmetry is an essential  ingredient
together with confinement.
Holographic QCD models, which are equipped with the correct structure for the problem, namely, chiral symmetry and confinement,
have suggested in  top-down  and bottom-up approaches.
They found to be rather successful for various hadronic observables and for certain processes dominated by large $N_c$.
Based on a D4/D8/$\overline{\rm D8}$ model, Sakai and Sugimoto
studied hadron phenomenology in the chiral limit $m_q=0$,
and the chiral symmetry breaking  geometrically \cite{Sakai:2004cn, Sakai:2005yt}.
More phenomenological holographic QCD models were proposed~\cite{Erlich:2005qh, Da_Rold:2005zs, Hirn:2005vk}.
In~\cite{Erlich:2005qh, Da_Rold:2005zs}, chiral symmetry breaking is realized
by a non-zero chiral condensate whose value is fitted to meson data from experiments.
Hadronic spectra and light-front wave functions were studied in \cite{Brodsky:2006uqa} based on the ``Light-Front Holography''
which maps amplitudes in extra dimension to a Lorentz invariant impact separation variable $\zeta$ in Minkowski space at
fixed light-front time. Light-Front Holography has led to many  successful applications in hadron physics
including light-quark hadron spectra, meson and baryon form factors, the nonperturbative QCD coupling, light-front wave-functions,
see \cite{Brodsky:2010px, deTeramond:2011yi, Brodsky:2011sk} for a review on this topic.
In \cite{Afonin:2010fr}, a relation between a bottom-up holographic QCD model
and QCD sum rules was analyzed.

\vskip 0.3cm
Now, we demonstrate how to construct a bottom-up holographic QCD model by looking at a low-energy QCD.
For illustration purposes, we compare our approach with the (gauged) linear sigma model.
The D3/D7 model is summarized in Appendix \ref{D3/D7} with some calculational details. For a review of the linear sigma model, we refer to \cite{Koch:1997ei}.
Some material in this section is taken from \cite{Kim:2010cpod}.
Suppose we are interested in two-flavor QCD at low energy, roughly below $1$ GeV.
In this regime usually we resort to the effective models or theories of QCD
for {\it analytic} studies since the QCD lagrangian does not help much.

To construct the holographic QCD model dual to two flavor low-energy QCD with chiral symmetry,
we first choose relevant fields.
To do this, we consider composites of quark fields that have the same quantum numbers with the
hadrons of interest.
For instance, in the linear sigma model we introduce pion-like and sigma-like fields: $\vec\pi \sim \bar q \vec\tau \gamma^5 q$ and
 $\sigma\sim \bar qq$, where $\vec\tau$ is the Pauli matrix for isospin.
In the AdS/CFT dictionary, this procedure may be dubbed operator/field correspondence: one-to-one mapping between gauge-invariant local
operators in gauge theory and bulk fields in  gravity sides. Then we introduce
\ba
&& \bar q_L\gamma^\mu t^a q_L \leftrightarrow A_{L\mu}^a (x,z)\, ,\no
&& \bar q_R\gamma^\mu t^a q_R \leftrightarrow A_{R\mu}^a (x,z)\, ,\no
&& \bar q_R^\alpha q_L^\beta \leftrightarrow (2/z) X^{\alpha\beta} (x,z)\, .
\ea
An interesting point here is that the 5D mass of the bulk field is not
 a free parameter of the model.
 This bulk mass is determined by the dimension $\Delta$ and spin $p$ of the dual 4D operator in AdS$_{d+1}$. For instance, consider a bulk field $X(x,z)$
dual to  $\bar q(x) q(x)$. The bulk mass of $X(x,z)$ is given by $m_X^2=(\Delta-p)(\Delta+p-d)$ with $\Delta=3$, $p=0$ and $d=4$, and so $m_X^2=-3$.
For more details, see Appendix \ref{Sec:MvsO}.

To write down the Lagrangian of the linear sigma model, we consider (global) chiral symmetry of QCD.
Since the mass of light quark $\sim 10$ MeV is negligible compared to  the QCD scale $\Lambda_{\rm QCD}\sim 200$ MeV,
we may consider the exact chiral symmetry of QCD and treat quark mass effect in a perturbative way.
Under the axial transformation, $q\rightarrow e^{-i\gamma_5 \vec\tau\cdot\vec\theta/2} q$,  the pion-like and sigma-like states
transform as $\vec\pi\rightarrow \vec\pi +\vec\theta\sigma$ and $\sigma\rightarrow \sigma-\vec\theta\cdot\vec\pi$. From this,
we can obtain terms that respect chiral symmetry such as $\vec\pi^2+\sigma^2$.
Similarly we ask the holographic QCD model to respect chiral symmetry of QCD.
In AdS/CFT, however, a global symmetry in gauge theory corresponds local symmetry in the bulk, and therefore the corresponding
holographic QCD  model should posses local chiral symmetry. This way vector and axial-vector fields naturally fit into chiral Lagrangian in the bulk
as the gauge boson of the local chiral symmetry.

We keep the chiral symmetry in the Lagrangian since it will be spontaneously broken.
Then we should ask how to realize the spontaneous chiral symmetry breaking.
In the linear sigma model, we have a potential term like  $((\vec\pi^2+\sigma^2)-c^2)^2$ that leads to spontaneous chiral symmetry
breaking due to a nonzero vacuum expectation value of the scalar field $\sigma$, $\la\sigma\ra=c$.
In this case the explicit chiral symmetry due to the small quark mass could be mimicked by adding a term $-\epsilon \sigma$ to the potential which
induces a finite mass of the pion, $m_\pi^2\sim \epsilon/c $.
In a holographic QCD model, the chiral symmetry breaking is encoded in the vacuum expectation value of a bulk scalar field
dual to $\bar q q$. For instance in the hard wall model~\cite{Erlich:2005qh,Da_Rold:2005zs}, it is given by $\la X\ra=m_q z+ \zeta z^3$, where $m_q$ and $\zeta$ are proportional to the quark mass and the chiral condensate in QCD. In the D3/D7 model,  chiral symmetry breaking can be realized by
the embedding solution as shown in Appendix \ref{D3/D7}.

The last step to get to the gravity dual to two flavor low-energy QCD is to ensure the confinement to have discrete spectra for hadrons.
The simplest way to realize it might be to truncate the extra dimension at $z=z_m$ such that the radial direction $z$ of dual gravity
runs from zero to $z_m$. Since the radial direction corresponds to an energy scale of a boundary gauge theory, $1/z_m$ maps to $\Lambda_{QCD}$.

Putting things together, we could arrive at the following bulk Lagrangian with local SU(2)$_{\rm L}\times$ SU(2)$_{\rm R}$, the hard wall model~\cite{Erlich:2005qh,Da_Rold:2005zs},
\begin{eqnarray}
 S_{\rm HW}
 = \int d^4 x \int dz \sqrt{g} ~{\rm Tr} \left[
-\frac{1}{4g_5^2}
(F_L^2+F_R^2)+|DX |^2 +3|X|^2~\right],\label{HW}
\end{eqnarray}
where $D_\mu X = \partial_\mu X - i A_{L\mu}\, X+ i X A_{R\mu}$ and
$A_{L,R}=A^a_{L,R} t^a$ with ${\rm Tr}(t^a t^b)=\frac{1}{2}
\delta^{ab}$. The bulk scalar field is defined by  $X = X_0\,
e^{2i\,\pi^at^a}$, where $X_0 \equiv \langle X \rangle$.
Here $g_5$ is the five dimensional gauge coupling, $g_5^2=12\pi^2/N_c$.
The background  is given by
\begin{equation}
ds^2=\frac{1}{z^2} (dt^2-d{\vec x}^2-dz^2)\, ,~ 0\le z\le z_m \, .\label{AdS}
\end{equation}
Instead of the sharp IR cutoff in the hard wall mode, we may introduce a bulk potential that plays a role of
a smooth cutoff.
In \cite{Karch:2006pv}, this smooth cutoff is introduced by a factor $e^{-\Phi}$ with $\Phi(z)=z^2$ in the bulk action, the soft wall model.
The form $\Phi(z)=z^2$ in the AdS would ensure the Regge-like behavior of the mass spectrum $m_n^2\sim n$.
The action is given by
\begin{eqnarray}
 S_{\rm SW}
 = \int d^4 x \int dz e^{-\Phi} \sqrt{g} ~{\rm Tr} \left[
-\frac{1}{4g_5^2}
(F_L^2+F_R^2)+|DX |^2 +3|X|^2~\right]\, .\label{SW}
\end{eqnarray}
Here we briefly show how to obtain the 4D vector meson mass in the soft wall model.
The vector field is defined by $V=A_L+A_R$.
With the Kaluza-Klein decomposition $V_\mu^a(x,z)= g_5\sum_n v_n(z)\rho_\mu^a (x)$, we obtain
\ba
\partial_z (e^{-B}\partial_z v_n) +m_n^2e^{-B}v_n=0,
\ea
where $B=\Phi(z)-A(z)=z^2+ \log z$ in the AdS geometry (\ref{AdS}).
With $v_n=e^B\psi_n$, we transform the equation of motion into
the form of a Schr\"odinger equation
\ba
\psi_n^{''}-V(z)\psi_n=-m_n^2\psi_n\, ,
\ea
where $V(z)=z^2+3/(4z^2)$.
Here $m_n$ is the mass of the vector resonances, and $\rho$ meson  corresponds to $n=0$.
The solution is well-known in quantum mechanics and the eigenvalue $m_n^2$ is given by~\cite{Karch:2006pv}
\ba
m_n^2=4c(n+1),
\ea
where $c$ is introduced to restore the energy dimension.

The finite temperature could be neatly introduced by a black hole in AdS$_{d+1}$, where $d$ is the dimension of the boundary gauge theory.
The background is given by
\begin{equation}
ds^2=\frac{1}{z^2} \left(f(z)dt^2-d{\vec x}^2-\frac{dz^2}{f(z)}\right)\, ,\label{AdSBH}
\end{equation}
where $f(z)=1-z^d/z_h^d$. The temperature of the boundary gauge theory is identified with the Hawking temperature of the black hole  $T=d/(4\pi z_h)$.
In Appendix \ref{BH_how_to} and \ref{BH}, we try to explain in a comprehensive manner how to calculate the Hawking temperature of a black hole.

Now we move on to dense matter.
According to the AdS/CFT dictionary, a
chemical potential in boundary gauge theory is encoded in the boundary value of the time component of
the bulk U(1) gauge field. To be more specific on this, we first consider the chemical potential term
in gauge theory,
\begin{equation}
{\cal L_\mu}=\mu_q q^\dagger q\, .
\end{equation}
Then, we introduce a bulk U(1) gauge field $A_\mu$ which is dual to $\bar q\gamma_\mu q$.
According to the dictionary, $A_0 (z\rightarrow 0)\sim c_1 z^{d-\Delta-p}+ c_2 z^{\Delta-p}$, we have $A_0 (z\rightarrow 0)\sim \mu_q$.
In the hard wall model, the solution of the bulk U(1) vector field is given by
\begin{equation}
A_t(z)=\mu + \rho z^2\, ,
\end{equation}
where $\mu$ and $\rho$ are related to quark chemical potential and quark (or baryon) number density
in boundary gauge theory.
It is interesting to notice that in chiral perturbation
theory, a chemical
potential is introduced as the time component of a gauge field
by promoting the global chiral symmetry to a local gauge one~\cite{chemicalP}.

\section{Vacuum structures\label{vacuumS}}

At low energy or momentum scales roughly smaller than $1$ GeV, $r>1$ fm,
QCD exhibits confinement and a non-trivial vacuum
structure with condensates of quarks and gluons. In this section, we discuss the gluon condensate and quark-gluon mixed condensate.

The gluon condensate $\la G_{\mu\nu}^a G_a^{\mu\nu} \ra$ was first introduced, at zero temperature, in \cite{Shifman:1978bx}
 as a measure for nonperturbative physics in QCD.
The gluon condensate characterizes the scale
symmetry breaking of massless QCD at quantum level.
Under the infinitesimal scale transformation
\ba
&&x_\mu^\prime= (1+\delta\lambda)x_\mu\, ,\no
&&A_\mu^\prime= (1-\delta\lambda)A_\mu\, ,\no
&&q^\prime= (1-\frac{3}{2} \delta\lambda)q\, ,
\ea
the trace of the energy momentum tensor reads schematically
\ba
\partial_\mu J_{\rm D}^\mu=T_\mu^\mu\simeq - \la\frac{\alpha_s}{\pi} G_{\mu\nu}^a G_a^{\mu\nu} \ra\, .
\ea
Here $J_{\rm D}^\mu$ is the dilatation current, $\alpha_s$ is the gauge coupling, $T_{\mu\nu}$ is the energy-momentum tensor of QCD.
Due to Lorentz invariance, we can write $\la T_{\mu\nu}\ra =-\epsilon_{vac}\eta^{\mu\nu}$, where $\epsilon_{vac}$ is the energy of the QCD vacuum.
Therefore, the value of the gluon condensate sets the scale of the QCD vacuum energy.
In addition, the gluon condensate is important in the QCD sum rule analysis since it
enters in the operator product expansion (OPE) of the hadronic correlators~\cite{Shifman:1978bx}.
At high temperature, the gluon condensate is useful to study the nonperturbative
nature of the QGP. For instance, lattice QCD results on the gluon condensate at
finite temperature \cite{Boyd:1996bx} indicate that the value of the gluon
condensate shows a drastic change around $T_c$ regardless of
the number of quark flavors. The change in the gluon
condensate could lead to a dropping of the heavy quarkonium
mass around $T_c$  \cite{Morita:2007pt}.

In holographic QCD, the gluon condensate figures in a dilaton profile according to the AdS/CFT since
the dilaton is dual to the scalar gluon operator Tr$( G_{\mu\nu} G^{\mu\nu})$.
The 5D gravity action with the dilaton is given by
\ba
S = \gamma \frac{1}{2 \k^2} \int d^5 x \sqrt{g} \lb {\cal R} + \frac{12}{R^2} - \half \pa_M \ph \pa^M \ph \rb ,\label{action2}
\ea
where $\gamma=+1$ for Minkowski metric, and $\gamma=-1$  for
Euclidean signature. We work with
Minkowski metric for most cases in this paper.
The solution of this system is discovered in
\cite{dAdS, Csaki:2006ji} by solving the
coupled dilaton equation of motion and the Einstein equation:
\ba
ds^2=\left ( \frac{R}{z} \right )^2  \left (
\sqrt{1-c^2z^8} ~\eta_{\mu\nu}dx^\mu dx^\nu +dz^2
\right )\, ,\label{dAdSb}
\ea
and the corresponding dilaton profile is given by
\ba
\phi (z) =\sqrt{\frac{3}{2}}\log \left
  (\frac{1+cz^4}{1-cz^4}
\right ) +\phi_0\, ,
\ea
where $\phi_0$ is a constant.
At $z=1/c^{1/4}$ there exists a naked singularity that might be resolved in a full string
theory consideration.
Near the boundary $z\rightarrow 0$,
\ba
\phi (z) \sim cz^4\, .
\ea
Therefore, $c$ is nothing but the gluon condensate up to a constant. Unfortunately, however, $c$ is an integration constant of
the coupled dilaton equation of motion and the Einstein equation and therefore,
it will be determined by matching with physical observables.
In \cite{Csaki:2006ji}, the value of the gluon condensate is estimated by the glueball mass.
An interesting idea based on the circular Wilson loop calculation in gravity side is proposed
to {\it calculate} the value of the gluon condensate $G_2$ $\equiv\la \frac{\alpha_s}{\pi} G_{\mu\nu}^a G_{\mu\nu}^a \ra$  \cite{Andreev:2007vn}.
The value is determined to be $G_2=0.010 \pm 0.0023$ GeV at zero temperature \cite{Andreev:2007vn}.
A phenomenological estimation of the gluon condensate in QCD sum rules gives
$ \la\frac{\alpha_s}{\pi} G_{\mu\nu}^a G_a^{\mu\nu} \ra\simeq 0.012$ GeV$^4$~\cite{Shifman:1978bx}.

Now we consider
quark-gluon mixed condensate $\la \bar q \sigma_{\mu\nu}G^{\mu\nu} q \ra$, which
can be regarded as an additional order parameter for the spontaneous chiral symmetry breaking
since the quark chirality flips via the quark-gluon operator.
Thus, it is naturally expressed in terms of the quark condensate as
\ba
\la \bar q \sigma_{\mu\nu}G^{\mu\nu} q \ra = m_0^2 \la\bar q q\ra\, .\label{mC}
\ea
In \cite{Kim:2008ff}, an extended hard wall model is proposed to calculate the value of $m_0^2$.
The bulk action of the extended model is given by
\begin{equation}
S\;=\;\int\,d^5x\,\sqrt{g}\,\mathrm{Tr}\left[|DX|^2+3|X|^2 -
  \frac1{4g_5^2}(F_L^2+F_R^2) + |D\Phi|^2 - 5\Phi^2
\right] ,
\end{equation}
where $\Phi$ is a bulk scalar field dual to the 4D operator on the left-hand-side of Eq. (\ref{mC}).
Then the chiral condensate and the mixed condensate are encoded in the vacuum expectation value of
the two scalar fields.
\begin{eqnarray}
&& \langle X(x,z)\rangle\;=\;  \frac12 (\hat{m} z + \sigma
z^3) ,\cr
&& \langle \Phi(x,z)\rangle\;=\; \frac16(c_1 z^{-1}
+\sigma_M z^5 ),
\end{eqnarray}
where $c_1$ is the source term for the mixed condensate and
$\sigma_M$ represents the mixed condensate
$\sigma_M\;=\;\langle\bar{q}_R \sigma_{\mu\nu}G^{\mu\nu} q_L\rangle$.
Taking $c_1=0$, a source-free condition to study only spontaneous symmetry breaking, we determine the value of the mixed condensate or $m_0^2$
by considering various hadronic observables. In this sense the mixed condensate is not calculated but fitted to
experimental data like the chiral condensate in the hard wall model.
The favored value of the $m_0^2$ in \cite{Kim:2008ff} is $0.72$ GeV$^2$.
A new  method to estimate the value of $m_0^2$ in is suggested in \cite{Andreev:2010zm}, where
a nonperturbative gauge invariant correlator (the nonlocal condensate) is calculated in dual gravity description
to obtain $m_0^2$. With inputs from
the slopes for the Regge trajectory of vector mesons  and the linear term of the Cornell potential, they obtained
$m_0^2=0.70$ GeV$^2$,
which is comparable to that from the QCD sum rules, $0.8$ GeV$^2$~\cite{Belyaev:1982sa}.

\section{Spectroscopy and form factors\label{Sec:SpecForm}}
Any newly proposed models or theories in physics are bound to confront experimental data,
for instance hadron masses, decay constants and form factors.
In this section, we consider the spectroscopy of  the glueball, light meson, heavy quarkonium, and hadron form factors in hard wall model, soft wall model, and
their variants.

\subsection{Glueballs}
Glueballs are made up of gluons with no constituent quarks in them.
The glueball states are in general mixed with conventional $\bar qq$ states,
so in experiments we may observe these mixed states only.
Their existence was expected from the early days of QCD~\cite{Fritzsch:1975tx, Jaffe:1975fd}.
For theoretical and experimental status of glueballs, we refer to \cite{Mathieu:2008me, Crede:2008vw}.

The spectrum of glueballs is one of the earliest QCD quantities calculated
based on the AdS/CFT duality.
In \cite{Witten:1998zw}, Witten confirmed that
the existence of the mass gap in
the dilaton equation of motion on a black hole  background, implying
 a discrete glueball spectrum with a finite gap. Extensive studies on the glueball
 spectrum were done in \cite{Csaki:1998qr, Brower:2000rp} and there also comparisons between the supergravity
results and lattice gauge theory results were made.

Now we consider a scalar glueball ($0^{++}$) on ${\bf R}^3\times {\bf S}^1$ as an example~\cite{Csaki:1998qr}.
When the radius of the circle ${\bf S}^1$ is very small $R\rightarrow 0$,  only the gauge
 degrees of freedom remains and the gauge theory is effectively the same as pure QCD$_3$~\cite{Witten:1998zw, Csaki:1998qr}.
 Using the operator/field correspondence, we first
find operators that have the quantum numbers with glueball states of interest
 and then introduce a corresponding bulk field to obtain the glueball masses.
In this case we are to solve an equation of motion for a bulk scalar field $\phi$, which is dual to
tr$F^2$ in the AdS$_5$ Euclidean black hole background.
The equation of motion for $\phi$ is given by
\ba
\partial_\mu \left(\sqrt{g} \partial_\nu \phi g^{\mu\nu} \right)=0\, ,
\ea
and the metric is
\ba
ds^2=\left(\rho^2-\frac{b^4}{\rho^2}\right)^{-1} d\rho^2 +\left(\rho^2-\frac{b^4}{\rho^2}\right)d\tau^2 +
\rho^2+(d\vec x)^2  +d\Omega_5^2\, ,
\ea
where $\tau$ is for the compactified imaginary time direction.
For simplicity, we assume that $\phi$ is independent of $\tau$ \cite{Witten:1998zw, Csaki:1998qr} and seek a solution
of the form $\phi(\rho,x)=f(\rho) e^{k\cdot x}$, where $\vec k$ is the momentum in  ${\bf R}^3$.
Then the equation of motion for $f(\rho)$ reads
\ba
\rho^{-1}\frac{d}{d\rho} \left( (\rho^4-b^4) \rho \frac{d f}{d\rho}\right )+m^2=0\, ,
\ea
where $m^2$ is the three-dimensional glueball mass, $m^2=-k^2$~\cite{Witten:1998zw, Csaki:1998qr}.
By solving this eigenvalue equation with suitable boundary conditions, regularity at the horizon ($\rho=b$) and
normalizability $f\sim \rho^{-4}$ at the boundary ($\rho\rightarrow \infty$), we can obtain discrete eigenvalues, the three-dimensional glueball masses.
In the context of a sliced AdS background of the Polchinski and Strassler set up~\cite{Polchinski:2001tt}
, which is dual to confining gauge theory, the mass ratios of glueballs
are studied in \cite{gball_FB}.

More realistic or phenomenology-oriented approaches follow the earlier developments.
In the soft wall model the mass spectra of scalar and vector glueballs,
and their dependence on the bulk geometry and the shape of the soft wall are studied in \cite{Colangelo:2007pt}.
The exact glueball correlators are calculated in \cite{Forkel:2007ru},
where the decay constants as well as the mass spectrum of the glueball are also obtained in both hard wall and soft wall models.
Here we briefly summarize the scalar glueball properties in the soft wall model~\cite{Colangelo:2007pt,Forkel:2007ru}.
Following a standard path to construct a bottom-up mode, we introduce a massless bulk scalar field $\phi$ dual to the scalar gluon operator Tr$( F_{\mu\nu} F^{\mu\nu})$
to write down the bulk action as~\cite{Colangelo:2007pt}
\ba
S\sim \int d^5 \sqrt{g} e^{-\Phi} g_{MN} (\partial_M \phi)(\partial^N\phi)\, ,
\ea
where $\Phi=z^2$ as in the soft wall model.
The equation of motion for $\phi(q,z)$ can be transformed to  a one dimensional Schr{\"o}dinger form
\ba
\psi^{\prime\prime} -V(z)\psi=q^2 \psi\, ,
\ea
where $\psi=e^{-(\Phi+3\ln z)/2}\phi$ with $q^2=m^2$~\cite{Colangelo:2007pt}.
The glueball mass spectrum is then given as the eigenvalue of the Schr{\"o}dinger type equation with regular eigenfunction
at $z=0$ and $z=\infty$
\ba
m_n^2=4(n +2)\tilde c\, ,
\ea
where $n$ is an integer, $n=0,1,2,\cdots$.
$\tilde c$ is introduced to make the exponent $\Phi$ dimensionless, $\Phi=\tilde c z^2$, and it will be fit to hadronic data.
Since the vector meson mass in the soft wall model is $m_n^2=4(n +1)\tilde c$, we calculate the ratio of
the lightest (n=0) scalar  glueball mass $m_{G0}^2$ to the $\rho$ meson mass to obtain $m_{G0}^2/m_\rho^2=2$~\cite{Colangelo:2007pt}.
The properties of the glueball at finite temperature is studied in the hard wall model \cite{Colangelo:2009ra}
and also in the soft wall model \cite{Colangelo:2009ra, Miranda:2009uw} by calculating the spectral function of the glueball in
the AdS black hole backgroudn.
The spectral function is related to various Green functions, and it can be defined by the two-point retarded Green function as
$\rho(\omega, \vec q)=-2{\rm Im}G^R(\omega,\vec q)$.
The retarded function can be computed in the real-time AdS/CFT, following the prescription proposed in \cite{Son:2002sd}.
Both studies using the soft wall model predicted that the dissociation temperature of scalar glueballs
is far below the deconfinement (Hawking-Page) transition temperature of the soft wall model.
See Section \ref{HPt_main} and in Appendix \ref{HP_app} for more on the Hawking-Page transition.
Note that below the  Hawking-Page transition temperature, the AdS black hole is unstable.
In \cite{Colangelo:2009ra}, the melting temperature of the scalar glueball from the spectral functions is
about $40-60$ MeV, while the deconfinement temperature of the soft wall model is about $190$ MeV~ \cite{Herzog:2006ra}.
This implies  that we have to build a more refined holographic QCD model to have
a realistic  melting temperature \cite{Colangelo:2009ra, Miranda:2009uw}.

\subsection{Light mesons}
There have been an armful of works in holographic QCD that studied light meson spectroscopy.
Here we will try to summarize results from the hard wall model, soft wall model and  their variants.

In Table \ref{tab:1}, we list some hadronic observables from hard wall models to see if the results are stable against some deformation of the model.
In the table, where
$^*$ means input data and the model with no $^*$ is a fit to all seven observables:
 Model A and Model B from the hard wall model \cite{Erlich:2005qh}, Model I from a hard wall model in a  deformed AdS geometry \cite{Shock:2006gt},
and Model II from a hard wall model with the quark-gluon mixed condensate~\cite{Kim:2008ff}.
In \cite{Da_Rold:2005zs}, the following deformed AdS background is considered
\begin{equation}
ds^2=\frac{\pi}{2z_m\sin[\pi z/(2z_m)]} (dt^2-dx_idx^i-dz^2)\, , ~0\le z\le z_m \, ,\label{deAdS1}
\end{equation}
and it is stated that the correction from the deformation is less than $10\%$.
 The backreaction on the AdS metric due to quark mass and chiral condensate
is investigated in \cite{Shock:2006gt}. One of the deformed backgrounds obtained  in
\cite{Shock:2006gt} phenomenologically reads
\begin{equation}
ds^2=\frac{1}{z^2}e^{-2B(z)} (dt^2-dx_idx^i-dz^2)\, , 0\le z\le z_m \,~ ,\label{deAdS2}
\end{equation}
where $B(z)= \frac{m_q^2}{24}z^2+\frac{m_q\sigma}{16}z^4+\frac{\sigma^2}{24} z^6$.
In Table \ref{tab:1}, we quote some results from this deformed background.
Dynamical (back-reacted) holographic QCD model with area-law confinement and linear Regge trajectories
was developed in~\cite{dePaula:2008fp}.

\begin{table}[ht]
  \centering
\begin{tabular}{cccccc}
\hline\hline
   &{\rm Model I } &{\rm Model II} &{\rm Model A }&{\rm Model
     B}&{\rm Experiment} \\
  \hline
 $m_\rho$ & $775.8^*$ & $775.8$ & $775.8^*$ &$832$ & $775.49\pm 0.34$  \\
$m_{a_1}$ & $1348$ & $1244$ & $1363$ &$1220$ & $1230\pm 40$ \\
$f_\pi$ & $92.4^*$ & $80.5$ & $92.4^*$ & $84.0$ & $92.4\pm 0.35$  \\
$F_\rho^{1/2}$ & $334$ & $330$ & $329$ & $353$ & $345\pm 8$  \\
$F_{a_1}^{1/2}$ & $481$ & $459$ & $486$ & $440$ & $433\pm 13$ \\
$m_\pi$ & $139.6^*$ & $139.3$ & $139.6^*$ & $141$ & $139.57\pm 0.00035$ \\
$g_{\rho\pi\pi}$ & $4.46$ & $4.87$ & $4.48$ & $5.29$ & $6.03\pm 0.07$ \\ \hline \hline
\end{tabular}
\caption{Meson spectroscopy from the hard-wall model and from its variations: Model I~\cite{Shock:2006gt},
Model II~\cite{Kim:2008ff}, Model A~\cite{Erlich:2005qh}, Model B~\cite{Erlich:2005qh}.
The experimental data listed in the
last column  are taken from the particle data
group~\cite{Amsler:2008zzb}.  All results are given  in units of MeV
except for the condensate and the ratio of two  condensates.}
 \label{tab:1}
\end{table}
We remark that the sensitivity of calculated hadronic observables to the details
of the hard wall model was studied in  \cite{Erlich:2008gp} by varying the infrared boundary
conditions, the 5D gauge coupling, scaling dimension of $\bar qq$ operator.
It turns out that predicted hadronic observables are not sensitive to varying scaling dimension
of $\bar qq$ operator, while they are rather sensitive to the IR boundary conditions and the 5D gauge coupling~\cite{Erlich:2008gp}.

In addition to mesons, baryons were also studied in the hard wall model \cite{Hong:2006ta,Maru:2009ux,Kim:2009bp,Zhang:2010bn}.
It is pointed out in \cite{Maru:2009ux,Kim:2009bp} that one has to use the same IR cutoff of the hard wall model $z_m$
for both meson  and baryon sectors.

Now we collect some results from the soft wall model \cite{Karch:2006pv}.
There were two non-trivial issues to be resolved in the original soft wall model.
Firstly, so called, the dilaton factor $\Phi\sim z^2$ is introduced phenomenologically to explain $m_n^2\sim n$.
The dilaton factor is supposed to be a solution of gravity-dilaton equations of motion.
Secondly, the chiral symmetry breaking in the  model is a bit different from QCD since
the chiral condensate is proportional to the quark mass in the soft wall model.
In QCD, in the chiral limit, where the quark mass is zero, the chiral condensate is finite that characterizes
spontaneous chiral symmetry breaking.
Several attempts have made to improve these aspects and to
fit experimental values better~\cite{Batell:2008zm,Gherghetta:2009ac,Colangelo:2008us,Sui:2009xe}.
In~\cite{Gherghetta:2009ac}, a quartic term in the potential for the bulk scalar $X$ dual to $\bar qq$ is introduced to
the soft wall model to incorporate chiral symmetry breaking with independent sources for spontaneous and explicit
breaking; thereby the chiral condensate remains finite in the chiral limit.
Then, the authors of~\cite{Gherghetta:2009ac}
 parameterized the vev of the bulk scalar $X_0$ such that it satisfies constraints from the AdS/CFT at UV
and from phenomenology at IR: $X_0\sim m_q z+ \sigma z^3$ as $z\to 0$ and $X_0\sim z$ as $z\to \infty$.
The constraint at IR is due to the observation~\cite{Shifman:2007xn} that chiral symmetry is not restored in the highly excited mesons.
Note that $X_0\sim z$ keeps the mass difference between vector and axial-vector
mesons  a constant as $z\to \infty$.
With the parameterized $X_0$, they obtained a dilaton factor $\Phi(z)$~\cite{Gherghetta:2009ac}.
We list some of results of \cite{Gherghetta:2009ac} in Table \ref{tab:2}.
An extended soft wall model with a finite UV cutoff was discussed in \cite{Evans:2006ea, Afonin:2011ff}.
In \cite{Forkel:2010gu}, the authors studied how a dominant tetra-quark component of the lightest scalar mesons
in the soft wall model, where a rather generic lower bound on the tetra-quark mass was derived.

\begin{table}[ht]
  \centering
\begin{tabular}{cccccc}
\hline\hline
  n & $\rho$-meson &{\rm $\rho$ experiment}& $a_1$-meson &{\rm $a_1$ experiment} \\
  \hline
 $1$ & $475$ &775.5&$1185$ & 1230   \\
$2$ & $1129$ & $1282$ & $1591$ &$1647$ \\
$3$ & $1429$ & $1465$ & $1900$ & $1930$ \\
$4$ & $1674$ & $1720$ & $2101$ & $2096$ \\
$5$ & $1884$ & $1909$ & $2279$ & $2270$ \\
$6$ & $2072$ & $2149$ & $ .$ & $.$ \\
$7$ & $2243$ & $2265$ & $.$ & $.$  \\ \hline \hline
\end{tabular}
\caption{Meson spectroscopy from the modified soft wall model~\cite{Gherghetta:2009ac}.
We show the center values of experimental data. In~\cite{Gherghetta:2009ac} the experimental data are mostly taken from the particle data
group~\cite{Amsler:2008zzb}, while $\rho$(1282) is from \cite{Bertin:1997vf}. All results are given  in units of MeV.}
 \label{tab:2}
\end{table}

As long as confinement and non-Abelian chiral symmetry are concerned, the Sakai-Sugimoto model \cite{Sakai:2004cn, Sakai:2005yt}
 based on a D4/D8/$\overline{\rm D8}$ brane configuration (see Appendix \ref{SSmodel}) is the only available stringy model. In this model, properties of light mesons and baryons have been greatly
studied~\cite{Sakai:2004cn, Sakai:2005yt, Harada:2006di, Nawa:2006gv, Hong:2007kx, Hata:2007mb, Hong:2007dq, Hashimoto:2008zw, Harada:2010cn, Harada:2011ur}.

In a simple bottom-up model with the Chern-Simons term,
it was also shown that baryons arise as stable solitons which are the 5D analogs of
4D skyrmions and the properties of the baryons are studied~\cite{Pomarol:2008aa}.

\subsection{Heavy quarkonium}
The properties of heavy quark system both at zero and at finite temperature have been the subject of intense investigation for
many years.  This is so because, at zero temperature, the
charmonium spectrum reflects detailed information about confinement and interquark potentials in QCD.
At finite temperature, due to
the small interaction cross section of the charmonium in hadronic matter, the charmonium spectrum is
expected to carry information about the early hot and dense stages of relativistic heavy ion collisions. In addition, the charmonium
states may remain bound even above the critical temperature $T_c$. This suggests that analyzing the charmonium data from
heavy ion collision inevitably requires more detailed information about the
properties of charmonium states in QGP. Therefore, it is very important to develop a consistent
non-perturbative QCD picture for the heavy quark system both below and above the phase transition temperature.
For a recent review on heavy quarkonium see, for example,~\cite{Brambilla:2010cs}.

Now we start with the hard wall model to discuss the heavy quarkonium in a bottom-up approach.
A simple way to deal with the heavy quarkonium in the hard wall model was proposed in \cite{Kim:2007rt}.
Since the typical energy scale involved for light mesons and heavy quarkonia are quite different,
we may introduce an IR cutoffs $z_m^H$ for heavy quarkonia in the hard wall model which is different from the IR cutoff
for light mesons,  $1/z_m^L\sim 300$ MeV.
Note that in the hard wall model there is a one-to-one correspondence between the IR cutoff and the vector meson mass $1/z_m\sim m_V$.
In \cite{Kim:2007rt}, the lowest  vector $c\bar c$ ($J/\psi$) mass, $\sim 3~{\rm GeV}$ is used as an input to fix
the IR cutoff for the charmonium, $1/z_m^H \simeq 1.32 {\rm GeV}$.
With this, the mass of
 the second resonance is predicted to be  $\sim 7.2~{\rm GeV}$, which
is quite different from the experiment $m_\psi'\sim 3.7~{\rm GeV}$.
This is in a sense generic limitation of the hard wall model whose predicted higher resonances are
quite different from experiments.
Moreover, having two different IR cutoffs in the hard wall model may cause a problem when we treat light quark and
heavy quark systems at the same time.
In the soft wall model, the mass spectrum of the vector meson is given by~\cite{Karch:2006pv}
\ba
m_n^2=4(n+1)c\, .
\ea
For charmonium system, again the
lowest mode ($J/\psi$) is used to fix $c$, $\sqrt{c}\simeq 1.55~{\rm
GeV}$. Then the mass of the second resonance $\psi'$ is $m_{\psi'}\simeq
4.38~{\rm GeV}$, which is ~$20\%$ away from the experimental value of
$3.686 ~{\rm GeV}$~\cite{Kim:2007rt}.
Additionally, the mass of heavy quarkonium such as  $J/\psi$ at finite temperature is calculated
to predict that the mass decreases suddenly at $T_c$ and above $T_c$ it
 increases with temperature. Furthermore,
 the dissociation temperature is determined to be around $494~{\rm
MeV}$ in the soft wall model~\cite{Kim:2007rt}.

To compare heavy quarkonium properties obtained in a holographic QCD study with lattice QCD,
the finite-temperature spectral function in the vector channel within the soft wall model
was explored in \cite{Fujita:2009wc}.
The spectral function is related to the two-point retarded Green function by
$\rho(\omega, \vec q)=-2{\rm Im}G^R(\omega,\vec q)$.
The retarded function can be computed following the prescription \cite{Son:2002sd}.
Thermal spectral functions  in a stringy set-up, D3/D7 model, were extensively
studied in \cite{Myers:2007we}.
To deal with the heavy quarkonium in the soft wall model,
two different scales ($c_\rho$ and $c_{J/\psi}$) are introduced.
It is observed in \cite{Fujita:2009wc} that a peak in the
spectral function melts with increasing temperature and
eventually is flattened at $T\simeq 1.2T_c$. It is also shown numerically
that the mass shift squared is approximately proportional to
the width broadening \cite{Fujita:2009wc}. Another interesting finding in \cite{Fujita:2009wc} is that
the spectral peak diminishes at high momentum, which could be interpreted
as the $J/\psi$ suppression under
the hot wind \cite{Liu:2006nn, Myers:2008cj}.
A generalized soft wall mode of charmonium is constructed by considering
not only the masses but also the decay constants of the charmonium, $J/\psi$ and $\psi^\prime$~\cite{Grigoryan:2010pj}.
They calculated the spectral function as well as the position of the complex singularities
(quasinormal frequencies) of the retarded correlator of the charm current at finite temperatures.
A predicted dissociation temperature is  $T\approx 540$ MeV, or $2.8T_c$~\cite{Grigoryan:2010pj}.

Alternatively,  heavy quarkonium properties can be studied
in terms of holographic heavy-quark potentials.
Since the mass of heavy quarks are much larger than the QCD scale parameter $\Lambda_{\rm QCD}\sim 200$ MeV,
the non-relativistic Schr\"odinger equation could be a useful tool to study heavy quark bound states.
\ba
\left (-\frac{ {\mathbf{\bigtriangledown}}^2}{2m_r} +V(r) \right)\Psi(r) =E\Psi(r)\, ,
\ea
where $m_r$ is the reduced mass, $m_r=m_Q/2$.
A tricky point with potential models for quarkonia is which potential
is to be used in the Schr\"odinger equation: the free energy or the internal energy.
In the context of the AdS/CFT, there have been a lot of works on holographic heavy quark potentials~\cite{HQpAdSCFT}.
Hou and ~Ren  calculated the dissociation temperature of heavy quarkonia  by solving the Schr\"odinger equation
with holographic potentials \cite{Hou:2007uk}. They used two ans\"atze of the potential model: the F-ansatz (U-ansatz) which identifies
the potential in the Schr\"odinger equation with the free energy (the internal energy), respectively.
With the F-ansatz, $J/\psi$ does not survive above $T_c$, while the dissociation temperature of $\Upsilon$ is $(1.3-2.1)T_c$.
For the U-ansatz, $J/\psi$ dissolves into open charm quarks around $(1.2-1.7)T_c$ and $\Upsilon$ dissociates at about
 $(2.5-4.2) T_c$.

We finish this subsection with a summary of the discussion in \cite{Mateos:2007vn} on the usefulness of Dq/Dp systems
in studying heavy quark bound states. A Dq/Dp system may be good for $s\bar s$ bound states
 at high temperature since the mesons in the Dq/Dp system are deeply bounded, while heavy quarkonia are shallow bound states.
However, there exist certain properties of heavy quarkonia in the quark-gluon plasma that could be understood in the D4/D6 model
such as dissociation temperature.

\subsection{Form-factors}
Form factors are a source of information about the
internal structure of hadrons such as the distribution of charge.
We take the pion electromagnetic form factor as an example.
Consider a pion-electron scattering process $\pi^{\pm}+e^-\rightarrow \pi^{\pm}+e^-$ through photon exchange.
The cross section of this process measured in experiments is different from that of Mott scattering which is for the Coulomb scattering
of an electron with a point charge.  This deviation is parameterized into the pion form factor $F_\pi(q^2)$, where $q^2$  is
given by the energy and momentum  of the photon $q^2=\omega^2-{\vec q}^2$. If the pion is a structureless point particle,
we have $F_\pi=1$.
The pion electromagnetic form factor is expressed by, with the use of Lorentz invariance, charge conjugation, and  electromagnetic gauge invariance,
\ba
(p_1+p_2)_\mu F_\pi (q^2)=\la\pi(p_2) |J_\mu |\pi(p_1) \ra\, ,
\ea
where $q^2=(p_2-p_1)^2$ and $J_\mu$ is the electromagnetic current, $J_\mu= \sum_f e_f \bar q_f \gamma_\mu q_f$.
The pion charge radius is determined by
\ba
\la r_\pi^2\ra =6 \frac{\partial F_\pi(q^2)}{\partial q^2}|_{q^2=0}\, .
\ea
In a vector meson dominance model, where the photon interacts with the pion only via vector mesons, especially $\rho$ meson,
the pion form factor is given by
\ba
F_\pi(q^2)=\frac{m_\rho^2}{m_\rho^2-q^2 -i m_\rho\Gamma_\rho (q^2)}\, .
\ea
Then we obtain the pion charge radius $\sqrt{\la r_\pi^2\ra}=\sqrt{6}/m_\rho\simeq 0.63$ fm.
The experimental value is $\sqrt{\la r_\pi^2\ra}= 0.672$ fm \cite{Eidelman:2004wy}.
To evaluate the form factor, we consider the three-point correlation function of
two axial vector currents which contains nonzero projection onto a one pion state  and the external electromagnetic current,
\ba
\Gamma_{\mu\alpha\beta}(p_1,p_2)= -\int dx\int dy e^{(-ip_1x+ip_2y)} \la 0|T \{J_{\alpha 5}^\dagger (x) J_\mu(0) J_{\beta 5}(y)\} |0\ra\, .
\ea
Alternatively, we can consider two pseudoscalar currents instead of the axial vector currents.
The three-point correlation function can be decomposed into several independent Lorentz structures.
Among them we pick up the Lorentz structure corresponding to the pion form factor,
\ba
\la 0|J_{\beta 5}|p_2\ra\la p_2|  J_\mu |p_1\ra\la p_1|J_{\alpha 5}^\dagger |0\ra \simeq f_\pi^2 F_\pi(q^2)p_1^\alpha p_2^\beta (p_1^\mu +p_2^\mu).
\ea
Note that $\la 0|J_{\alpha 5}|p\ra =if_\pi p_\alpha$, where $|p\ra$ is a one pion state.
For more details on the form factor, we refer to \cite{FormFactor}.

In a holographic QCD approach, we can easily evaluate the three-point correlation function of
two axial vector currents (or two pseudoscalar currents) and the external electromagnetic current.
In \cite{Grigoryan:2007vg}, the form factors of vector mesons were calculated in the hard wall model and
the electric charge radius of the $\rho$-meson was evaluated to be $\la r_\rho^2\ra=0.53$ fm$^2$.
The number from the soft wall model is $\la r_\rho^2\ra=0.655$ fm$^2$~\cite{Grigoryan:2007my}.
The approach based on the Dyson-Schwinger equations predicted  $\la r_\rho^2\ra=0.37$ fm$^2$ \cite{Hawes:1998bz}
and $\la r_\rho^2\ra=0.54$ fm$^2$~\cite{Bhagwat:2006pu}. The quark mass (or pion mass) dependence of
the charge radius of the $\rho$-meson was calculated in lattice QCD: for instance, with $m_\pi\simeq 300$ MeV,
 $\la r_\rho^2\ra=0.55$ fm$^2$ \cite{Hedditch:2007ex}.
The pion form factor were studied in the hard wall model \cite{Grigoryan:2007wn} and in a model that interpolates between the hard wall and soft wall models \cite{Kwee:2007nq}.
The results obtained are $\sqrt{\la r_\pi^2\ra}= 0.58$ fm \cite{Grigoryan:2007wn} and in \cite{Kwee:2007nq}
$\sqrt{\la r_\pi^2\ra}= 0.500$ fm, $\sqrt{\la r_\pi^2\ra}= 0.576$ fm, depending on their parameter choice.
The gravitational form factors of mesons were calculated in the hard wall model \cite{Abidin:2008ku, Abidin:2008hn}.
The gravitational form factor of the pion is defined by,
\ba
\la \pi^b (p^\prime) | \Theta^{\mu\nu}(0) |\pi^a(p)\ra =
\frac{1}{2}\delta^{ab}[(g^{\mu\nu} q^2-q^\mu q^\nu) \Theta_1(q^2) +4P^\mu P^\nu\Theta_2 (q^2)]\, ,
\ea
where $\Theta_{\mu\nu}$ is the energy momentum tensor, $q=p^\prime-p$, and $P=(p^\prime+p)/2$.
There are also interesting works that studied various form factors in holographic QCD
\cite{Grigoryan:2008up,Abidin:2009hr,Brodsky:2011xx, Zuo:2011sk}.
Form factors of vector and axial-vector mesons were calculated in the Sakai-Sugimoto model \cite{BallonBayona:2009ar}.

\section{Phases of QCD\label{Sec:QCDphase}}

Understanding the QCD phase structure is one of the important problems
in modern theoretical physics, see~\cite{QCDphaseDreviews} for some
recent reviews. However, a quantitative calculation of the phase diagram from the first principle
is extraordinarily difficult.

\begin{figure}[!ht]
\begin{center}
  \includegraphics[width=10cm]{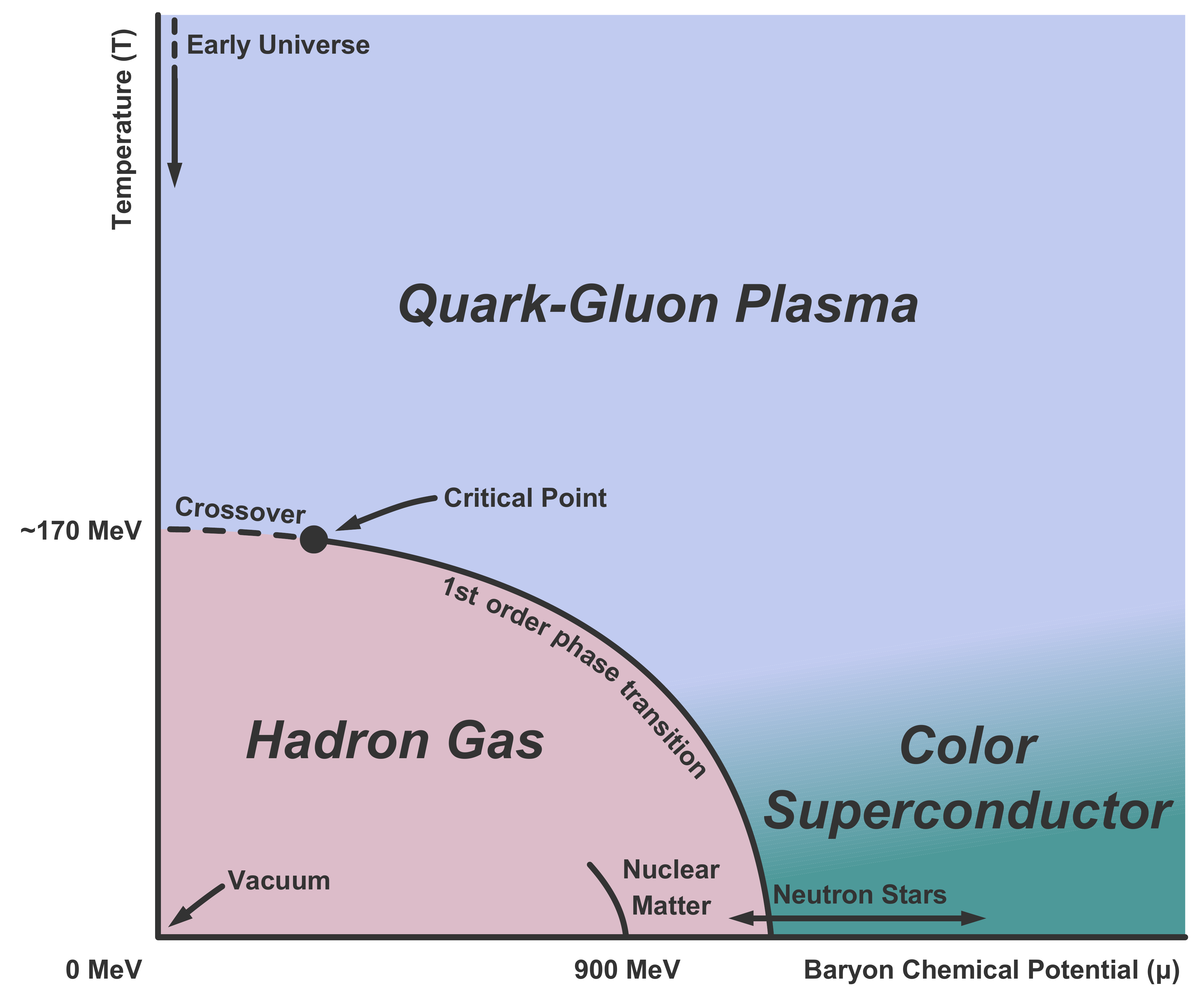}
  \caption{QCD phase diagram.}\label{fig:QCDphase}
\end{center}
\end{figure}

Basic order parameters for the QCD phase transitions
are the Polyakov loop which characterizes the deconfinement transition in the limit of infinitely large quark mass
and the chiral condensate for chiral symmetry in the limit of zero quark mass.
The expectation value of the Polyakov loop is loosely given by
\ba
\la L \ra\approx \lim_{r\rightarrow\infty} e^{-\beta V(r)}\, ,\nonumber
\ea
where $V(r)$ is the potential between a static quark-antiquark pair at a distance r, and $\beta\sim 1/T$.
The expectation value of the Polyakov loop is zero in confined phase, and it is finite in deconfined phase.
While the chiral condensate, which is the simplest order parameter
for the chiral symmetry, is non-zero with  broken chiral symmetry, vanishing with a restored chiral symmetry.
Apart from these order parameters, there are thermodynamic quantities that are relevant to study
the QCD phase transition.
The equation of state is one of them.
The energy density, for instance, has been found
to rise rapidly at some critical temperature.
This is usually interpreted as
deconfinement; liberation of many new degrees of
freedom.
The fluctuations
of conserved charges such as baryon number or electric charge~\cite{McLerran:1987pz, chargeF}
is also an important signal of  the quark-hadron phase transition.
The quark (or baryon) number susceptibility, which measures the
response of QCD to a change of the quark chemical potential
is one of such fluctuations~\cite{McLerran:1987pz, Gottlieb:1987ac}.

The nature of the chiral transition of QCD depends
on the number of quark flavors and the value of the quark mass.
For pure $SU(3)$ gauge theory with no quarks, it is first order.
In the case of two massless and one massive quarks,
the transition is the second order at zero or small quark chemical
potentials,
and it becomes the first order as we increase the chemical potential.
The point where the second order transition becomes
the first order is called tricritical point.
With physical quark masses of up, down, and strange,
the second order at zero or low chemical potential
becomes the crossover, and the tricritical point turns into
the critical end point.

\subsection{Confinement/deconfinement transition\label{HPt_main}}
We first discuss  the deconfinement transition.
In holographic QCD,
the confinement to deconfinement phase transition
is described by the Hawking-Page transition~\cite{Hawking:1982dh}, a phase transition between the Schwarzschild-AdS black hole and thermal AdS backgrounds.
This identification was made in \cite{Witten:1998zw}. One simple reasoning for this identification is from the observation that the Polyakov expectation
value is zero on the thermal AdS geometry, while it is finite on the AdS black hole.
See Appendix \ref{HP_app} for some more description of the Hawking-Page transition
and the Polyakov expectation in thermal AdS and AdS black hole.
In low-temperature confined phase, thermal
AdS, which is nothing but the AdS metric in Euclidean space, dominates the partition function, while at high temperature, AdS-black hole geometry
does. This was first discovered in the finite volume boundary case in \cite{Witten:1998zw}.
In the bottom-up model, it is shown  that the same phenomena happen also for infinite boundary
volume if there is a finite scale associated with the fifth direction \cite{Herzog:2006ra}.

Here we briefly summarize the Hawking-Page analysis of ~\cite{Herzog:2006ra} done in the hard wall model.
The Euclidean gravitational action given by
\begin{equation}
S_{grav} ~=~ -\frac{1}{2\kappa^2} \int d^5x \sqrt{g}\left(\textrm{R}+\frac{12}{L^2}\right)\, ,
\end{equation}
where $\kappa^2 = 8\pi G_5$ and $L$ is the length scale of the $AdS_5$, there are two solutions for the equations of motion derived from the gravitational action.
The one is the sliced thermal AdS (tAdS)
\begin{equation}
ds^2=\frac{L^2}{z^2}\left(d\tau^2+dz^2+d\vec{x}^2_3\right)\, ,
\end{equation}
where the radial coordinate runs from the boundary of tAdS space $z=0$ to the
cut-off $z_m$. Here $\tau$ is for the compactified Euclidean time-direction with periodicity $\beta^\prime$.
The other solution is the  AdS black hole (AdSBH) with the horizon $z_h$
\begin{equation}
ds^2=\frac{L^2}{z^2}\left(f(z)d\tau^2+\frac{dz^2}{f(z)}+d\vec{x}^2_3\right)\, ,
\end{equation}
where $f(z)=1-(z/z_h)^4$.
The Hawking temperature of the black hole solution is $T = 1/(\pi z_h)$, which is given by regularizing the metric near the horizon.
At the boundary $z=\epsilon$ the periodicity of the time-direction in  both backgrounds is the same and so
 the time periodicity of the tAdS is given by
\begin{equation}
\beta = \pi z_h \sqrt{f(\epsilon)}.
\end{equation}
Now we calculate the action density $V$, which is defined by the action divided by the common volume factor of $R^3$.
The regularized action density of the tAdS is given by
\begin{equation}
V_1(\epsilon) = \frac{4L^3}{\kappa^2} \int^{\beta '}_{0} d\tau \int^{z_{IR}}_{\epsilon}\frac{dz}{z^5}\, ,
\end{equation}
and that of the AdSBH is given by
\begin{equation}
V_2(\epsilon) = \frac{4L^3}{\kappa^2}\int^{\pi z_h}_{0} d\tau \int^{\bar{z}}_{\epsilon}\frac{dz}{z^5}
\end{equation}
where $\bar{z} = {\rm min}(z_{m},z_h)$.
Then, the difference of the regularized actions is given by
\begin{equation}
\Delta V_g = \lim_{\epsilon\rightarrow 0}\left[V_2(\epsilon) -V_1(\epsilon)\right]= \left\{\begin{array}{ll} \frac{L^3 \pi z_h}{\kappa^2}
\frac{1}{2z_h^4} & z_{m} < z_h\\ \\
\frac{L^3 \pi z_h}{\kappa^2} \left( \frac{1}{z_{m}^4} - \frac{1}{2z_h^4}\right)
& z_{m} > z_h.\end{array}
\right.
\end{equation}
When $\Delta V_g$ is positive (negative), tAdS (AdSBH) is stable.
Thus, at $\Delta V_g=0$ there exists a Hawking-Page transition. In the first case $z_m < z_h$, there is no Hawking-Page transition and the tAdS is always stable.
In the second case $z_m > z_h$, the Hawking-Page transition occurs at
\begin{equation}    \label{tempads}
T_c = 2^{1/4}/(\pi z_m)\, ,
\end{equation}
and at low temperature $T < T_c$ (at high temperature $T > T_c$) the
thermal AdS (the AdS black hole) geometry becomes a dominant background.
When we fix the IR cutoff by the $\rho$ meson mass, we obtain $1/z_m=323$ MeV
and $T_c= 122$ MeV. In the soft wall model, $T_c=191$ MeV \cite{Herzog:2006ra}.

This work has been extended in various directions.
The authors of \cite{BallonBayona:2007vp} revisited the thermodynamics of the hard wall and soft wall model.
They used holographic renormalization to compute the finite actions of the relevant
supergravity backgrounds and verify the presence of a Hawking-page type phase transition.
They also showed that the entropy, in the gauge theory side, jumps from $N^0$ to $N^2$ at the transition point~\cite{BallonBayona:2007vp}.
In \cite{Cai:2007zw}, the extension was done by studying the thermodynamics of AdS black holes with spherical or
negative constant curvature horizon, dual to a non-supersymmetric Yang-Mills theory on
a sphere or hyperboloid respectively.
They also
studied charged AdS black holes \cite{Chamblin:1999tk}
 in the grand canonical ensemble, corresponding to a
Yang-Mills theory at finite chemical potential, and found that there is always a gap for the
infrared cutoff due to the existence of a minimal horizon for the charged AdS black holes
with any horizon topology~\cite{Cai:2007zw}.
With an assumption that the gluon condensate melts out at finite temperature, a Hawking-Page type transition between
the dilaton AdS geometry in Eq. (\ref{dAdSb}) and the usual AdS black hole has studied in \cite{Evans:2008tu}.

The effect of the number of quark flavors $N_f$ and baryon number density on the critical temperature was investigated by
considering a bulk meson action together with the gravity action in \cite{Kim:2007em}. It is shown that
the critical temperature decreases with increasing $N_f$. As the number density was raised, the critical temperature
begins to drop, but it saturates to a constant value even at very large density. This is mostly due to the absence of
the back-reaction from number density~ \cite{Kim:2007em}. The back-reaction due to the number density has included in \cite{bRdnesity}.
In \cite{Cai:2007bq}, deconfinement transition of AdS/QCD with $\mathcal{O} ( {\alpha^\prime}^3 )$  corrections were investigated.
In \cite{Gursoy:2008za}, thermodynamics of the asymptotically-logarithmically-AdS black-hole solutions of 5D dilaton
gravity with a monotonic dilaton potential are analyzed in great detail, where it is shown
that in a special case, where the asymptotic geometry in the string frame reduces to flat space with a linear dilaton,
the phase transition could be {\it second order}.
The renormalized Polyakov loop in the deconfined phase of a pure SU(3) gauge theory was computed in \cite{Andreev:2009zk}
based on a soft wall metric model. The result obtained in this work is in good agreement with the one from lattice QCD simulations.

Due to this Hawking-Page transition, we are not to use the black hole in the confined phase, and so we
are not to obtain the temperature dependence of any hadronic observables.
This is consistent with large $N_c$ QCD at leading order.
For instance it was shown in~\cite{Gocksch:1982en, Neri:1983ic} that the Wilson loops, both time-like and space-like,
and the chiral condensate are independent of the
temperature in confining phase to leading order in $1/N_c$.  This means that the chiral and deconfinement transitions
are first order. The  deconfinement and chiral phase transitions of an SU(N) gauge theory at large $N_c$ were also discussed
in \cite{Pisarski:1983db}. However, in reality we observe temperature dependence of hadronic quantities, and therefore we have to
include large $N_c$ corrections in holographic QCD in a consistent way.
A quick fix-up for this might be to use the temperature dependent chiral condensate as an input
in a holographic QCD model and study how this temperature dependence conveys into other hadronic quantities~\cite{Kim:2008hx}.

\subsection{Chiral transition}

Now we turn to the chiral transition of QCD based on the chiral condensate.
In the hard wall model, the chiral symmetry is broken, in a sense, by the IR boundary condition.
In case we have a well-defined IR boundary condition at the wall $z=z_m$, we could calculate the
value of chiral condensate by solving the equation of motion for the bulk scalar $X$.
In the case of the AdS black hole we could have a well defined IR boundary condition at the black hole  horizon, which
allows us to calculate the chiral condensate.
For instance in \cite{Kim:2006ut},
  it is shown that with the AdS black hole background the chiral condensate together with the current quark mass
is zero both in the hard wall and soft wall models.
This is easy to see from the solution of $X_0$ in the AdS black hole background~\cite{Kim:2006ut, Ghoroku:2005kg}
\ba
X_0(z)=z\biggl(m_q ~{}_2F_1
(\frac{1}{4},\frac{1}{4},\frac{1}{2},\frac{z^4}{z_h^4})
+  \sigma_q z^2 ~{}_2F_1
(\frac{3}{4},\frac{3}{4},\frac{3}{2},\frac{z^4}{z_h^4}) \biggr)\, .
\ea
At $z=z_h$, both terms in $X_0(z)$ diverges logarithmically,
which requires to set  both  of them  zero: $m_q=0$, $\sigma=0$.
This is  different from real QCD, where current quark mass can be non-zero in the regime
$T > T_c$.

The finite temperature phase structure of the Sakai-Sugimoto model was analyzed in \cite{Aharony:2006da}
to explore  deconfinement and chiral symmetry restoration. Depending on a value of the model parameter,
 it is predicted that deconfinement and chiral symmetry restoration happens at the same temperature or
the presence of a deconfined phase with broken chiral symmetry~\cite{Aharony:2006da}.
Phase structure of a stringy D3/D7 model has extensively studied in \cite{Evans:2010iy, Evans:2011mu}.

\subsection{Equation of state and susceptibility}

Apart from the chiral condensate, various thermodynamic quantities
could serve as an indicator for a transition from hadron to quark-gluon phase.
Energy density, entropy, pressure, and susceptibilities are such examples.
We first consider energy density and pressure.
Schematically, based on the ideal gas picture  we discuss how the energy density and pressure tell hadronic matter to
quark-gluon plasma.
At low temperature thermodynamics of hadron gas will be dominated by pions which are almost massless,
while in QGP quarks and gluons are the relevant degrees of freedom.
Energy density and pressure of massless pions are
\ba
\epsilon=\gamma\frac{\pi^2}{30}T^4,~~~p=\gamma\frac{\pi^2}{90}T^4\, ,
\ea
where the number of degrees of freedom $\gamma$ is three.
In the QGP, they are given by
\ba
\epsilon=\gamma\frac{\pi^2}{30}T^4+B,~~~p=\gamma\frac{\pi^2}{90}T^4 -B \, ,
\ea
where $\gamma=37$, and $B$ is the bag constant.
Apart from the bag constant, the degeneracy factor $\gamma$ changes from $3$ to $37$, and therefore we can expect that
the energy density and pressure will increase rapidly at the transition point.
Since the dual of the boundary energy-momentum tensor $T_{\mu\nu}$ is the metric, we can obtain the energy density and
pressure of a boundary gauge theory from the near-boundary behavior of the gravity solution.
To demonstrate how-to, we follow \cite{de_Haro:2000xn, Janik:2005zt}.
We first rewrite the gravity solution in the Fefferman-Graham coordinate~\cite{FGc}
\ba
ds^2=\frac{1}{z^2}\left(g_{\mu\nu}dx^\mu dx^\nu -dz^2 \right)\, .
\ea
Next, we expand the metric $g_{\mu\nu}$ at the boundary $z\to 0$,
\ba
g_{\mu\nu}= g_{\mu\nu}^{(0)}+z^2 g_{\mu\nu}^{(2)}+ g_{\mu\nu}^{(4)}+\dots\, .
\ea
Now we consider flat 4D metric such that $g_{\mu\nu}^{(0)}=\eta_{\mu\nu}$.
Then $ g_{\mu\nu}^{(2)}=0$ and the vacuum expectation value of the energy momentum tensor is given by
\ba
\la T_{\mu\nu}\ra ={\rm const}  \cdot g_{\mu\nu}^{(4)}\, .
\ea
For example, we consider an AdS black hole in the  Fefferman-Graham coordinate
\ba
ds^2=\frac{1}{z^2}\left( \frac{(1-z^4/z_h^4)}{(1+z^4/z_h^4)}dt^2 -(1+z^4/z_h^4)d\vec x^2 -dz^2
\right)\, .
\ea
Here the temperature is defined by $T=\sqrt{2}/(\pi z_h)$.
Then, we read off
 \ba
 \la T_{\mu\nu}\ra\approx {\rm diag} (3/z_h^4, 1/z_h^4,1/z_h^4,1/z_h^4)\, ,
 \ea
 which satisfies $\epsilon=3p$.
There have been many works on the equations of state for a holographic matter at finite temperature~\cite{EoST}.
In \cite{Gursoy:2008bu},  the energy density, pressure, and entropy of a deconfined pure Yang-Mills matter  were evaluated in the improved holographic QCD model~\cite{Gursoy:2007cb}.
The energy density and pressure vanish at low temperature, and at the critical temperature, $T_c\sim 235$ MeV, they jump up to a finite value, showing
the first order phase transition. It is interesting to note that in \cite{Panero:2009tv} some high-precision lattice QCD simulations
were performed with increasing $N_c$ at finite temperature, and the results were compared with those from holographic QCD studies.

Various susceptibilities are also useful quantities to characterize phases of QCD.
For instance, the quark number susceptibility has been calculated in holographic QCD
in a series of works~\cite{Kim:2006ut,Kim:2010zg, Kim:2010ag}.
The quark number susceptibility was originally proposed as a probe of the
QCD chiral phase transition at zero chemical potential~\cite{McLerran:1987pz, Gottlieb:1987ac},
\begin{equation}
\chi_q=\frac{\partial n_q}{\partial\mu_q}.
\end{equation}
In terms of the retarded Green function $G_{\mu  \nu}^R(\omega, k)$,
the quark number susceptibility can be written as \cite{Kunihiro:1991qu},
\begin{equation}
\chi_q(T, \mu)
=-\lim_{k\to 0} {\mbox{Re}}\Big(G_{t t}^R(\omega=0, k)\Big).
\label{suss}
\end{equation}
In \cite{Kim:2010zg}, it is claimed that quark number susceptibility will show a sudden jump at $T_c$ in high density regime,
and so QCD phase transition in low temperature and high density regime will be always first order.
Thermodynamics of a charged dilatonic black hole, which  is asymptotically RN-AdS black hole in
the UV and AdS$_2\times \mathbf{R}^3$ in the IR, including the quark number susceptibility was extensively studied in \cite{Stoffers:2010sp}.
The critical end point of the QCD phase diagram was studied in \cite{DeWolfe:2010he} by considering
the critical exponents of the specific heat, number density, quark number susceptibility, and the relation between the number density and chemical potential at finite chemical potential and temperature. It is shown that the critical end point located at
 $T=143$ MeV and $\mu= 783$ MeV in the QCD phase diagram~\cite{DeWolfe:2010he}.

\subsection{Dense baryonic matter}
Understanding  the properties of dense QCD is of key
importance for laboratory physics such as heavy ion collision and for our understanding of the physics of
stable/unstable nuclei, and of various astrophysical objects such as neutron stars.

To expose an essential physics of dense nuclear matter,  we take the Walecka model~\cite{Walecka:1974qa}, which describes nuclear matter properties rather well, as an example.
The simplest version of the model contains the nucleon $\psi$, omega meson $\omega$, and an isospin singlet, Lorentz scalar meson $\sigma$
whose minimal Lagrangian is
\ba
{\mathcal L}=\bar\psi (i\not\!\partial   +g_\sigma\sigma -g_\omega \not\! \omega)\psi
+\frac{1}{2}(\partial_\mu\sigma \partial^\mu\sigma-m_\sigma^2\sigma^2)-\frac{1}{4}F_{\mu\nu}F^{\mu\nu}
+\frac{1}{2}m_\omega^2\omega_\mu\omega_\mu\, .
\ea
Within the mean field approximation, the properties of nuclear matter are mostly determined by
the scalar mean field $\bar\sigma=(g_\sigma/m_\sigma^2)n_s$ and the mean field of the time component of the
$\omega$ field $\bar\omega_0=(g_\omega/m_\omega^2)n$, where $n$ is the baryon number density and
$n_s$ is the scalar density.
For instance, the pressure of the nuclear matter described by the Walecka is
\ba
P&=&\frac{1}{4\pi^2}\biggl[ \frac{2}{3} E_F^* p_F^3 -{m_N^*}^2 E_F^*p_F+{m_N^*}^4 \ln (
\frac{E_F^*+p_F}{m_N^*} )\biggr]\\
&&+\frac{1}{2}\frac{g_\omega^2}{m_\omega^2}n^2 -\frac{1}{2}\frac{g_\sigma^2}{m_\sigma^2}n_s^2\, ,
\ea
where
\ba
E_F^*=\sqrt{p_F^2+{m_N^*}^2},~~~ m_N^*=m_N-\frac{g_\sigma^2}{m_\sigma^2}n_s \, .
\ea
Further, many successful  predictions based on the Walecka model and its
generalized versions, Quantum Hadrodynamics,  require large scalar and vector fields in nuclei.
This implies that to gain a successful description of nuclear matter or nuclei, having both  scalar
 and vector mean fields in the model seem crucial.
The importance of the interplay between the scalar and vector fields can be also seen in the
static non-relativistic potential between two nucleons.
The nucleon-nucleon potential from single $\sigma$-exchange and single $\omega$-exchange is given by
\ba
V(r)=\frac{g_\omega^2}{4\pi}\frac{1}{r}e^{-m_\omega r} -\frac{g_\sigma^2}{4\pi}\frac{1}{r}e^{-m_\sigma r}\, .\label{NNpOBE}
\ea
Note that  single $\sigma$-exchange can be replaced by two pion exchange.
If $g_\omega> g_\sigma$ and $m_\omega> m_\sigma$, then the potential in Eq. (\ref{NNpOBE})
captures some essential features of the two nucleon potential
to form stable nuclear matter: repulsive at short distance and attraction at intermediate and long distance.
We remark here that the scalar field in the Walecka model may not be the scalar associated with a
linear realization of usual chiral symmetry breaking in QCD, see for instance \cite{Furnstahl:2000in}.

The hard wall model or soft wall model in its original form does not do much in dense matter.
This is primarily due to its simple  structure and chiral symmetry.
Suppose we turn on the time component of a U(1) bulk vector field dual to a boundary number operator,
$V_t(z)=\mu+\rho z^2$. To incorporate this U(1) bulk field into the hard wall mode, we consider U(2) chiral symmetry.
The covariant derivative with U(1) vector and axial-vector is given by
$D_\mu X = \partial_\mu X - i A_{L\mu}\, X+ i X A_{R\mu}$ and it becomes
$D_\mu X = \partial_\mu X - i X ( A_{L\mu}- A_{R\mu})$.
Therefore, the U(1) bulk field  $V_\mu= A_{L\mu}+ A_{R\mu}$
does not couple to the scalar $X$,
 meaning that the physical properties of $X$ are not affected by the chemical potential or number density.
Note, however, that the vacuum energy of the hard wall or soft wall model should depend on the chemical potential and number density
 by the AdS/CFT.
One simple way to study the physics of dense matter  in the hard or soft wall
model is to work with higher dimensional terms in the action.
For instance, the role of dimension six terms in the hard wall model was studied in free space \cite{Grigoryan:2007iy}.
If we turn on the number density through the U(1) bulk field, we  have a term like $X_0^2F_V^2$, where $F_V$ is the field strength of the bulk U(1) gauge field~\cite{Kim:2010an}.
Then we may see interplay between number density and chiral condensate encoded in $X_0$.
In \cite{Domokos:2007kt}, based on the hard wall model with the Chern-Simons term it is shown that there exists
a Chern-Simons coupling between
vector and axial-vector mesons at finite baryon density.
This mixes transverse $\rho$  and $a_1$ mesons and leads to the condensation of
the vector and axial-vector mesons.
The role of the scalar density or the scalar field in the hard wall model was explored in~\cite{Kim:2007xi}.
In~\cite{Lee:2009bya}, a  back-reaction due to the density is studied in the hard wall model.

Physics of dense matter in Sakai-Sugimoto model has been developed with/without the source term for baryon charge~\cite{denseD4D8}.
For instance, in \cite{Rozali:2007rx} localized and smeared source terms are introduced and a Fermi sea has been observed, though
there are no explicit fermionic modes in the model.
A deficit with the Sakai-Sugimoto model for nuclear matter might be the absence of the scalar field which is quite important together with U(1)
vector field.
The phase structure of the D3/D7 model at finite density is studied in~\cite{denseD3D7}.
The nucleon-nucleon potential is playing very important role in understanding the properties of nuclear matter.
For example, one of the conventional methods to study nuclear matter is to work with the independent-pair approximation, Brueckner's theory,
where two-nucleon potentials are  essential inputs.
 Holographic nuclear forces were studied in \cite{Kim:2008iy, Hashimoto:2009ys, Kim:2009sr, Kim:2010zzv}.

\section{Closing remarks\label{Sec:remarks}}

The holographic QCD model
has proven to be a successful and promising analytic tool to study non-perturbative nature of low energy QCD.
However, its success should always come with ``qualitative'' since it is capturing only large $N_c$ leading physics.
To have any transitions from ``qualitative''  to ``quantitative'', we have to invent a way to calculate
subleading corrections in a consistent manner.
A bit biased, but the most serious defect of the approach based on the gauge/gravity duality might be
that it offers inherently macroscopic descriptions of a physical system.
For instance, we may understand the QCD confinemnt/deconfinement transition through the Hawking-Page transition, qualitatively.
Even though we accept generously the word ``qualitatively'', we are not to be satisfied completely since we don't know how gluons and quarks
bound together to form a color singlet hadron or how hadrons dissolve themselves into quark and gluon degrees of freedom.
In this sense, the holographic QCD can not be stand-alone.
Therefore, the holographic QCD should go together with conventional QCD-based models or theories to guide them qualitatively and to
gain microscopic pictures revealed by the conventional approaches.

Finally, we collect some interesting works done in bottom-up models that are not yet properly discussed in this review.
Due to our limited knowledge, we could not list all of the interesting works and most results from top-down models
will not be quoted.
To excuse this defect we refer to recent review articles on holographic QCD~ \cite{review_hQCD}.

Deep inelastic scattering has been studied in gauge/gravity duality
\cite{Polchinski:2002jw, Andreev:2002aw, Hatta:2007he, Cornalba:2008sp, Albacete:2008ze, Gao:2009ze, Avsar:2009xf, BallonBayona:2009uy,Cornalba:2009ax, Gao:2010qk,Cornalba:2010vk, Brower:2011dx}.
Light and heavy mesons were studied in the soft-wall holographic approach \cite{Branz:2010ub}.

Unusual bound states of quarks are also interesting subjects to work in holographic QCD.
In \cite{Andreev:2008tv}, the multi-quark potential was calculated and tetra-quarks were discussed in AdS/QCD.
Based on holographic  quark-antiquark potential in the static limit, the masses of the states X(3872) or Y(3940) were predicted
and also tetra-quark masses with open charm and strangeness were computed in \cite{Carlucci:2007um}.
A hybrid exotic meson, $\pi_1$(1400), in \cite{Kim:2008qh}. The spectrum of baryons with two heavy quarks was predicted in
\cite{Giannuzzi:2009gh}.

Low-energy theorems of QCD and spectral density of the Dirac operator were studied in the soft wall model \cite{Kopnin:2009ny}.
A holographic model of hadronization was suggested in \cite{Evans:2007sf}.

The equation of state for a cold quark matter was
calculated  in the soft wall metric model with a U(1)
gauge field. The result is in agreement with phenomenology~\cite{Andreev:2010bv}.

\vskip 1cm \noindent {\large\bf Acknowledgments}\\

We thank Jihun Kim, Yumi Ko, Ik Jae Shin, and Takuya Tsukioka for useful comments on the manuscript.
We are slso grateful to Sergey Afonin, Oleg Andreev, Stanley J. Brodsky, Miguel Costa, Guy F. de Teramond,
Hilmar Forkel, Jian-Hua Gao, Marco Panero for their comments on the the manuscript.
YK  expresses his gratitude  to Hyun-Chul Kim, Kyung-il Kim, Yumi Ko, Bum-Hoon Lee , Hyun Kyu Lee, Sangmin Lee, Chanyong Park,
Ik Jae Shin, Sang-Jin Sin, Takuya Tsukioka,
Xiao-Hong Wu, Ulugbek Yakhshiev, Piljin Yi, and Ho-Ung Yee for collaborations in hQCD.
We acknowledge the Max Planck Society(MPG), the Korea Ministry of Education, Science and
Technology(MEST), Gyeongsangbuk-Do and Pohang City for the support of the Independent Junior
Research Group at APCTP.

\vskip 1cm

\appendix

\section{Bulk mass and the conformal dimension of boundary operator\label{Sec:MvsO}}

In this Appendix, we summarize the relation between the conformal dimension of a boundary operator and
the bulk mass of dual bulk field.
We work in the Euclidean version of $AdS_{d+1}$,
\begin{equation}
ds^2=\frac{1}{(x^0)^2}\sum_{\mu=0}^d (dx^\mu)^2.
\label{metric}
\end{equation}

\subsection{Massive scalar case}
We first consider a free massive scalar field whose  action is given by
\begin{equation}
S= \frac{1}{2}\int d^{d+1}x \sqrt{g}\left(\del_\mu\phi\del^\mu\phi+m^2\phi^2\right).
\end{equation}
Let the propagator of $\phi$ be $K(x^0,\vec{x};\vec{x}')$.
To solve for $\phi$ in terms of its boundary function $\phi_0$,
we look for a propagator of $\phi$, a solution $K(x^0,\vec{x};\vec{x}')$ of
the Laplace equation on $B_{d+1}$ whose boundary value is a delta function
at a point $P$ on the boundary.
We take $P$ to be the point at $x_0\rightarrow \infty$.
The boundary conditions and metric are invariant under translations of the $x_i$,
then we can consider $K$ as a function of only $x_0$, thus $K(x^0,\vec{x};P)=K(x^0)$.
Then, the equation of motion is
\begin{equation}
\left(-(x^0)^{d+1}\frac{d}{dx^0}(x^0)^{-d+1}\frac{d}{dx^0}+m^2 \right)K(x^0)=0\, ,
\end{equation}
where we used
\begin{equation}
\frac{1}{\sqrt{g}}\del_\mu\sqrt{g}\del^\mu
=(x^0)^{d+1}\frac{d}{dx^0}(x^0)^{-d+1}\frac{d}{dx^0}.
\end{equation}
We analyze the equation of motion near the boundary, $x^0\rightarrow 0$, and take $K(x^0)\propto (x^0)^{\lambda+d}$.
From the equation of motion, we have
\begin{equation}
-(\lambda+d)\lambda+m^2=0\, ,
\label{massive scalar rel}
\end{equation}
where $\lambda$ is the larger root $\lambda=\lambda_+$.
The conformal dimension $\Delta$ of the boundary operator is related to the mass m on $AdS_{d+1}$ space by
$\Delta=d+\lambda_+$.
Thus, we obtain
\begin{equation}
(\Delta-d)\Delta=m^2\, ,
\end{equation}
or
\begin{equation}
\Delta = \frac{1}{2}\left(d+\sqrt{d^2+4m^2}\right).
\end{equation}

\subsection{Massive p-form field case}
Consider a massive $p$-form potential \cite{l'Yi:1998eu}
\begin{equation}
{\cal A}={1\over p!} {\cal A}_{ \mu_1 \ldots \mu_p }dx^{\mu_1}\cdots dx^{\mu_p}\, .
\end{equation}
The free action of ${\cal A}$ is
\begin{equation}
S={1\over 2}\int_{AdS_{d+1}} \left(
 {\cal F}\wedge{}^{*}{\cal F} + m^2{\cal A}\wedge{}^{*}{\cal A}\right),
  \label{the_action}
\end{equation}
where ${\cal F} = d{\cal A}$ is the field strength $p+1$ form.
The variation of this action is
\begin{eqnarray*}
\delta S=\int_{AdS_{d+1}}
\left(-(-1)^p \delta{\cal A}\wedge{}d^{*}{\cal F} + m^2\delta{\cal A}\wedge{}^{*}{\cal A}\right)\, ,
\end{eqnarray*}
and then the classical equation of motion for ${\cal A}$ from (\ref{the_action}) is
\begin{equation}
(-1)^p d^{*}d{\cal A} -m^2\;{}^{*}{\cal A} = 0.
\label{em_in_cal_A}
\end{equation}
In addition, ${\cal A}$ satisfies $d^{*}{\cal A}=0$.
By using the metric (\ref{metric}), the equation of motion (\ref{em_in_cal_A})
can be written as
\begin{eqnarray}\label{eq_Ao}
&&\left[ (x^0)^2\partial_\mu^2 - (d+1-2p) x^0 \partial_0 + (d+1-2p-m^2)
    \right]{\cal A}_{0 {i_2} \ldots {i_p} } =0,\\
&&\left[ (x^0)^2\partial_\mu^2 - (d-1-2p) x^0 \partial_0 -m^2
  \right]  {\cal A}_{ {i_1} \ldots {i_p} } \\
&&\hskip1.5cm = 2x^0\left( \partial_{i_1}\omega_{0i_2\ldots i_p}
    + (-1)^{p-1} \partial_{i_2}\omega_{0i_3\ldots i_pi_1} + \cdots \right).
   \nonumber
\end{eqnarray}
Now from the vielbein $e_a^{\mu}=x^0\delta_a^\mu$,
we introduce fields with flat indices
\begin{equation}
A_{0i_2\ldots i_p} = (x^0)^{p-1}{\cal A}_{0i_2\ldots i_p},\quad
A_{i_1\ldots i_p} = (x^0)^{p}{\cal A}_{i_1\ldots i_p}. \label{flat_A}
\end{equation}
Then the equations of motion (\ref{eq_Ao}) of $A_{0 i_2\ldots i_p}$ becomes
\begin{eqnarray}
&&\left[ (x^0)^2\partial_\mu^2 - (d-1) x^0 \partial_0 - (m^2+p^2-pd)
    \right]A_{0 {i_2} \ldots {i_p} } =0.
\label{eq_Ao final}
\end{eqnarray}
We consider
\begin{equation}
A_{0 i_2\ldots i_p}\sim (x^0)^{-\lambda}
\end{equation}
as $x^0 \rightarrow 0$.
Then substituting this in (\ref{eq_Ao final}) gives
\begin{eqnarray*}
0&\!=\!&\left[(x^0)^2\del_0^2-(d-1)x^0\del_0-(m^2+p^2-pd)\right](x^0)^{-\lambda}\\
&\!=\!&\left[(x^0)^2\del_0(-\lambda(x^0)^{-\lambda-1})-(d-1)x^0(-\lambda(x^0)^{-\lambda-1})-(m^2+p^2-pd)(x^0)^{-\lambda}\right]\\
&\!=\!&\left[\lambda(\lambda+1)+\lambda(d-1)-(m^2+p^2-pd)\right](x^0)^{-\lambda}\\
&\!=\!&\left[\lambda(\lambda+d)-(m^2+p^2-pd)\right](x^0)^{-\lambda}
\end{eqnarray*}
and therefore we obtain the relation
\begin{equation}
\lambda(\lambda+d) = m^2 +p^2 -pd.
\label{eq_lambda}
\end{equation}
With $\Delta=d+\lambda$, we have
\begin{eqnarray*}
&&\quad(\Delta-d)\Delta = m^2 +p^2 -pd\\
&&\Rightarrow(\Delta-p)p+(\Delta-p)(\Delta-d) = m^2
\end{eqnarray*}
and we finally arrive at
\begin{equation}
(\Delta-d+p)(\Delta-p)=m^2,
\end{equation}
or
\begin{equation}
\Delta = \frac{1}{2}\left(d+\sqrt{(d-2p)^2+4m^2}\right).
\end{equation}

\subsection{General cases}
Now for completeness, we list the relations between the conformal dimension $\Delta$
and the mass for the various bulk fields in $AdS_{d+1}$.
\begin{enumerate}
\item
scalars \cite{Witten:1998qj}: $\Delta_{\pm} = \frac{1}{2}(d \pm \sqrt{d^2+4m^2})$,

\item
spinors \cite{Henningson:1998cd}:  $\Delta = \frac{1}{2}(d + 2|m|)$,

\item
vectors\footnote{Entries 3. and 4. are for forms with Maxwell type actions.}:
$ \Delta_{\pm} = {1 \over 2} (d \pm \sqrt{(d-2)^2 + 4m^2})$,

\item
$p$-forms \cite{l'Yi:1998eu}:
$ \Delta_{\pm} = {1 \over 2} (d \pm \sqrt{(d-2p)^2 + 4m^2})$,

\item
first-order $(d/2)$-forms ($d$ even)\footnote{
See \cite{Arutyunov:1998xt} for $d=4$ case.}:
$\Delta={1\over 2}(d + 2|m|)$,

\item
spin-3/2 \cite{Volovich:1998tj}\cite{Koshelev:1998tu}:
$\Delta = \frac{1}{2}(d + 2|m|)$,

\item
massless spin-2 \cite{Polishchuk:1999nh}:
$\Delta = d$.
\end{enumerate}

\section{ D3/D7 model and U(1) axial symmetry \label{D3/D7}}
In the original AdS/CFT, the duality between type IIB superstring theory on $AdS_5\times S^5$
and $\CN=4$ super Yang-Mills theory with gauge group $SU(N_c)$ can be embodied by the low-energy dynamics of a stack
of $N_c$ D3 branes in Minkowski space.
All matter fields in the gauge theory produced by the D3 branes are
in the adjoint representation of the gauge group.
To introduce the quark degrees of freedom in the fundamental representation,
we  introduce some other branes in this supersymmetry theory on top of the D3 branes.

\subsection{Adding flavour}\label{sec:flavour}

It was shown in \cite{Karch:2002sh} that by introducing $N_f$ D7 branes into $AdS_5 \times S^5$,
$N_f$ dynamical quarks can be added to the gauge theory, breaking the supersymmetry to $\CN=2$.
The simplest way to treat D3/D7 system is to work in the limit where the D7 is a probe brane,
which means that only a small number of D7 branes are added, while the number of D3 branes $N_c$ goes to infinity.
In this limit $N_f\ll N_c$  we may neglect the back-reaction of the D7 branes on $AdS_5 \times S^5$ geometry.
In field theory side, this corresponds to ignoring the quark loops,  quenching the gauge theory.

\begin{table}[!ht]
\begin{center}
\begin{tabular}{|r|c|c|c|c|c|c|c|c|c|c|}
\hline
& 0 & 1 & 2 & 3 & 4 & 5 & 6 & 7 & 8 & 9 \\
\hline
D3    & $\circ$ & $\circ$ & $\circ$ & $\circ$ & & & & & & \\
\hline
D7 & $\circ$ & $\circ$ & $\circ$ & $\circ$ & $\circ$ & $\circ$ & $\circ$ & $\circ$ & & \\
\hline
\end{tabular}
\caption{The D3/D7-brane intersection in $9+1$ dimensional flat space.}
\label{tb:D3D7setup}
\end{center}
\end{table}

The D7 branes are added in such a way that they extend parallel in Minkowski space and extend in spacetime as
given in Table \ref{tb:D3D7setup}.
The massless modes of open strings both ends on the $N_c$ D3 branes give rise to
 $\CN=4$ degrees of freedom of supergravity on $AdS_5\times S^5$ consisting of the $SU(N_c)$ vector bosons, four fermions and six scalars.
In the limit of large $N_c$ at fixed but large 't~Hooft coupling $\lambda=g_{YM}^2N_c=g_sN_c \gg 1$ ,
the D3 branes can be replace with near horizon geometry that is given by
\begin{align}\label{eq:AdS5S5metric}
ds^2&= \frac{r^2}{R^2}(-dt^2+dx_1^2+dx_2^2+dx_3^2)+\frac{R^2}{r^2}d\vec{y}^2\no
&= \frac{r^2}{R^2}(-dt^2+dx_1^2+dx_2^2+dx_3^2)+\frac{R^2}{r^2}\left(d\rho^2+ \rho^2d\Omega_3^2+dy_5^2+dy_6^2\right)
\end{align}
where $\vec{y}=(y_1,\ldots,y_6)$ parameterize the 456789 space and $r^2\equiv\vec{y}^2$.
$R$ is the radius of curvature $R^2=\sqrt{4\pi g_sN_c}\alpha'$ and $d\Omega_3^2$ is the three-sphere metric.
The dynamics of the probe D7 brane is described by the
combined DBI and Chern-Simons actions \cite{Leigh:1989jq,Polchinski:1998rq},
\begin{equation}
S_{D7} = -T_7\int d^8x
\sqrt{-\det\left(P[g]_{ab}+2\pi\alpha' F_{ab}\right)}
+  \frac{(2\pi\alpha')^2}{2} T_7\int P[C^{(4)}] \wedge F \wedge F\, .
\label{eq:D7action}
\end{equation}
where $g$ is the bulk metric (\ref{eq:AdS5S5metric})
and $C^{(4)}$ is the four form potential. $T_7=1/((2\pi)^7g_s\alpha'4)$ is the D7 brane tension
and $P$ denotes the pullback. $F_{ab}$ is the world-volume field strength.

The addition of D7 branes to this system as in Table \ref{tb:D3D7setup} breaks the supersymmetry to $\CN=2$.
The lightest modes of the 3-7 and 7-3 open strings
corresponds the quark supermultiplets in the field theory.
If the D7 brane and the D3 brane overlap then SO(6) symmetry is broken into
$SO(4)\times SO(2) \sim SO(2)_R\times SO(2)_L\times U(1)_R$
in the transverse directions to D3 and so preserves 1/4 of the supersymmetry.
The $SO(4)$ rotates in 4567, while the $SO(2)$ group acts on 89 in \ref{tb:D3D7setup}.
The induced metric on D7 takes the form, in general, as
\begin{equation}\label{eq:D7metric general}
ds_{D7}^2=\frac{r^2}{R^2}\eta_{\mu\nu}dx^\mu dx^\nu +\frac{R^2}{r^2}\left( (1+y_5'^2+y_6'^2)d\rho^2+ \rho^2d\Omega_3^2\right)
\end{equation}
where $y_5'=dy_5/d\rho$ and $y_6'=dy_6/d\rho$.
When the D7 brane and the D3 brane overlap, the embedding is
\begin{equation}
y_5=0,\qquad y_6=0
\end{equation}
and the induced metric on the D7 brane is replaced by
\begin{equation}\label{eq:D7metric 0 0}
ds_{D7}^2=\frac{\rho^2}{R^2}\eta_{\mu\nu}dx^\mu dx^\nu +\frac{R^2}{\rho^2}\left(d\rho^2+ \rho^2d\Omega_3^2\right)\, .
\end{equation}
 The D7 brane fills $AdS_5$ and is wrapping a three sphere
of $S^5$. In this case the quarks are massless and the R-symmetry of the theory is $SU(2)_R\times U(1)_R$ and
we have an extra $U(1)_R$ chiral symmetry.

If the D7 brane is separated from the D3 branes in the 89-plane direction by distance $L$,
then the minimum length string has non-zero energy and the quark gains a finite mass, $m_q=L/2 \pi \alpha'$.
It is known that the R-symmetry is then only $SU(2)_R$ and separation of D7 and D3 breaks
the $SO(2)\sim U(1)_R $ that acts on the 89-plane.
In this case, we can set for the embedding as
\begin{equation}
y_5=0, \qquad y_6=y_6 (\rho)\,.
\end{equation}
Then, the action for a static D7 embedding (with $F_{ab}$ zero on its world volume) becomes
\begin{align}
S_{D7} &= -T_7\int d^8x \sqrt{-\det\left(P[g]_{ab}\right)}
= -T_7\int d^8x \sqrt{-\det g_{ab}}\sqrt{1+g^{ab}\partial_a y_i \partial_b y_j g_{ij} }\no
&= -T_7\int d^8x \epsilon_3 \rho^3 \sqrt{1+(\partial_\rho y_5)^2+(\partial_\rho y_6)^2}\, ,
\end{align}
where $i,j=5,6$ and $\epsilon_3$ is the determinant from the three sphere.
The ground state configuration
of the D7 brane is given by the equation of motion with $y_5=0$
\begin{equation}
\frac{d}{d\rho}\left[ \frac{\rho^3\partial_\rho y_6}{\sqrt{1+(\partial_\rho y_6)^2}} \right]=0\, .
\end{equation}
The solution of this equation has an asymptotic behavior at UV ($\rho \rightarrow \infty$) as
\begin{equation}
y_6\simeq m+\frac{c}{\rho^2}+\ldots\, .
\end{equation}
Now we can identify \cite{Polchinski:2002jw} that
$m$ corresponds to the quark mass and $c$ is for the quark condensate $\langle\bar{\psi}\psi\rangle$
in agreement with the AdS/CFT dictionary.

\subsection{Chiral symmetry breaking}
One of the significant feature of QCD is chiral symmetry breaking by a quark condensate $\bar{\psi}\psi$.
The $U(1)$ symmetry under which $\psi$ and $\bar{\psi}$ transform as
$\psi \rightarrow e^{-i\alpha}\psi$, $\bar{\psi} \rightarrow e^{i\alpha}\bar{\psi}$ in the gauge theory
corresponds to a $U(1)$ isometry in the $y_5y_6$ plane transverse to the D7 brane.
This $U(1)$ symmetry can be explicitly broken by a non-vanishing quark mass
due to the separation of the D7 brane from the stack of D3 branes in the $y_5+iy_6$ direction.
Assume the embedding as $y_5=0$ and $y_6\sim c/\rho^2$, then by a small rotation $e^{-i\epsilon}$
on $y_5+iy_6$ generates $y_5'\simeq \epsilon c/\rho^2$ and $y_6'\simeq y_6$ up to the ${\cal O}(\epsilon^2)$ order.

In \cite{Babington:2003vm} the embedding of a D7 probe brane is embodied in the Constable-Myers background and
the regular solution $y_6\sim m+c/\rho^2$ of the embedding $y_5=0$, $y_6=y_6(\rho)$ shows the behavior
$c\neq 0$ as $m\rightarrow 0$ which corresponds to the spontaneous chiral symmetry breaking by a quark condensate.
In \cite{Ghoroku:2006nh}, the chiral symmetry breaking comes from a cosmological constant with a constant dilaton
configuration which is dual to the ${\cal N}=4$ gauge theory in a four-dimensional AdS space.

\subsection{Meson mass spectrum}

The open string modes with both ends on the flavour D7 branes are in the adjoint of the $U(N_f)$ flavour symmetry
of the quarks and hence can be interpreted as the mesonic degrees of freedom.
As an example, we discuss the fluctuation modes for the scalar fields (with spin 0) following the argument of \cite{Kruczenski:2003be}.
The directions transverse to the D7 branes are chosen to be $y_5$ and $y_6$ and the embedding is
\begin{equation}
y_5=0+\chi,\qquad y_6=L+\varphi
\end{equation}
where $\delta y_5=\chi$ and $\delta y_6=\varphi$ are the scalar fluctuations of the transverse direction.
To calculate the spectra of the world-volume fields it is sufficient to work to quadratic order.
For the scalars, we can write the relevant Lagrangian density as
\begin{align}
\CL_{D7} &= -T_7 \sqrt{-\det P[g]_{ab}}\no
&= -T_7 \sqrt{-\det g_{ab}}
\sqrt{1+g^{ab}\left(\partial_a\chi\partial_b\chi g_{55}+\partial_a\varphi\partial_b\varphi g_{66}\right)} \no
&= -T_7 \sqrt{-\det g_{ab}}
\sqrt{1+g^{ab}\frac{R^2}{r^2} \left(\partial_a\chi\partial_b\chi +\partial_a\varphi\partial_b\varphi \right)} \no
&\simeq-T_7 \sqrt{-\det g_{ab}}
\left(1+\frac{1}{2} \frac{R^2}{r^2}{g^{ab}} \left(\partial_a \chi \partial_b \chi+\partial_a \varphi \partial_b \varphi \right) \right)
\end{align}
where $P[g]_{ab}$ is the induced metric on the D7 world-volume.
In spherical coordinates with $r^2=\rho^2+L^2$, this  can be written as
\begin{eqnarray}
\CL_{D7} \simeq -T_7 \rho^3\epsilon_3 \left(1+\frac{1}{2} \frac{R^2}{\rho^2+L^2}
{g^{ab}} \left(\partial_a \chi \partial_b \chi+\partial_a \varphi \partial_b \varphi \right) \right)
\end{eqnarray}
where $\epsilon_3$ is the determinant of the metric on the three sphere.
Then the equations of motion becomes
\begin{equation}
\partial_a\left(\frac{\rho^3\epsilon_3}{\rho^2+L^2}g^{ab}\partial_b \Phi\right)=0\, ,
\end{equation}
where $\Phi$ is used to denote the real fluctuation either $\chi$ or $\varphi$.
Evaluating a bit more, we have
\begin{equation}\label{eq:expanded Phi eq}
\frac{R^4}{(\rho^2+L^2)^2}\partial_\mu\partial^\mu\Phi
+\frac{1}{\rho^3}\partial_\rho(\rho^3\partial_\rho\Phi)
+\frac{1}{\rho^2}\nabla_i \nabla^i \Phi=0\ ,
\end{equation}
where $\nabla_i$ is the covariant derivative on the three-sphere.
We apply the separation of variables to write the modes as
\begin{equation}\label{eq:wavefunc}
\Phi = \phi(\rho) e^{ik \cdot x} \CY^{\ell}(S^3)\ ,
\end{equation}
where $\CY^\ell(S^3)$ are the scalar spherical harmonics on $S^3$,
which transform in the $(\ell/2,\ell/2)$ representation of $SO(4)$ and satisfy
\begin{equation}
\nabla^i \nabla_i \CY^{\ell} = -\ell(\ell+2) \CY^{\ell}\ .
\end{equation}
The meson mass is defined by
\begin{equation}\label{eq:meson mass def}
M^2 \equiv -k^2\, .
\end{equation}
Now we define
$\varrho = \rho/L$ and  $\bar{M}^2 = -k^2R^4/L^2$, and
then the equation for $\phi(\rho)$ is
\begin{equation}\label{eq:phi eq}
\partial_\varrho^2\phi+\frac{3}{\varrho}\partial_\varrho\phi
+\left(\frac{\bar{M}^2}{(1+\varrho^2)^2}
-\frac{\ell(\ell+2)}{\varrho^2}\right)\phi=0\ .
\end{equation}
This equation was solved in \cite{Kruczenski:2003be} in terms of the hypergeometric function.
To solve the equation, we first set
\begin{equation}
\phi(\varrho)=\varrho^\ell(1+\varrho^2)^{-\alpha}P(\varrho),
\end{equation}
where
\begin{equation}
2\alpha=-1+\sqrt{1+\bar{M}^2}\geq 0\, .
\end{equation}
With a new variable $y=-\varrho^2$, (\ref{eq:phi eq}) becomes
\begin{equation}
y(1-y)P''(y)+[c-(a+b+1)y]P'(y)-abP(y)=0
\end{equation}
where $a=-\alpha$, $b=-\alpha +\ell+1$ and $c=\ell+2$. The general solution is taken
by $\alpha\geq 0$ and by noting that the scalar fluctuations are real
for $-\infty < y \leq 0$, one finds, up to a normalization constant, the solution of $\phi$
\begin{equation}
\phi(\rho)=\frac{\rho^\ell}{(\rho^2+L^2)^{\alpha}}
F\left(-\alpha ,\ -\alpha+\ell+1 ;\ \ell+2\ ;\ -\rho^2/L^2\right).
\end{equation}
Imposing the nomalizability at $\rho\to \infty$, we obtain
\begin{equation}\label{eq:quantization}
-\alpha+\ell+1=-n,\qquad n=0,1,2,\ldots \, .
\end{equation}
The solution is then
\begin{equation}
\phi(\rho)=\frac{\rho^\ell}{(\rho^2+L^2)^{n+\ell+1}}\
F\left(-(n+\ell+1)\ ,\ -n\ ;\ \ell+2\ ;\ -\rho^2/L^2\right)\
\end{equation}
and from the condition (\ref{eq:quantization}) we get
\begin{equation}
\bar{M}^2=4(n+\ell+1)(n+\ell+2)\ .
\end{equation}
Then by the definition of meson mass (\ref{eq:meson mass def}),
we derive the four-dimensional mass spectrum of the scalar meson
\begin{equation}
M_s(n,\ell)=\frac{2L}{R^2}\sqrt{(n+\ell+1)(n+\ell+2)}\ .
\end{equation}

\subsection{Mesons at finite temperature}

In previous sections, we have focused on gauge theories and their gravity dual
at zero temperature. To understand the thermal properties of gauge theories
using the holography, we work with the AdS-Schwarzschild
black hole which is dual to  $\CN=4$ gauge theory is at finite temperature \cite{Witten:1998qj,Witten:1998zw}.
The Euclidean AdS-Schwarzschild solution is given by
\begin{equation}\label{eq:AdSSchwarzschildmetric}
ds^2 = \frac{K(r)}{R^2} d\tau^2 + R^2 {\frac{dr^2}{ K(r)}} +
\frac{r^2}{R^2} d\vec{x}^2 + R^2 d\Omega_5^2 \, ,
\end{equation}
where
\begin{equation}
K(r) = r^2\left(1 - {\frac{r_H^4}{r^4}}\right) \,.
\end{equation}
For $r \gg r_H$, this approaches $AdS_5 \times S^5$ and the AdS radius $R$
is related to the 't~Hooft coupling by $R^2= \sqrt{4 \pi \lambda} \alpha'$.
Note that the $S^1$ parameterized by $\tau$ collapses at $r=r_H$, which
 is responsible for the existence of an area law of the Wilson loop
and a mass gap in the dual field theory.
This geometry is smooth and complete if the imaginary time $\tau$ is periodic with
the period $\beta= R^2\pi/r_H$. The temperature of the field theory corresponds to
the Hawking temperature is given by $T=1/\beta=r_H/(R^2\pi)$.
At finite temperature,
the fermions have anti-periodic boundary conditions in $\tau$ direction
\cite{Witten:1998zw} and the supersymmetry is broken. In addition, the adjoint scalars also become massive
at one loop. Thus, fermions and scalars decouple.

We now introduce  D7 branes in this background. It is convenient to change the variable
in the metric (\ref{eq:AdSSchwarzschildmetric}) such that the it possesses an explicit flat 6-plane.
To this end,  we change the variable from $r$ to $w$ as
\begin{equation}
\frac{dr^2}{K(r)}=\frac{r^2dr^2}{r^4-r_H^4}\equiv\frac{dw^2}{w^2}\, ,
\end{equation}
and take $R\equiv1$.
One of the solutions is
\begin{equation}
2 w^2=r^2+\sqrt{r^4-r_H^4}\quad\textrm{ or }\quad r^2=\frac{w^4+w_H^4}{w^2}
\end{equation}
and $w_H=r_H/\sqrt{2}$. Then the metric (\ref{eq:AdSSchwarzschildmetric}) becomes
\begin{align}\label{eq:AdSSchwarzschildmetric2}
ds^2
&=
\frac{(w^4-w_H^4)^2}{w^2(w^4 +w_H^4)}dt^2+
\left(w^2+\frac{w_H^4}{w^2}\right)d\vec{x}^2+
\frac{1}{w^2}(dw^2 + w^2 d\Omega_5^2) \no
&=
\frac{(w^4-w_H^4)^2}{w^2(w^4 +w_H^4)}dt^2+
\left(w^2+\frac{w_H^4}{w^2}\right)d\vec{x}^2+
\frac{1}{w^2}(d\rho^2 + \rho^2d\Omega_3^2 + dw_5^2 + dw_6^2)\, ,
\end{align}
where $d\Omega_3^2$ is the three-sphere metric. The D7 brane is embedded in the
static gauge, which worldvolume coordinates are now identified with $x_{0,1,2,3}$ and
$w_{1,2,3,4}$ and the transverse fluctuations will be parameterized by $w_5$ and $w_6$.
The radial coordinate is given by
\begin{eqnarray}
w^2=\sum_{i=1}^6w_i^2=\rho^2+w_5^2+w_6^2.
\end{eqnarray}
At large $w$, this geometry asymptotically approaches $AdS_5 \times S^5$
and the D7 embedding should approach the constant solutions $w_5 = 0, w_6 = const$
which is the same as the exact solution in section~\ref{sec:flavour}.
To consider the deformation, we take the following ansatz for the embedding
\begin{equation}
w_5=0,\qquad w_6 = w_6(\rho)
\end{equation}
Then, the action of D7 becomes
\begin{align}
S_{D7}
&=-T_7\int d^8x \sqrt{-\det P[g]_{ab}}
 = -T_7 \int d^8x \sqrt{-\det g_{ab}}
\sqrt{1+g^{ab}\left(\partial_aw_5\partial_bw_5 g_{55}+\partial_aw_6\partial_bw_6 g_{66}\right)} \no
&=-T_7\int d^8x ~\epsilon_3 ~ {\cal G}(\rho,w_5,w_6)
\sqrt{1 + (\partial_\rho w_5)^2+ (\partial_\rho w_6)^2}\, ,
\end{align}
where $\epsilon_3$ is the determinant of the three-sphere metric and the function ${\cal G}(\rho,w_5,w_6)$ is given by
\begin{align}
{\cal G}(\rho,w_5,w_6)
= \sqrt{\frac{(w^4-w_H^4)^2(w^4+w_H^4)^2\rho^6}{w^{16}}}
= \rho^3\left(1-\frac{w_H^8}{(\rho^2+w_5^2+w_6^2)^4}\right).
\end{align}
With the assumption $w_5=0$, the equation of motion takes the form
\begin{equation}\label{eq:eom w6}
\frac{d}{d\rho} \left[ \rho^3\left(1-\frac{w_H^8}{(\rho^2+w_6^2)^4}\right)\frac{\partial_\rho w_6}{\sqrt{1+(\partial_\rho w_6)^2}}\right]
- \frac{8 w_H^8 \rho^3 w_6}{( \rho^2 + w_6^2)^5}\sqrt{1+(\partial_\rho w_6)^2} = 0\,.
\end{equation}
With the solution of this equation, the induced metric on the D7 brane  is given by
\begin{equation}\label{eq:AdSBHD7metric}
ds_{D7}^2 =
\frac{(\tilde w^4 -  w_H^4)^2}{\tilde w^2 (\tilde w^4 +w_H^4)} dt^2
+ \left( \tilde w^2 +\frac{w_H^4}{\tilde w^2} \right) d\vec{x}^2
+ \frac{1+(\partial_\rho w_6)^2}{\tilde w^2} d\rho^2
+ \frac{\rho^2}{\tilde w^2} d\Omega_3^2 \, ,
\end{equation}
with $\tilde w^2=\rho^2+w_6^2(\rho)$ and the D7-brane metric becomes
$AdS_5 \times S^3$ for $\rho \gg w_H^{}, w_6^{}$.

\begin{figure}[!ht]
\begin{center}
\includegraphics[height=6cm,clip=true,keepaspectratio=true]{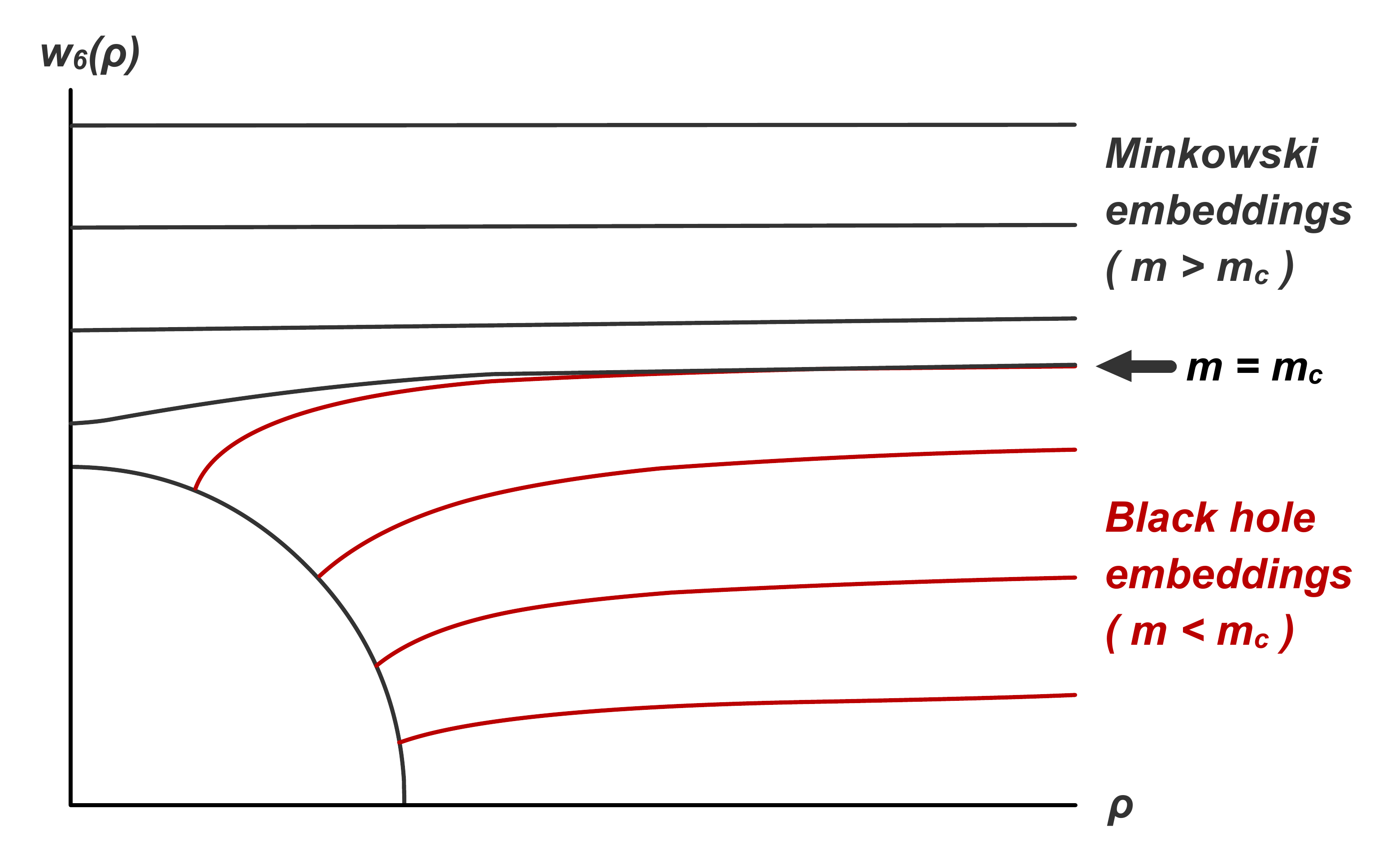}
\caption{Two classes of regular solutions in the AdS black hole background.}\label{fig:finiteTempBH}
\end{center}
\end{figure}

The asymptotic solution at large $\rho$ is of the form
\begin{equation}
w_6(\rho) \sim m + \frac{c}{\rho^2} \,.
\end{equation}
As discussed in section~\ref{sec:flavour},  the parameters $m$ and $c$
are interpreted as a quark mass and bilinear quark condensate $\langle \bar \psi \psi \rangle$, respectively.
With suitable boundary conditions
for the second order equation (\ref{eq:eom w6}), one can solve it numerically by the shooting method.
The results are shown in Figure~\ref{fig:finiteTempBH}.

We see that there can be two different classifications for the D7 brane embeddings.
First, for large quark masses the D7 brane ends outside the horizon.
It can be interpreted that the D7 brane tension is stronger than the attractive force of the black hole.
Such a D7 brane solution is called a {\em Minkowski embedding}.
They behave similarly to the supersymmetric solutions in $AdS_5 \times S^5$.
Second, for small masses the D7 brane ends at the horizon $w=w_H$ at which
the $S^1$ of the black hole geometry collapses. This is called a {\em black hole embedding}.
These two classes of embeddings also differ by their topology.
The D7 brane topology is $R^3 \times B^4 \times S^1$ for Minkowski embedding and $R^3 \times S^3 \times B^2$ for black hole embedding.
The change in the topology is related to a phase transition in the
dual field theory.

\section{ D4/D8/$\overline{\rm D8}$ model and non-Abelian chiral symmetry \label{SSmodel}}
The D3/D7 system is a supersymmetric configuration which gives gauge theories in the ultra-violet
and only has a $U(1)_A$ symmetry. To realize more realistic non-Abelian chiral symmetry,
Sakai and Sugimoto proposed D4/D8/$\overline{\rm D8}$
brane configuration \cite{Sakai:2004cn, Sakai:2005yt} which is intrinsically non-supersymmetric.
They placed probe D8 and $\overline{\rm D8}$ branes into the $N_c$ D4 background
of the fundamental representation of the $SU(N_c)$ gauge group.
The D4/D8/$\overline{\rm D8}$ intersection in 9+1 dimensional flat space is given in Table \ref{tb:D4D8setup}.

\begin{table}[!ht]
\begin{center}
\begin{tabular}{|r|c|c|c|c|c|c|c|c|c|c|}
\hline
& 0 & 1 & 2 & 3 & (4) & 5 & 6 & 7 & 8 & 9 \\
\hline
D4    & $\circ$ & $\circ$ & $\circ$ & $\circ$ & $\circ$ & & & & & \\
\hline
D8-$\overline{\rm D8}$ & $\circ$ & $\circ$ & $\circ$ & $\circ$ & & $\circ$ & $\circ$ & $\circ$ & $\circ$& $\circ$ \\
\hline
\end{tabular}
\caption{The D4/D8/$\overline{\rm D8}$ brane intersection in $9+1$ dimensional flat space.}
\label{tb:D4D8setup}
\end{center}
\end{table}

\begin{figure}[!ht]
\begin{center}
\includegraphics[width=8cm]{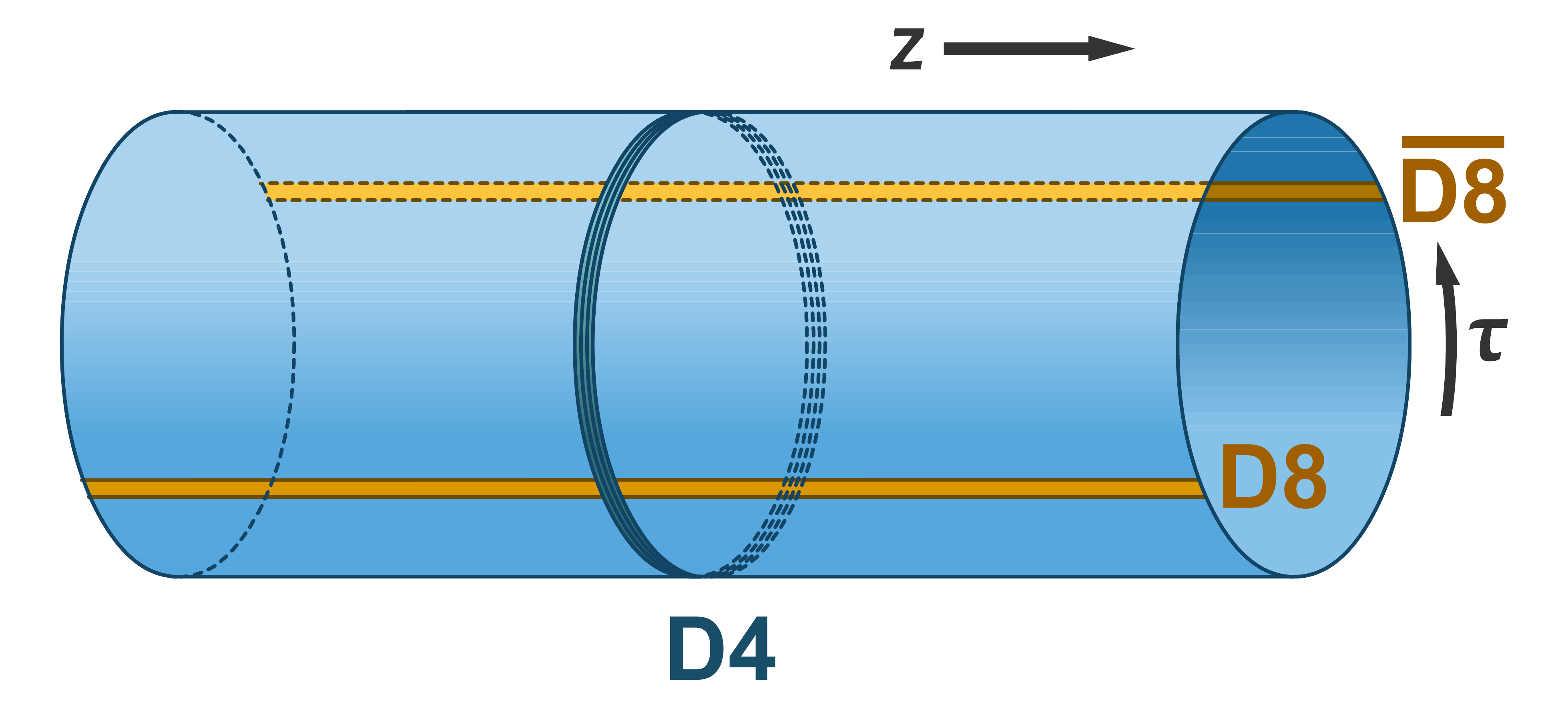}\hspace{.5cm}
\includegraphics[width=8cm]{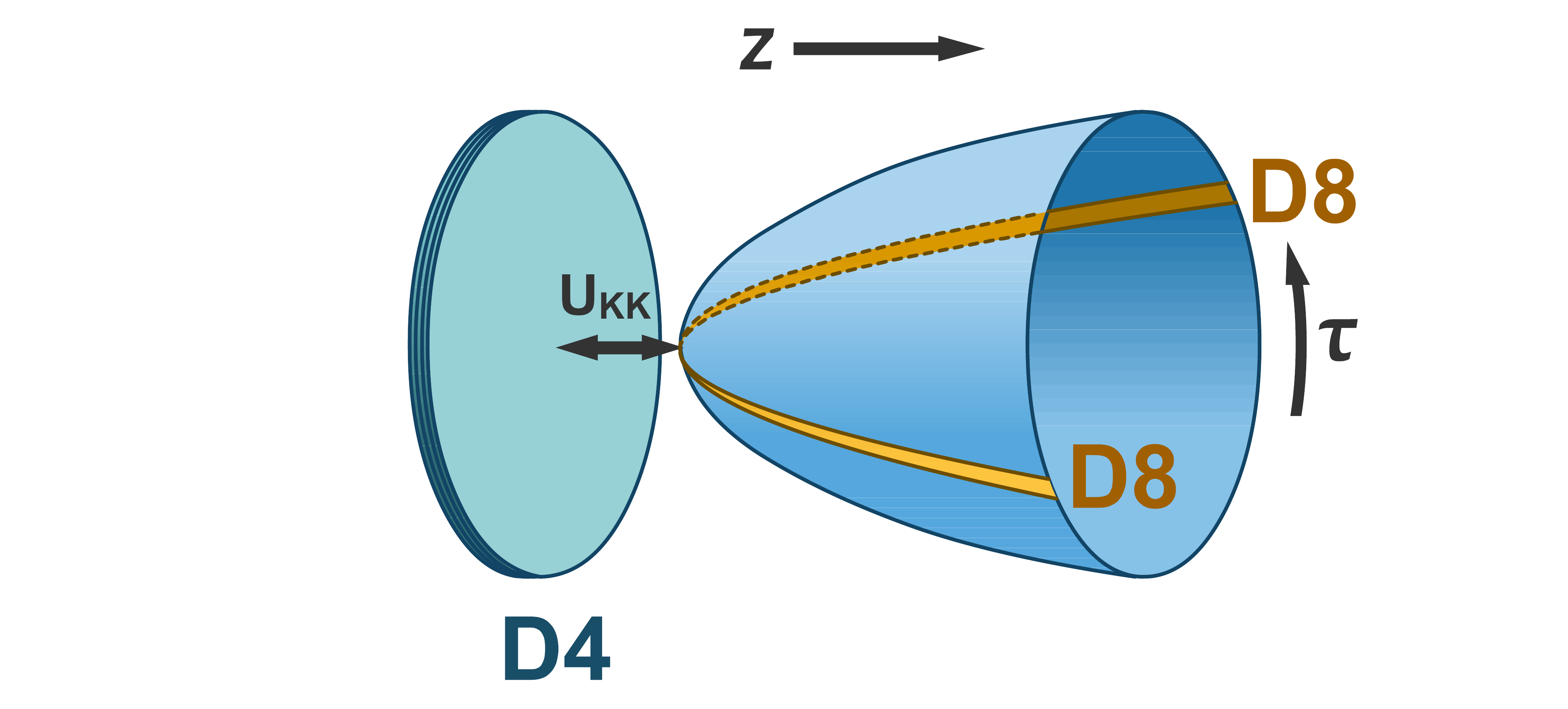}
\vspace{-0.3cm}
\end{center}
\caption{Sketch of the D4/D8/$\overline{\rm D8}$ configuration.}
\label{fig:D4D8}
\end{figure}

In this configuration, $N_c$ D4-branes are compactified on a direction ($x^4$) wrapped on  $S^1$.
We impose anti-periodic boundary conditions on this $S^1$ in order to break the supersymmetry.
$N_f$ D8 and $\overline{\rm D8}$ pairs are put in the transverse to the $S^1$.
Then, from the D4-D8 and D4-$\overline{\rm D8}$ open strings we obtain $N_f$ flavors
of massless chiral and anti-chiral quark fields of the $U(N_c)$ gauge group.
If we take the strong coupling limit to take the large $N_c$ dual,
 the D8 and $\overline{\rm D8}$ will overlap into a single curved D8 brane, which is interpreted
  as non-Abelian chiral symmetry breaking.
The $U(N_f)_{D8} \times U(N_f)_{\overline{D8}}$ gauge symmetry of the D8 and $\overline{\rm D8}$ branes
is interpreted as the $U(N_f)_{L} \times U(N_f)_{R}$ chiral symmetry.
These configurations are sketched in Figure~\ref{fig:D4D8}.

\subsection{Background D4  and probe D8 branes}

In order to obtain a holographic dual of the the large $N_c$ gauge theory with non-Abelian chiral symmetry,
we consider the SUGRA description
of the D4/D8/$\overline{\rm D8}$ system as discussed above. Assuming $N_f\ll N_c$, we treat
D8-$\overline{\rm D8}$ pairs as probe D8 branes embedded in the D4 background.
By taking the near horizon limit of the geometry of  $N_c$ stack of D4 branes  wrapped on a circle,
we obtain the D4 brane solution geometry
\begin{equation}
ds^2 = \left(\frac{U}{R}\right)^{3/2}
\left(\eta_{\mu\nu}dx^\mu dx^\nu +f(U) d\tau^2 \right) + \left(\frac{R}{U}\right)^{3/2}
\left(\frac{dU^2}{f(U)} +U^2 d\Omega_4^2 \right)
\end{equation}
with $f(U) \equiv 1- U_{KK}^3/U^3$.
Here $x^\mu$ ($\mu=0,1,2,3$) and $\tau$ are the directions along which the D4-brane is extended and
$R$ is the radius of curvature related to the string coupling $g_s$ and string length $l_s$ as $R^3=\pi g_s N_c l_s^3$.
There is a nonzero four-form flux $F_4=dC_3=2\pi N_c \epsilon_4/V_4$ with the volume form $\epsilon_4$
and a dilaton $e^{-\phi}=g_s \left(U/R\right)^{-3/4}$.
$U$ is the holographic direction and bounded from below by the condition $U\geq U_{KK}$.
As $U \rightarrow U_{KK}$ the radius of $S^1$ parameterized by $\tau$ shrinks into zero and
then the D8 and $\overline{\rm D8}$ branes are connected at some point $U=U_0$. In this case the
gauge symmetry is $U(N_f)$.
The configuration is then interpreted as a theory with massless quarks with chiral symmetry breaking, which
 is sketched on the right hand side of Figure~\ref{fig:D4D8}.
In order to avoid a singularity at $U=U_{KK}$, $\tau$ must be periodic with
period $\delta\tau\equiv=4\pi R^{3/2}/3U_{KK}^{1/2}$, which defines the Kaluza-Klein mass,
$M_{KK}=2\pi/\delta\tau=3U_{KK}^{1/2}/2R^{3/2}$.

Next, we consider the induced metric on the D8 probe brane in the D4 background with an ansatz $U=U(\tau)$.
Then, $dU^2=(dU/d\tau)^2\tau^2=U'(\tau)^2d\tau^2$ and the induced metric is given by
\begin{equation}\label{eq:D8metric}
ds^2_{D8} = \left(\frac{U}{R}\right)^{3/2}\eta_{\mu\nu}dx^\mu dx^\nu
+\left(\left(\frac{U}{R}\right)^{3/2}f(U)+\left(\frac{R}{U}\right)^{3/2}\frac{U'^2}{f(U)}\right)d\tau^2
+\left(\frac{R}{U}\right)^{3/2}U^2d\Omega_4^2\, ,
\end{equation}
The D8 brane DBI action now becomes
\begin{equation}\label{eq:D8DBIrough}
S_{D8}\sim\int d^4x d\tau \epsilon_4 e^{-\phi}\sqrt{-\det(g_{D8})}
\sim\int d^4x d\tau U^4 \sqrt{f(U)+\left(\frac{R}{U}\right)^{3} \frac{U'^2}{f(U)}}
\end{equation}
Since the integrand of (\ref{eq:D8DBIrough}) does not explicitly depend on $\tau$, we obtain
the energy conservation
\begin{equation}\label{eq:tau eom}
\frac{d}{d\tau}\left(\frac{\partial\CL}{\partial U'}\partial U'-\CL\right)
=\frac{d}{d\tau}\left(\frac{U^4f(U)}{f(U)+R^3U'^2/U^3f(U)}\right)=0.
\end{equation}
Let $U(0)=U_0$ and assume $U'(0)=0$ at $\tau=0$, then the solution of (\ref{eq:tau eom}) will be
\begin{equation}
\tau(U)=U_0^4f(U_0)^{1/2}\int_{U_0}^{U} \frac{dU}{(U/R)^{3/2}f(U)\sqrt{U^8f(U)-U_0^8f(U_0)}}.
\end{equation}
It can be shown that $\tau(U\rightarrow\infty)$ is a monotonically decreasing function of $U_0$,
roughly $\tau(\infty)\sim U_0^{-9/2}$, that is varying from $\tau(\infty)|_{U=U_{KK}}=\delta/4$ to
$\tau(\infty)|_{U\rightarrow\infty}=0$. In the limit $U_0=U_{KK}$, the D8 and $\overline{\rm D8}$ branes are
at antipodal points on the $S^1$ parameterized by $\tau$. And when $U_0\rightarrow \infty$,
the D8 and $\overline{\rm D8}$ separate far away from each other. Here we concentrate on the case
$U_0=U_{KK}$. For the sake of convenience, we perform the change of variables
\begin{equation}\label{eq:U theta}
U^3=U_{KK}^3+U_{KK}r^2,\qquad \theta=\frac{2\pi}{\delta\tau}\tau=\frac{3U_{KK}^{1/2}}{2R^{3/2}}\tau\, .
\end{equation}
Now, the induced metric (\ref{eq:D8metric}) on a D8 brane becomes
\begin{equation}\label{eq:D8metric2}
ds^2_{D8} = \left(\frac{U}{R}\right)^{3/2}\eta_{\mu\nu}dx^\mu dx^\nu
+\frac{4}{9}\left(\frac{U}{R}\right)^{3/2}\left(\frac{U_{KK}}{U}dr^2+r^2d\theta\right)
+\left(\frac{R}{U}\right)^{3/2}U^2d\Omega_4^2.
\end{equation}
We change the variables once more as
\begin{equation}
y=r\cos\theta,\qquad z=r\sin\theta\, ,
\end{equation}
accordingly, the $dr$ and $d\theta$ part of (\ref{eq:D8metric2}) read
\begin{align}
\frac{U_{KK}}{U}dr^2+r^2d\theta=
(1-h(r)z^2)dz^2+(1-h(r)y^2)dy^2-2h(r)zydzdy
\end{align}
with $h(r)=1/r^2(1-U_{KK}/U)$.
Near $U=U_{KK}$, (\ref{eq:D8metric2}) approaches a flat two-dimensional plane and $h(r)$ is a
regular function in the neighborhood of $r=0$. Then, $y(x^\mu,z)=0$ can be a solution of the
equation of motion of the probe D8 brane's world-volume theory and the stability of the solution
can be shown by examination of small fluctuations around it \cite{Sakai:2004cn}.
Thus, from (\ref{eq:U theta}) we have
\begin{equation}
U^3=U_{KK}^3+U_{KK}z^2
\end{equation}
and the induced metric on the D8 brane
\begin{equation}\label{eq:D8metric U z}
ds^2_{D8} = \left(\frac{U}{R}\right)^{3/2}\eta_{\mu\nu}dx^\mu dx^\nu
+\frac{4}{9}\left(\frac{U}{R}\right)^{3/2}\frac{U_{KK}}{U}dz^2
+\left(\frac{R}{U}\right)^{3/2}U^2d\Omega_4^2.
\end{equation}

\subsection{Gauge field and meson spectrum}
Now, we consider the gauge field on the probe D8 brane configuration.
The gauge field on the D8 brane $A_M$ ($M=0,1,2,3,5,6,7,8,z$) has nine components, and among them we are interested in the
$SO(5)$ singlet states, so we can set $A_\alpha=0$ for $\alpha=5,6,7,8$. We assume that $A_\mu$ and $A_z$
are independent of the coordinates on the $S^4$. Then the DBI action for $D8$ branes becomes
\begin{align}\label{eq:D8action}
S_{D8} &= -T_8\int d^9x \sqrt{-\det\left(g_{ab}+2\pi\alpha' F_{ab}\right)}+S_{CS}\no
&= -\tilde{T}(2\pi\alpha')^2\int d^4xdz
\left[\frac{R^3}{4U}\eta^{\mu\nu}\eta^{\rho\sigma}F_{\mu\nu}F_{\rho\sigma}+\frac{9}{8}\frac{U^3}{U_{KK}}\eta^{\mu\nu}F_{\mu z}F_{\nu z}\right]+{\cal O}(F^3)\,,
\end{align}
where $\tilde{T}=(2/3)R^{3/2}U_{KK}^{1/2}T_8 V_4 g_s^{-1}$.
We expand the gauge fields $A_{\mu}$ and $A_z$  in terms of the complete  sets
of the Kaluza-Klein (KK) modes profile functions $\{\psi_n(z)\}$ and $\{\phi_n(z)\}$, respectively,
\begin{align}
A_\mu(x,z) &=\sum_{n\geq 0}v_\nu^{(n)}(x)\psi_{n}(z), \\
A_z(x,z) &=\sum_{n\geq 0}\varphi^{(n)}(x)\phi_{n}(z).
\end{align}
 and consequently, the field strengths read
\begin{align}
F_{\mu\nu}(x,z) &=\sum_{n\geq 0}\left(\partial_\mu v_\nu^{(n)}(x)\psi_n(z)\right)
\equiv \sum_{n\geq 0} F_{\mu\nu}^{(n)}(x)\psi_n(z),\\\label{eq:Fmuz}
F_{\mu z}(x,z) &=\sum_{n\geq 0}\left(\partial_\mu \varphi^{(n)}(x)\phi_n(z)-v_\mu^{(n)}(x)\partial_z\psi_n(z)\right).
\end{align}
Then,  the action (\ref{eq:D8action}) becomes
\begin{align}\label{eq:D8action ver2}
S_{D8} =
& -\tilde{T}(2\pi\alpha')^2\int d^4xdz
\sum_{n,m\geq 0}
\Bigg[
\frac{R^3}{4U}F_{\mu\nu}^{(n)}F^{\mu\nu (n)}\psi_n\psi_m \no
&+\frac{9}{8}\frac{U^3}{U_{KK}}
\left(
v_\mu^{(n)}v^{\mu(n)}\partial_z\psi_n\partial_z\psi_m +\partial_\mu\varphi^{(n)}\partial^\mu\varphi^{(m)}\phi_n\phi_m
-2\partial_\mu \varphi^{(n)}v^{\mu(m)}\phi_n\partial_z\psi_m
\right)
\Bigg]
\end{align}
For convenience, we introduce a new variable $Z$ and the function $K(Z)$ as
\begin{equation}\label{eq:Z K(Z)}
Z\equiv z/U_{KK}\,, \quad \textrm{ and } \quad\,
K(Z)\equiv 1+Z^2
\end{equation}
With these, (\ref{eq:D8action ver2}) changes into
\begin{align}\label{eq:D8action ver3}
S_{D8} =&
-\tilde{T}(2\pi\alpha')^2R^3\int d^4x\,dZ \no
& \qquad \times
\sum_{n,m\geq 0}
\left[
\frac{1}{4}K^{-1/3}(Z)F_{\mu\nu}^{(n)}F^{\mu\nu (n)}\psi_n\psi_m
+\frac{1}{2}M_{KK}^2K(Z)v_\mu^{(n)}v^{\mu(n)}\partial_Z\psi_n\partial_Z\psi_m
\right] \no
&-\tilde{T}(2\pi\alpha')^2R^3\int d^4x\,dZ \,
\frac{1}{2}M_{KK}^2K(Z) \no
& \qquad \times
\sum_{n,m\geq 0}
\left(
U_{KK}^3\partial_\mu\varphi^{(n)}\partial^\mu\varphi^{(m)}\phi_n\phi_m
-2U_{KK}^2\partial_\mu \varphi^{(n)}v^{\mu(m)}\phi_n\partial_Z\psi_m
\right).
\end{align}
Now we choose $\psi_n(z)$ ($n\geq1$) as the eigenfunctions
\begin{equation}\label{eq:psi eigenfunc}
-K^{1/3}\partial_Z(K\partial_Z\psi_n)=\lambda_n\psi_n
\end{equation}
with the normalization condition
\begin{equation}\label{eq:psi normal cond}
\tilde{T}(2\pi\alpha')^2 R^3\int dZ K^{-1/3}\psi_m\psi_n=\delta_{mn},
\end{equation}
to arrive at
\begin{equation}\label{eq:delZpsi normal cond}
\tilde{T}(2\pi\alpha')^2 R^3\int dZ K \partial_Z\psi_m\partial_Z\psi_n=\lambda_n\delta_{mn}.
\end{equation}
Then, the action (\ref{eq:D8action ver3}) becomes
\begin{align}\label{eq:D8action ver4}
S_{D8} =
\int d^4x \sum_{n\geq 0}
\left[
-\frac{1}{4}F_{\mu\nu}^{(n)}F^{\mu\nu (n)}
-\frac{1}{2}M_{KK}^2\lambda_n v_\mu^{(n)}v^{\mu(n)}
\right] + \left(\,\varphi^{(n)} \textrm{ parts}\,\right).
\end{align}
Similarly, we can normalize the profile functions $\{\phi_n\}$ of $\varphi^{(n)}$.
By observation (\ref{eq:delZpsi normal cond}), we can choose
\begin{equation}\label{eq:phi_n phi_0}
\phi_n=\frac{1}{\lambda_n^{1/2} M_{KK}}\partial_Z\psi_n\quad(n\geq 1)\,,\quad \textrm{ and } \quad \phi_0=\frac{C}{K(Z)}.
\end{equation}
for some constant $C$. Then from (\ref{eq:delZpsi normal cond})
the orthonormal condition for $\{\phi_n\}$ becomes
\begin{equation}\label{eq:delZphi normal cond}
\tilde{T}(2\pi\alpha')^2 M_{KK}^2 R^3 \int dZ\, K \phi_m\phi_n = \delta_{mn}.
\end{equation}
The definitions in (\ref{eq:phi_n phi_0}) imply that  $\phi_0$ is orthogonal
to $\phi_n$ for $n \geq 1$ since $\int dZ\,K(Z)\phi_0\phi_n \sim \int dZ\,\partial_Z\psi_n=0$ from (\ref{eq:delZphi normal cond}).
Taking $n=m=0$ in (\ref{eq:delZphi normal cond}),  we  determine $C$ as
$C^{-2}=\tilde{T}(2\pi\alpha')^2 M_{KK}^2R^3 \pi$ since $\int dZ\,K(Z)^{-1}=\int dZ\,(1+Z^2)^{-1}=\pi$.
Note that $\psi_0$ can be determined as
\begin{equation}
\psi_0(z)=\int_0^zd\hat{z}\frac{\hat{C}}{1+\hat{z}^2}=\hat{C}\tan^{-1}z
\end{equation}
which is not normalizable. However, the field strength is normalizable and $\psi_0$ can be considered as the
zero mode of the eigenfunction equation (\ref{eq:psi eigenfunc}). Now from (\ref{eq:phi_n phi_0}), $F_{\mu z}$ becomes
\begin{align}\label{eq:Fmuz ver2}
F_{\mu z}(x,z)&=\sum_{n\geq 0}\left(\partial_\mu \varphi^{(n)}\phi_n-v_\mu^{(n)}\partial_z\psi_n\right) \no
&=\partial_\mu \varphi^{(0)}\phi_0 + \sum_{n\geq1}\left(\lambda_n^{-1/2}M_{KK}^{-1}\partial_\mu \varphi^{(n)} - v_\mu^{(n)} \right)\partial_z\psi_n.
\end{align}
We can absorb $\lambda_n^{-1/2}M_{KK}^{-1}\partial_\mu\varphi^{(n)}$ into $v_\mu^{(n)}$, and then we obtain
\begin{align}\label{eq:Fmuz ver3}
F_{\mu z}(x,z)=\partial_\mu \varphi^{(0)}\phi_0 - \sum_{n\geq1}v_\mu^{(n)}\partial_z\psi_n.
\end{align}
Then again with the change of variable (\ref{eq:Z K(Z)}), the action (\ref{eq:D8action}) can be written as
\begin{align}\label{eq:D8action with redefined v}
S_{D8} =&
-\tilde{T}(2\pi\alpha')^2R^3\int d^4x\,dZ \no
& \qquad \times
\sum_{n,m\geq 0}
\left[
\frac{1}{4}K^{-1/3}(Z)F_{\mu\nu}^{(n)}F^{\mu\nu (n)}\psi_n\psi_m
+\frac{1}{2}M_{KK}^2K(Z)v_\mu^{(n)}v^{\mu(n)}\partial_Z\psi_n\partial_Z\psi_m
\right] \no
&-\tilde{T}(2\pi\alpha')^2R^3\int d^4x\,dZ \,
\frac{1}{2}M_{KK}^2K(Z) U_{KK}^2 \partial_\mu\varphi^{(0)}\partial^\mu\varphi^{(0)}\phi_0^2.
\end{align}
With the normalization conditions (\ref{eq:psi normal cond}), (\ref{eq:delZpsi normal cond}), and $C\rightarrow C/U_{KK}$,
 the action (\ref{eq:D8action with redefined v}) becomes
\begin{align}
S_{D8} =& \int d^4x \left[ -\frac{1}{2}\partial_\mu\varphi^{(0)}\partial^\mu\varphi^{(0)}
+\sum_{n\geq 1} \left( -\frac{1}{4}F_{\mu\nu}^{(n)}F^{\mu\nu (n)} -\frac{1}{2}m_n^2v_\mu^{(n)}v^{\mu(n)}\right)\right],
\end{align}
where
\begin{equation}
m_n^2\equiv M_{KK}^2\lambda_n\, .
\end{equation}
The KK modes of $A_\mu$ are $v_\mu^{(n)}$ ($n\geq 1$) and they are regarded as the massive vector meson field.
Also, we interpret $\varphi^{(0)}$, which is the KK mode of $A_z$, as the pion field or the Nambu-Goldstone boson of the chiral symmetry breaking.

By solving (\ref{eq:psi eigenfunc}) numerically, we find the eigenvalues corresponding to
the masses of vector mesons
\begin{equation}
\lambda_n=0.67, \quad 1.6, \quad 2.9, \quad 4.5, \quad \ldots\, ,
\end{equation}
and we compare the meson mass ratio obtained in this model with the experimental data
\begin{align}
\frac{\lambda_2}{\lambda_1}=2.4 & \longleftrightarrow \frac{m_{a_1}^2}{m_\rho^2}=2.51, \no
\frac{\lambda_3}{\lambda_1}=4.3 & \longleftrightarrow \frac{m_{\rho(1450)}^2}{m_\rho^2}=3.56.
\end{align}

\section{Surface gravity of a Schwarzschild black hole\label{BH_how_to}}

The surface gravity is the gravitational acceleration experienced by a test body
(with negligible mass) close to the surface of an object.
For a black hole, the surface gravity is defined as the acceleration of gravity
at the horizon.  The acceleration of a test body at a black hole event horizon
is infinite in relativity, therefore one defines the surface gravity
in a different way, corresponding to the Newtonian surface gravity
in the non-relativistic limit. Thus for a black hole, the surface gravity is defined
in terms of the Killing vector which is orthogonal to the horizon
and here its event horizon is a Killing horizon.
For the Schwarzschild case this value is well defined.
Alternatively, one can derive the same value as a period of the imaginary time in the
Euclidean signature. we will discuss both points of view.

\subsection{Orthogonal Killing Vector}
The horizon of a black hole is a null surface.
It means that any vector normal to the surface is a null vector.
Let us consider the Killing vector that generates time translations,
$\mathbf{\xi}=\xi^\mu\bfe_\mu$. In the Schwarzschild spacetime this vector is simply
$\mathbf{\xi}=e_t$. This Killing vector is normal to the horizon, so that
$\xi^\mu\xi_\mu=0$ and this is why the event horizon is called a Killing horizon.
More specifically, $\xi^\mu\xi_\mu$ is constant on the horizon
thus the gradient $\nabla^\alpha(\xi^\mu\xi_\mu)$ is also normal to the horizon.
Hence, there exists a function $\kappa$ such that
\begin{equation}
\nabla^\alpha(\xi^\mu\xi_\mu)=-2\kappa\xi^\alpha.
\end{equation}
Since the field $\xi$ is a Killing vector it satisfies
$\nabla_\mu\xi_\nu+\nabla_\nu\xi_\mu=0$, then the above equation
can be rewritten as
\begin{equation}
\xi^\nu\nabla_\mu\xi_\nu=-\kappa\xi_\mu.
\end{equation}
By using the hypersurface orthogonal vector property $\xi_{[\mu}\nabla_\nu\xi_{\rho]}=0$, one can find the relation
\begin{equation}
\kappa^2=-\frac{1}{2}(\nabla_\mu\xi_\nu)(\nabla^\mu\xi^\nu)
\end{equation}
evaluated at the horizon. The Schwarzschild metric is diagonal
so we have $\xi^\mu=\delta^\mu_t$ and $\xi_\mu=\delta_{\mu t}g_{tt}$
and for evaluating the covariant derivative, the only nonzero $\xi_{\nu,\mu}$
is $\xi_{t,r}=g_{tt,r}$ since the metric components  dependent only on r.
Thus we get \cite{Wald:1984rg}
\begin{equation}
\label{kappa killing}
\kappa=\sqrt{-\frac{1}{4}g^{rr}g^{tt}(g_{tt,r})^2}
=\frac{1}{2}\frac{|\del_rg_{tt}|}{\sqrt{-g_{tt}g_{rr}}}.
\end{equation}

\subsection{Conical Singularity}
The expression (\ref{kappa killing}) can be obtained from another method.
Notice that the horizon property implies that near $r=r_H$,
the matric takes the following form
\begin{equation}
ds^2\sim -A(r-r_H)dt^2+\frac{dr^2}{B(r-r_H)}+r^2d\Omega_{n-1}^2
\end{equation}
This metric has a coordinate singularity at the horizon, but only if the Euclidean
time $\tau=it$ is periodic with a particular period $\tau\sim \tau+\beta$,
which one then define the inverse temperature $\beta=1/T$.
The metric read in Euclidean signiture
\begin{equation}
\label{euclidean metric}
ds^2\sim A(r-r_H)d\tau^2+\frac{dr^2}{B(r-r_H)}+r^2d\Omega_{n-1}^2\, .
\end{equation}
With
\begin{equation}
d\rho^2=\frac{dr^2}{B(r-r_H)},\qquad \rho=\frac{2}{\sqrt{B}}\sqrt{r-r_H}\, .
\end{equation}
 the metric (\ref{euclidean metric}) becomes
\begin{equation}
ds^2\sim \rho^2\kappa^2d\tau^2 +d\rho^2+r^2d\Omega_{n-1}^2,
\end{equation}
where $\kappa=\sqrt{AB}/2$.
This geometry is regular if the Euclidean time $\tau$ is periodic with a
period
\begin{equation}
\beta=\frac{2\pi}{\kappa}.
\end{equation}
Here we have defined $\kappa$ and by the first law of the (stationary)
black hole thermodynamics, this $\kappa$ corresponds to the surface gravity
and the black hole temperature is given by
\begin{equation}
T=\frac{1}{\beta}=\frac{\kappa}{2\pi}.
\end{equation}
This will give the same result as the one obtained by direct calculation
of the surface gravity (\ref{kappa killing}), but this
Euclidean signature formulation has
the advantage of showing subtleties for the noncanonical horizon cases,
and verifying the relation between Hawking temperature and surface gravity.

\section{Black hole temperature and entropy}\label{BH}
\subsection{AdS black holes}
The $AdS_5$ black hole  is
\begin{equation}
\label{AdS metric}
ds^2=\frac{L^2}{z^2}\left(-fdt^2+d\vx^2+\frac{dz^2}{f}\right)
\end{equation}
where $f(z)=1-z^4/z_m^4$. We expand the denominator of $g_{zz}$ near $z\simeq z_m$ as
\begin{eqnarray}
z^2f(z)&\simeq& z_m^2f(z_m)+\left(2z_mf(z_m)+z_m^2f'(z_m)\right)(z-z_m)\no
&=&z_m^2f'(z_m)(z-z_m)\, ,
\end{eqnarray}
and for $g_{tt}$ part, we have
\begin{eqnarray}
\frac{f}{z^2}&\simeq& \frac{f(z_m)}{z_m^2}+\frac{f'(z_m)z_m^2-2f(z_m)z_m}{z_m^4}(z-z_m)\no
&=& \frac{f'(z_m)}{z_m^2}(z-z_m).
\end{eqnarray}
Thus the near horizon (Euclidean) metric is
\begin{equation}
ds^2\sim L^2\frac{f'(z_m)}{z_m^2}(z-z_m) d\tau^2+\frac{L^2}{z_m^2}d\vx^2
+\frac{L^2}{z_m^2f'(z_m)(z-z_m)}dz^2.
\end{equation}
If we take
\begin{equation}
d\rho=\frac{L}{z_m\sqrt{f'(z_m)}}\frac{dz}{\sqrt{z-z_m}}
=\frac{2L}{z_m\sqrt{f'(z_m)}}d\left(\sqrt{z-z_m}\right),
\end{equation}
then, the near horizon metric becomes
\begin{equation}
ds^2\sim \kappa^2 \rho^2 d\tau^2+d\rho^2+\frac{L^2}{z_m^2}d\vx^2
\end{equation}
where
\begin{equation}
\kappa=\frac{|f'(z_m)|}{2}=\frac{4}{z_m}.
\end{equation}
This also can be calculated by the relation (\ref{kappa killing})
directly from (\ref{AdS metric}) as
\begin{equation}
\kappa=\lim_{z\rightarrow z_m}\frac{1}{2}\frac{|\del_zg_{tt}|}{\sqrt{-g_{tt}g_{zz}}}
=\lim_{z\rightarrow z_m}\frac{|2zf-z^2f'|}{2z^2}=\frac{|f'(z_m)|}{2}.
\end{equation}
Now, the temperature is given by
\begin{equation}
T=\frac{\kappa}{2\pi}=\frac{1}{\pi z_m}.
\end{equation}
The area of the horizon is
\begin{equation}
A=\int_{z=z_m\textrm{, t fixed}}d^3\vx\sqrt{g_\bot}=\frac{L^3}{z_m^3}V_3
\end{equation}
where $g_\bot$ is the determinant of the transverse part of the metric
and $V_3$ is the volume.
Then by the first law of the black hole thermodynamics, the entropy is
\begin{equation}
S=\frac{A}{4G_5}=\frac{L^3V_3}{4G_5z_m^3}.
\end{equation}

\subsection{Another ansatz}
Now, we consider the following metric ansatz \cite{Gubser:2008ny}
\begin{equation}
\label{ansatz metric}
ds^2=e^{2A}\left(-hdt^2+d\vx^2\right)+e^{2B}\frac{dr^2}{h},
\end{equation}
where $A$, $B$ and $h$ are some functions of $r$. We assume that
the geometry is asymptotically $AdS$. A regular horizon $r=r_H$ arises when $h$
has a simple zero. It is also assumed that $A(r)$ and $B(r)$
are finite and regular functions at $r=r_H$.
We consider the metric near the horizon.
For  $1/g_{rr}$,
\begin{equation}
\frac{h(r)}{e^{2B(r)}}\simeq
\frac{h(r_H)}{e^{2B(r_H)}}+
\left(\frac{h'(r_H)}{e^{2B(r_H)}}-\frac{2h(r_H)B'(r_H)}{e^{2B(r_H)}}\right)(r-r_H).
\end{equation}
Since $B$ is finite and $h(r_H)=0$, we obtain
\begin{equation}
\frac{h(r)}{e^{2B(r)}}\simeq
\frac{h'(r_H)}{e^{2B(r_H)}}(r-r_H).
\end{equation}
Similarly, for $g_{tt}$ part we get
\begin{equation}
h(r)e^{2A(r)}\simeq h'(r_H)e^{2A(r_H)}(r-r_H).
\end{equation}
Then, the near horizon geometry with Euclidean time $\tau=it$ becomes
\begin{equation}
ds^2\simeq
h'(r_H)e^{2A(r_H)}(r-r_H)d\tau^2+e^{2A(r_H)}d\vx^2+\frac{e^{2B(r_H)}}{h'(r_H)(r-r_H)}dr^2.
\end{equation}
With
\begin{equation}
d\rho=\frac{e^{B(r_H)}dr}{\sqrt{h'(r_H)}\sqrt{r-r_H}}
=\frac{2e^{B(r_H)}}{\sqrt{h'(r_H)}}d\left(\sqrt{r-r_H}\right)\, ,
\end{equation}
we arrive at
\begin{equation}
ds^2\simeq
\rho^2\frac{|h'(r_H)|^2e^{2A(r_H)}}{4e^{2B(r_H)}}d\tau^2+d\rho^2+e^{2A(r_H)}d\vx^2.
\end{equation}
Thus, we obtain the surface gravity
\begin{equation}
\kappa=\frac{|h'(r_H)|}{2}e^{A(r_H)-B(r_H)}.
\end{equation}
Again, using (\ref{kappa killing}),
this process also can be done directly from (\ref{ansatz metric}) as
\begin{equation}
\kappa=\lim_{r\rightarrow r_H}\frac{1}{2}\frac{|\del_rg_{tt}|}{\sqrt{-g_{tt}g_{rr}}}
=\lim_{r\rightarrow r_H}
\frac{1}{2}\left|\frac{d}{dr}he^{2A}\right|\frac{1}{\sqrt{e^{2A}e^{2B}}}
=\frac{|h'(r_H)|}{2}e^{A(r_H)-B(r_H)}.
\end{equation}
Now we define the temperature
\begin{equation}
T=\frac{e^{A(r_H)-B(r_H)}|h'(r_H)|}{4\pi}.
\end{equation}
Now, the area is $A=e^{3A(r_H)}$ for unit volume
and $G_N=\kappa_5^2/8\pi$.
Thus the entropy density takes the form
\begin{equation}
S=\frac{A}{4G_N}=\frac{2\pi}{\kappa_5^2}e^{3A(r_H)}.
\end{equation}

\section{Hawking-Page transition and deconfinement \label{HP_app}}

The partition function for canonical ensemble is given by
\begin{equation}
Z \simeq e^{-I(\phi_{cl})},
\end{equation}
where $\phi_{cl}$ is a classical solution of the equation of motion
with suitable boundary conditions. When there are multiple classical configurations,
we should sum over the all contributions or may take the absolute minimum
which globally minimizes $I_{SUGRA}$ and dominates the path integral
for the leading contribution.
If there are two or more solutions to minimize $I_{SUGRA}$, there may be
a phase transition between them. This is the Hawking-Page transition \cite{Hawking:1982dh},
a thermal phase transition to a black hole geometry in asymptotically AdS space.
Through the AdS/CFT correspondence, it was generalized in \cite{Witten:1998qj,Witten:1998zw}
that the corresponding dual of this transition is confinement-deconfinement transition
in boundary gauge theories.

\subsection{Hawking-Page phase transition}

The gravity action we study here is
\begin{equation}\label{eq:Iaction}
I=-\frac{1}{16\pi G_5}\int d^5 x \sqrt{g}\left(\CR + \frac{12}{R^2}\right)\, ,
\end{equation}
where $G_5$ is the five-dimensional Newton's constant and $R$ is the radius of $AdS_5$.
For a solution of the equation of motion,
\begin{equation}
\CR_{\mu\nu}-\frac{1}{2}\CR g_{\mu\nu}-\frac{6}{R^2} g_{\mu\nu}= 0,
\end{equation}
the action (\ref{eq:Iaction}) becomes
\begin{equation}\label{eq:Isolved}
I = \frac{1}{2\pi G_5 R^2} \int d^5 x \sqrt{g}= \frac{1}{2\pi G_5 R^2}V(\epsilon)
\end{equation}
where $V(\epsilon)$ is the volume of the space time.

There is a solution into AdS space which minimize (\ref{eq:Isolved}).
In this case, the $AdS_5$ metric is given by
\begin{equation}\label{eq:AdSmetric}
ds^2=
\left(1+\frac{r^2}{R^2}\right) d\tau^2
+\left(1+\frac{r^2}{R^2}\right)^{-1} dr^2
+ r^2 d\Omega_3^2 ,
\end{equation}
where $d\Omega_3^2$ is a metric on the three sphere $S^3$.
We call this solution as $X_1$. The Euclidean time direction is chosen that
the asymptotic boundary at $r \rightarrow \infty$ becomes $R^3 \times S^1$.
Thus the dual gauge theory lives on the spatial manifold $S^3$.

There is another solution, $X_2$, on a space with a certain temperature.
The geometry is now the AdS-Schwarzschild black hole and the metric is given by
\begin{equation}\label{eq:AdSBHmetric}
ds^2 = \left(1+\frac{r^2}{R^2} -\frac{\mu}{r^2}\right) d\tau^2
+\left(1+\frac{r^2}{R^2} -\frac{\mu}{r^2}\right)^{-1} dr^2
+ r^2 d\Omega_3^2 ,
\end{equation}
where $\mu=16\pi G_5 M/3V(S^3)$.
The radial direction is restricted to $r \geq r_+$,
where $r_+$ is the largest root of the equation
\begin{equation}
1+\frac{r^2}{R^2} -\frac{\mu}{r^2} =0.
\end{equation}
The metric is smooth and complete if $\tau$ is periodic with the period
\begin{equation}\label{eq:b0}
\beta_0 =\frac{2\pi R^2 r_+}{2r_+^2+R^2}.
\end{equation}
The topology of $X_2$ $R^2\times S^3$ and the boundary is $R^3\times S^1$.

The geometry $S^3\times S^1$ for at large $r$ from both $X_1$ and $X_2$ configurations are explained.
Let the $S^1$ radius be $\beta= (r/R)\beta_0$ and the $S^3$ radius be $\tilde{\beta}= r/R$.
If we wish to make the topology $S^3\times S^1$, we should take $\beta/\tilde{\beta}=\beta_0\rightarrow 0$
and this is the limit of large temperatures.
From (\ref{eq:b0}) it seems that this can be done with either $r_+ \rightarrow 0$ or $r_+\rightarrow\infty$,
but the $r_+\rightarrow 0$ branch is thermodynamically excluded \cite{Hawking:1982dh} and
we have the large $r_+$ branch, corresponding to large $M$. Therefore for either $X_1$ or $X_2$,
the topology becomes $S^3\times S^1$ at large $r$.

Actually, both $I(X_1)$ and $I(X_2)$ are infinite, so
we compute $I(X_2)-I(X_1)$ to get a finite result.
We put a cut-off $R_0$ in the radial direction $r$.
Then, the regularized volume of the AdS spacetime for $X_1$ is given by
\begin{equation}
V_1(R_0)=\int_0^{\beta'} d\tau\int_0^{R_0} dr \int_{S^3}\, d\Omega r^3.
\end{equation}
And for $X_2$ case of the AdS-Schwarzschild black hole, it is
\begin{equation}
V_2(R_0)=\int_0^{\beta_0} d\tau\int_{r_+}^{R_0} dr \int_{S^3}\, d\Omega r^3.
\end{equation}
In order to compare the $I(X_1)$ and $I(X_2)$, we match period of $\tau$
so that the proper circumference of the Euclidean time direction at $r=R_0$
is the same each other. This can be done by setting
\begin{equation}
\beta' \sqrt{\frac{r^2}{R^2}+1}=\beta_0 \sqrt{\frac{r^2}{R^2}+1-\frac{\mu}{r^2}}
\end{equation}
at $r=R_0$ and we determine $\beta'$. Then the difference $I(X_2)-I(X_1)$ becomes
\begin{equation}\label{eq:I1I2diff}
I=\frac{1}{2\pi G_5 R^2}
\lim_{R_0\rightarrow \infty}
(V_2-V_1)=
\frac{V(S^3)r_+^3(R^2  - r_+^2)}{4G_5(4 r_+^2 + 2 R^2)}.
\end{equation}
Then (\ref{eq:I1I2diff}) changes its sign at $r_+ = R$ and the phase transition
to the AdS-Schwarzschild black hole geometry takes place.

\subsection{Confinement and deconfinement}

Deconfinement at high temperature can be understood by the spontaneous breaking
of the center of the gauge group.  The corresponding order parameter  is the Polyakov loop
that is defined by a Wilson loop wrapping around $\tau$ direction as
\begin{equation}
{\cal P} = \frac{1}{N_c} \tr P e^{i \int_0^\beta d\tau A_0},
\end{equation}
where $P$ denotes the path ordered configuration.
In our case the gauge group is $SU(N_c)$ and its center is $\mathbb{Z}_{N_c}$.

The gravity dual calculation for the expectation value $\langle{\cal P}\rangle$
was performed in \cite{Rey:1998ik,Maldacena:1998im}
from the regularized area of the minimal surface ending on the loop,
\begin{equation}
\langle{\cal P}\rangle \sim e^{-\mu{\cal A}}
\end{equation}
where $\mu$ is the fundamental string tension, and  ${\cal A}$ is the area of the minimal surface
ending on the loop.
The boundary of the spaces $X_1$ and $X_2$ is $S^3 \times S^1$
and $\tau$ direction wraps around the loop $C$.
To compute $\langle {\cal P}(C)\rangle$, we should evaluate the partition function
of strings with its worldsheet $D$ which is bounded by the loop $C$.

\begin{figure}
\centering
\mbox{%
\subfigure[thermal AdS]{%
\includegraphics[width=5cm]{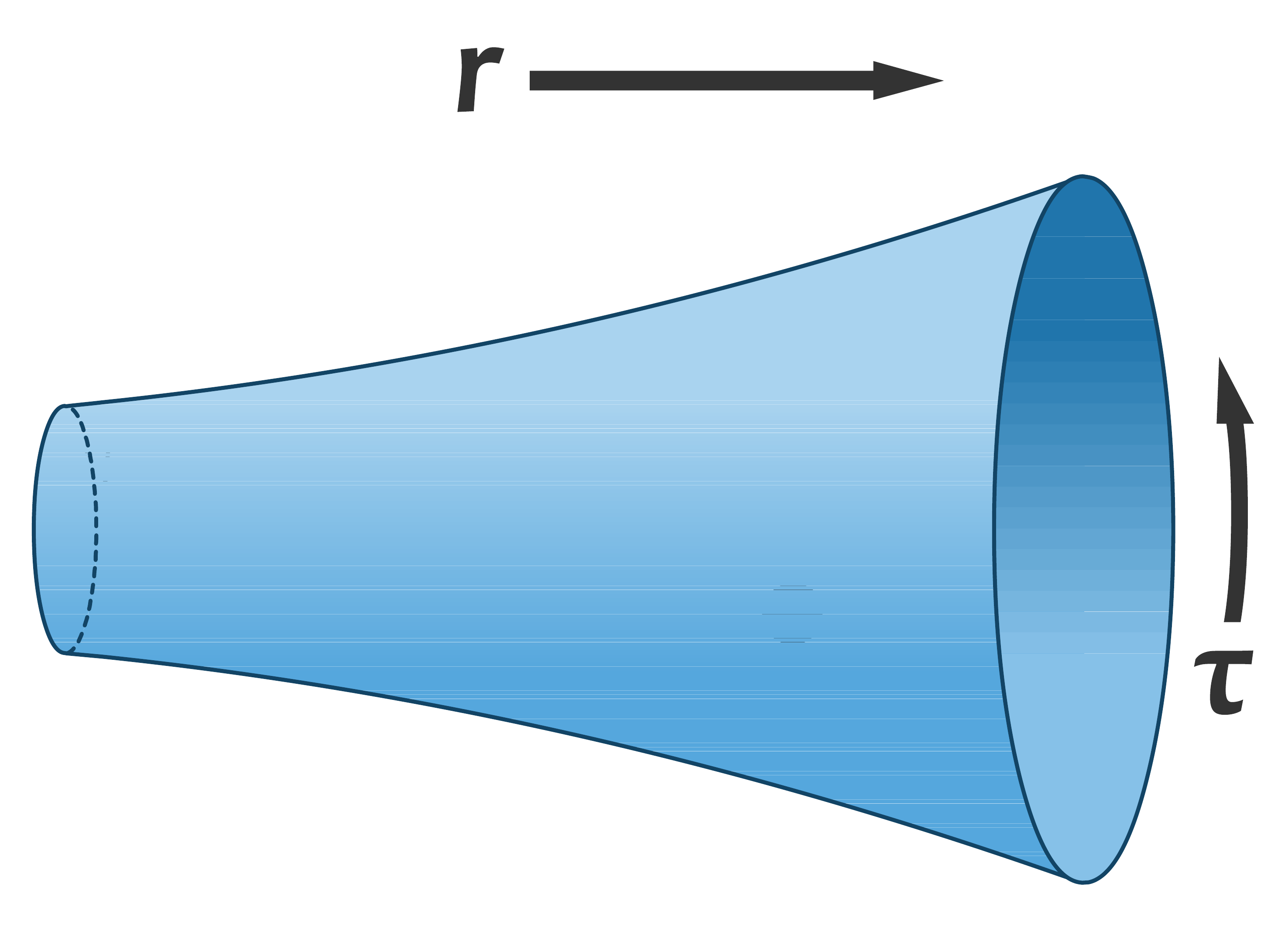}
\label{fig:tAdS}
}\hspace{1.5cm}
\subfigure[AdS-Schwarzschild]{%
\includegraphics[width=5cm]{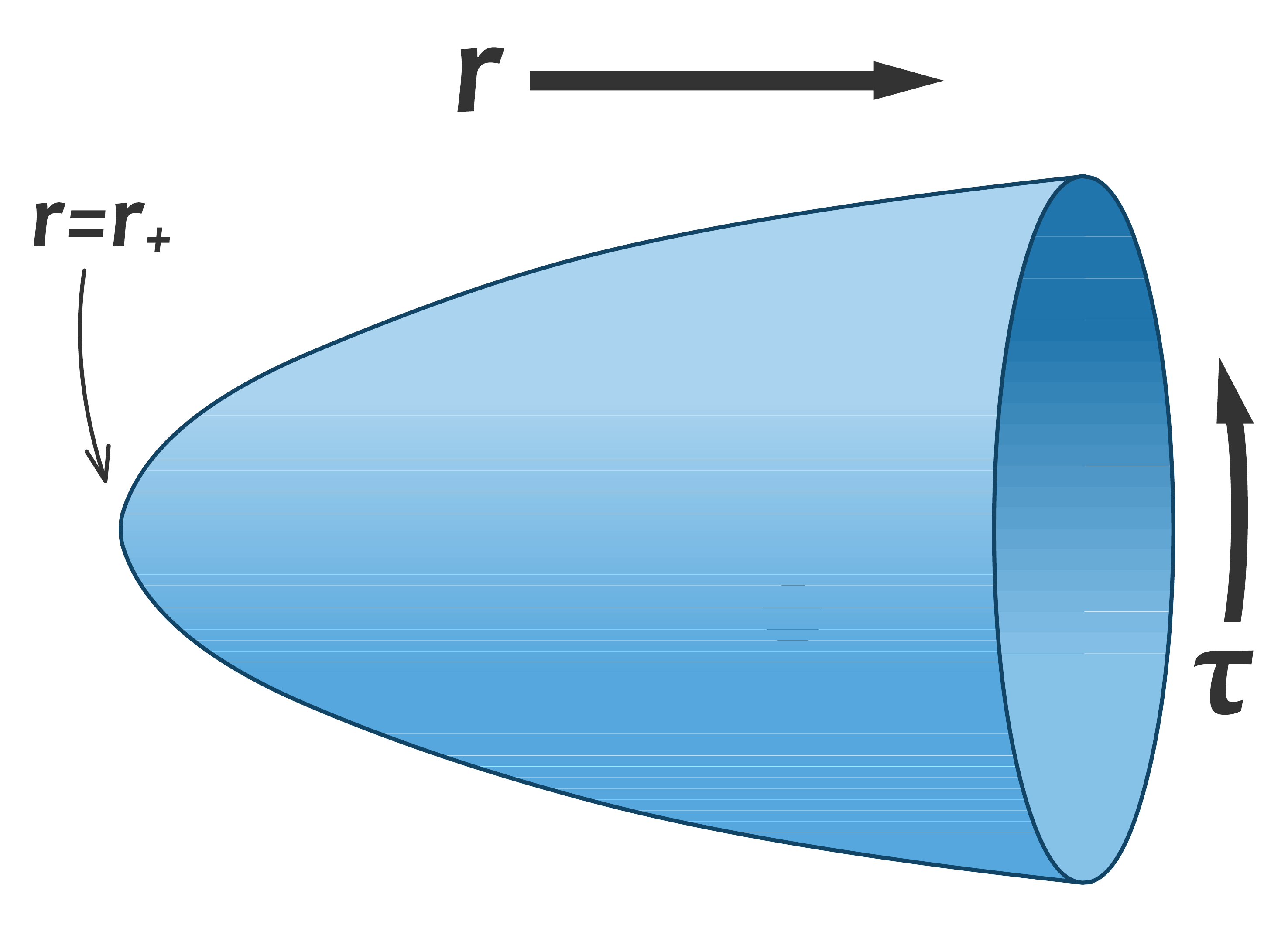}
\label{fig:AdSBH}
}}
\caption{The topology of the thermal AdS geometry and the AdS-Schwarzschild geometry.}
\end{figure}

For the low-temperature phase, the space is $X_1$ and the geometry is the thermal AdS with topology $S^1\times B^4$.
Then the loop $C$ is never contracted to zero in $X_1$ and $C$ is not a boundary of any string worldsheet $D$, see Figure~\ref{fig:tAdS}.
This immediately implies that
\begin{equation}
\langle{\cal P}\rangle=0
\end{equation}
in the large $N_c$ limit.
Thus at the low-temperature, the thermal AdS geometry is stable and the configuration $X_1$
corresponds to the confined phase.

On the other hand,
for the high-temperature phase, the relevant space is $X_2$ and
the geometry is the AdS-Schwarzschild black hole with topology $R^2\times S^3$.
In this case, the loop $C$ can be a boundary of a string worldsheet $D=B^2$, see Figure~\ref{fig:AdSBH}.
Then the (regularized) area of the surface gives the non-zero Polyakov loop expectation value,
\begin{equation}
\langle{\cal P}\rangle\neq0.
\end{equation}
Therefore, at the high-temperature, the the AdS-Schwarzschild black hole geometry is more stable
and the configuration $X_2$ corresponds to the deconfined phase.

\end{document}